\documentclass[tightenlines,eqsecnum,floats,aps,amsmath,amssymb,nofootinbib,
prd,showpacs]{revtex4}
\usepackage{amsmath,amssymb,amsfonts}
\usepackage{graphicx}
\usepackage{enumerate} % advanced enumerate environment
\usepackage{colordvi} % for color text
\usepackage{bm}% bold math
\newcommand{\bfv}{\mbox{\boldmath$v$}}
\newcommand{\bfx}{\mbox{\boldmath$x$}}
\newcommand{\bfk}{\mbox{\boldmath$k$}}
\newcommand{\bfp}{\mbox{\boldmath$p$}}
\newcommand{\bfq}{\mbox{\boldmath$q$}}
\newcommand{\bfr}{\mbox{\boldmath$r$}}
\newcommand{\bfs}{\mbox{\boldmath$s$}}
\newcommand{\calF}{\mathcal{F}}
\begin{document}
\title{Non-linear Evolution of Baryon Acoustic Oscillations from 
  Improved Perturbation Theory in Real and Redshift Spaces}
\vfill
\author{Atsushi Taruya$^{1,2}$, Takahiro Nishimichi$^3$, Shun Saito$^3$, 
Takashi Hiramatsu$^4$}
\bigskip
\address{$^1$Research Center for the Early Universe, School of Science, 
University of Tokyo, Bunkyo-ku, Tokyo 113-0033, Japan}
\address{$^2$Institute for the Physics and Mathematics of the Universe, 
University of Tokyo, Kashiwa, Chiba 277-8568, Japan}
\address{$^3$Department of Physics, University of Tokyo, 
113-0033, Japan}
\address{$^4$Institute for Cosmic Ray Research, University of Tokyo, Kashiwa, Chiba 277-8582, Japan}
\bigskip
\date{\today}
%
%%%%%%%%%%%%%%%%%%%%%%%%%%%%%%%%%%%%%%%%%%%%%%%%%%%%%%%%%%%%%%%%%%%%%%%
%%%%%%%%%%%%%%%%%%%%%%%%%%%%%%%%%%%%%%%%%%%%%%%%%%%%%%%%%%%%%%%%%%%%%%%
\begin{abstract}
We study the non-linear evolution of baryon acoustic oscillations 
in the matter power spectrum and correlation function from the improved 
perturbation theory (PT). Based on the framework of renormalized PT,  
which provides a non-perturbative way to treat the gravitational clustering 
of large-scale structure, we apply the {\it closure approximation} that 
truncates the infinite series of loop contributions at one-loop order, and 
obtain a closed set of integral equations for power spectrum and non-linear 
propagator. The resultant integral expressions are basically equivalent to 
those previously derived in the form of evolution equations, and they 
keep important non-perturbative properties which can dramatically improve 
the prediction of non-linear power spectrum. Employing the Born approximation, 
we then derive the analytic expressions for non-linear power spectrum 
and the predictions are made for non-linear evolution of baryon acoustic 
oscillations in power spectrum and correlation function. We find that 
the improved PT possesses a better convergence property compared with 
standard PT calculation. A detailed comparison between improved PT results 
and N-body simulations shows that a percent-level agreement is achieved 
in a certain range in power spectrum and in a rather wider range in 
correlation function. 
Combining a model of non-linear redshift-space distortion, we also 
evaluate the power spectrum and correlation function in 
redshift space. In contrast to the results in real space, the agreement 
between N-body simulations and improved PT predictions tends to be worse, 
and a more elaborate modeling for redshift-space distortion needs to 
be developed. Nevertheless, with currently existing model, we find that
the prediction of correlation function has a sufficient accuracy 
compared with the cosmic-variance errors for future galaxy surveys  
with volume of a few $h^{-3}$Gpc$^3$ at $z\gtrsim0.5$. 
\end{abstract}
%%%%%%%%%%%%%%%%%%%%%%%%%%%%%%%%%%%%%%%%%%%%%%%%%%%%%%%%%%%%%%%%%%%%%%%
%%%%%%%%%%%%%%%%%%%%%%%%%%%%%%%%%%%%%%%%%%%%%%%%%%%%%%%%%%%%%%%%%%%%%%%

\maketitle

%\vspace{0.6cm}
%-----------------------------------------------------------
\maketitle
%\vfill

%%%%%%%%%%%%%%%%%%%%%%%%%%%%%%%%%%%%%%%%%%%%%%%%%%%%%%%
%%%%%%%%%%%%%%%%%%%%%%%%%%%%%%%%%%%%%%%%%%%%%%%%%%%%%%%
\section{Introduction}
\label{sec:intro}
%%%%%%%%%%%%%%%%%%%%%%%%%%%%%%%%%%%%%%%%%%%%%%%%%%%%%%%
%%%%%%%%%%%%%%%%%%%%%%%%%%%%%%%%%%%%%%%%%%%%%%%%%%%%%%%

In the last decade, systematic measurements of the cosmic 
    microwave background anisotropies as well as large-scale structure 
    of the Universe have led to the establishment of the ``standard 
    cosmological model'' (e.g., \cite{Spergel:2003cb,Spergel:2006hy,
      Komatsu:2008hk,Tegmark:2003ud,Tegmark:2006az}). 
    The Universe is close to a flat geometry, 
    and is filled with the hypothetical cold dark matter (CDM) particles,  
    together with a small fraction of baryons, which serve as 
    the seeds of 
    structure formation of the Universe.  The most striking 
    feature in the standard cosmological model is that the energy 
    contents of the Universe is dominated by 
    the mysterious energy component called dark energy, which is 
    supposed to drive the late-time cosmic acceleration discovered by 
    the observation of distant supernovae (e.g., 
    \cite{Perlmutter:1998np,Riess:1998cb}). 

Currently, our understanding of the nature of dark energy is still 
    lacking. Although the observation is roughly consistent with cosmological 
    constant and with no evidence for time dependence of dark energy,  
    long-distance modifications of general relativity 
    have been proposed alternative to the dark energy and these reconcile 
    with the observation of late-time acceleration 
    (see \cite{Nojiri:2006ri,Durrer:2007re,Durrer:2008in,Koyama:2007rx} for 
    reviews). While a fully consistent model of modified gravity 
    has not yet been constructed (
    see \cite{Dvali:2000hr,Hu:2007nk,Starobinsky:2007hu} for popular models), 
    a possibility of break-down of general 
    relativity still remains and should be tested. 

To understand deeply the nature of dark energy or origin of cosmic 
    acceleration, a further observational study is definitely 
    important. There are 
    two comprehensive ways to distinguish between many 
    models of dark energy and discriminate the dark energy from modified
    gravity. One is to precisely measure the expansion history of the 
    Universe, and the other is to observe the growth of structure. 

Among various observational techniques, baryon acoustic oscillations 
    (BAOs) imprinted on the matter power spectrum or 
    two-point correlation function can be used as a standard ruler 
    to measure the cosmic expansion history 
    (e.g., \cite{Seo:2003pu,Blake:2003rh}, 
    see also \cite{Eisenstein:2005su,Huetsi:2005tp,Cole:2005sx, 
      Percival:2006gs,Percival:2007yw} for recent BAO measurements). 
    The characteristic scale of BAOs, which is determined by the 
    sound horizon scale of primeval baryon-photon fluid at the last 
    scattering surface \cite{Hu:1995en,Eisenstein:1997ik}, 
    is thought to be a robust measure and it 
    lies on the linear to quasi-linear regimes of the gravitational 
    clustering of large-scale structure \cite{Meiksin:1998ra,Seo:2005ys}. 
    With a percent-level 
    determination of the characteristic scale of BAOs, the expansion 
    history can be 
    tightly constrained, and the equation-of-state parameter of the 
    dark energy, $w_{\rm de}$, defined by the ratio of pressure to energy 
    density of dark energy, would be precisely determined 
    within the precision of a few \% level \cite{Albrecht:2006um,
      Peacock:2006kj}. 
    This is the basic reason why most of the planned and ongoing 
    galaxy redshift surveys aim at precisely measuring the BAOs 
    (e.g., \cite{Bassett:2005kn,Schlegel:2009hj,Hill:2008mv,Wang:2008hgb}).

While the robustness of the BAOs as a standard ruler has been  
repeatedly stated and emphasized in the literature,  in order to pursue an 
order-of-magnitude improvement, a precise theoretical modeling of BAOs  
    definitely plays an essential role for precision measurement of BAO 
    scale, and it needs to be investigated taking account of 
    the various systematic effects. Among these, the non-linear 
    clustering and redshift-space distortion effects as well as 
    the galaxy biasing cannot be neglected,  
    and affect the characteristic scale, although their effects 
    are basically moderate at the relevant wavenumber, 
    $k\lesssim0.3h\,$Mpc$^{-1}$.  

Recently, several analytic approaches to deal with the non-linear 
    clustering have been developed, complementary to the N-body simulations 
    \cite{Crocce:2005xy,Crocce:2005xz,Crocce:2007dt,Matsubara:2007wj,
      Matsubara:2008wx,McDonald:2006hf,Izumi:2007su, 
      Pietroni:2008jx,Matarrese:2007wc,Valageas:2003gm,Valageas:2006bi}. 
    In contrast to the standard analytical calculation with 
    perturbation theory (PT), these have been formulated in a non-perturbative 
    way with techniques resumming a class of infinite series of higher-order 
    corrections in perturbative calculation. 
    Thanks to its non-perturbative formulation, the applicable 
    range of the prediction is expected to be greatly improved,  
    and the non-linear evolution of baryon acoustic oscillations would 
    be accurately described with a percent-level precision.

The purpose of this paper is to investigate the viability of this analytic 
approach, focusing on a specific improved treatment. In the previous paper 
\cite{Taruya:2007xy}, we have applied a non-linear statistical method, which is 
widely accepted in the statistical theory of turbulence \cite{Leslie1973}, 
to the cosmological perturbation theory of large-scale structure. We have 
derived the non-perturbative expressions for the power spectrum, coupled with 
non-linear propagator, which effectively contain the information on 
the infinite series of higher-order corrections in the standard PT 
expansion. Based on this formalism, the analytic treatment of the 
non-perturbative expression is developed employing the Born approximation, 
and  the leading-order calculation of power spectrum is compared with 
N-body simulations in real space \cite{Nishimichi:2008ry}, finding that  
a percent-level agreement is achieved in a mildly non-linear regime 
(see also \cite{Carlson:2009it}). 
Here, we extend the analysis to those including the next-to-leading 
order corrections of Born approximation. 
In addition to the power spectrum, we will 
consider the two-point correlation function, paying a special attention  
on the baryon acoustic peak, i.e., a Fourier counterpart of BAOs in power 
spectrum. Further, we also discuss the non-linear clustering in redshift 
space, and the predictions of improved PT are compared with N-body 
results, combining a non-linear model of redshift-space distortion.   
We examine how well the present non-linear model accurately describe the 
systematic effects on BAOs and/or baryon acoustic peak.

This paper is organized as follows. In Sec.~\ref{sec:preliminaries},  
we briefly mention the basic equations for cosmological PT as our 
fundamental basis to deal with the non-linear gravitational clustering. 
We then discuss in some details in Sec.~\ref{sec:CLA} how to  
compute the non-linear power spectrum or two-point correlation functions. 
Starting from the discussions on standard treatment of perturbative 
calculation and its non-perturbative reformulation called renormalized PT, 
we introduce the closure approximation,  
which gives a consistent non-perturbative scheme to treat the 
infinite series of renormalized PT expansions, 
and obtain a closed set of non-perturbative expressions 
for power spectrum. Based on this, we present 
a perturbative treatment of the closed set of equations 
while keeping important non-perturbative properties.  
Section \ref{sec:PT_vs_N-body} gives the main result of this paper, 
in which a detailed comparison between 
improved PT calculation and N-body simulation is made, especially 
focusing on the non-linear evolution of BAOs. 
We compute the power spectrum and two-point correlation function
in both real and redshift spaces, and investigate the accuracy of 
both predictions by comparing improved PT with N-body results. 
Finally, section \ref{sec:conclusion} is devoted to the discussion 
and conclusion.

%%%%%%%%%%%%%%%%%%%%%%%%%%%%%%%%%%%%%%%%%%%%%%%%%%%%%%%
%%%%%%%%%%%%%%%%%%%%%%%%%%%%%%%%%%%%%%%%%%%%%%%%%%%%%%%
\section{Preliminaries}
\label{sec:preliminaries}
%%%%%%%%%%%%%%%%%%%%%%%%%%%%%%%%%%%%%%%%%%%%%%%%%%%%%%%
%%%%%%%%%%%%%%%%%%%%%%%%%%%%%%%%%%%%%%%%%%%%%%%%%%%%%%%

Throughout the paper, we consider the 
evolution of cold dark matter (CDM) plus baryon systems neglecting 
the tiny fraction of (massive) neutrinos. 
Owing to the single-stream approximation of the collisionless Boltzmann
equation, which is thought to be a quite accurate approximation 
on large scales, the evolution of 
the CDM plus baryon system can be treated as the irrotational and 
pressureless fluid system whose governing equations are continuity 
and Euler equations in addition to the Poisson equation (see 
Ref.~\cite{Bernardeau:2001qr} for review). In the Fourier 
representation, these equations are further reduced to a 
more compact form. Let us introduce the two-component vector 
(e.g.,\cite{Crocce:2005xy}):  
%%%%%%%%%%%%%%%%%%%%%%%%%%%%%%%%%%%%%%%%%%%%%%%%%%%%%%%%%%%%%%%%%%%%%%
\begin{equation}
\Phi_a(\bfk;t)=\Bigl(\delta(\bfk;t), \,\,
-\frac{\theta(\bfk;t)}{f(t)} \Bigr),  
\end{equation}
%%%%%%%%%%%%%%%%%%%%%%%%%%%%%%%%%%%%%%%%%%%%%%%%%%%%%%%%%%%%%%%%%%%%%%
where the subscript $a=1,\,2$ selects the density and the velocity 
components of CDM plus baryons, with $\delta$ and 
$\theta(\bfx)\equiv\nabla\cdot \bfv(\bfx)/(a\,H)$, where 
$a$ and $H$ are the scale factor of the Universe and the Hubble 
parameter, respectively.    
The function $f(t)$ is given by $f(t)\equiv d\ln D(t)/d\ln a$ and 
the quantity $D(t)$ being the linear growth factor. 
Then, in terms of the new time variable 
$\eta\equiv\ln D(t)$, the evolution equation for the vector quantity 
$\Phi_a(\bfk;t)$ becomes 
%%%%%%%%%%%%%%%%%%%%%%%%%%%%%%%%%%%%%%%%%%%%%%%%%%%%%%%%%%%%%%%%%%%%%%
\begin{equation}
\left[\delta_{ab}\,\frac{\partial}{\partial \eta}+\Omega_{ab}(\eta)\right]
\Phi_b(\bfk;\eta)=\int\frac{d^3\bfk_1\,d^3\bfk_2}{(2\pi)^3}
\delta_D(\bfk-\bfk_1-\bfk_2)\,\gamma_{abc}(\bfk_1,\bfk_2)\,
\Phi_b(\bfk_1;\eta)\,\Phi_c(\bfk_2;\eta),
\label{eq:vec_fluid_eq}
\end{equation}
%%%%%%%%%%%%%%%%%%%%%%%%%%%%%%%%%%%%%%%%%%%%%%%%%%%%%%%%%%%%%%%%%%%%%%
where $\delta_D$ is the Dirac delta function. 
Here and in what follows, 
we use the summation convention that the repetition of the same 
subscripts  indicates the sum over the whole vector components. 
The time-dependent matrix $\Omega_{ab}(\eta)$ is given by
%%%%%%%%%%%%%%%%%%%%%%%%%%%%%%%%%%%%%%%%%%%%%%%%%%%%%%%%%%%%%%%%%%%%%%
\begin{equation}
\Omega_{ab}(\eta)=\left(
\begin{array}{cc}
{\displaystyle 0} & {\displaystyle -1 }
\\
{\displaystyle -\frac{3}{2f^2}\Omega_{\rm m}(\eta)} \quad&\quad 
{\displaystyle \frac{3}{2f^2}\Omega_{\rm m}(\eta)-1}
\end{array}
\right).
\label{eq:matrix_M}
\end{equation}
%%%%%%%%%%%%%%%%%%%%%%%%%%%%%%%%%%%%%%%%%%%%%%%%%%%%%%%%%%%%%%%%%%%%%%
The quantity $\Omega_{\rm m}(\eta)$ is the density parameter of CDM plus 
baryons at a given time. Each component of the vertex function 
$\gamma_{abc}$ becomes
%%%%%%%%%%%%%%%%%%%%%%%%%%%%%%%%%%%%%%%%%%%%%%%%%%%%%%%%%%%%%%%%%%%%%%
\begin{eqnarray}
\gamma_{abc}(\bfk_1,\bfk_2) =
\left\{
\begin{array}{ccl} 
\frac{1}{2}\left\{1+\frac{\bfk_2\cdot\bfk_1}{|\bfk_2|^2}\right\} &;& 
(a,b,c)=(1,1,2) 
\\
\\
\frac{1}{2}\left\{1+\frac{\bfk_1\cdot\bfk_2}{|\bfk_1|^2}\right\} &;& 
(a,b,c)=(1,2,1) 
\\
\\
\frac{(\bfk_1\cdot\bfk_2)|\bfk_1+\bfk_2|^2}{2|\bfk_1|^2|\bfk_2|^2} &;& 
(a,b,c)=(2,2,2) 
\\
\\
     0                       &;&  \mbox{otherwise}
\end{array}
\right..
\label{eq:def_Gamma}
\end{eqnarray}
%%%%%%%%%%%%%%%%%%%%%%%%%%%%%%%%%%%%%%%%%%%%%%%%%%%%%%%%%%%%%%%%%%%%%%

Note that the formal solution of $\Phi_a$ can be obtained from 
Eq.~(\ref{eq:vec_fluid_eq}) and is expressed as (e.g., 
\cite{Bernardeau:2001qr,Crocce:2005xy})
%%%%%%%%%%%%%%%%%%%%%%%%%%%%%%%%%%%%%%%%%%%%%%%%%%%%%%%%%%%%%%%%%%%%%%
\begin{equation}
\Phi_a(\bfk;\eta)=g_{ab}(\eta,\eta_0)\,u_b\,\delta_0(\bfk) +
\int_{\eta_0}^{\eta} d\eta' g_{ab}(\eta,\eta')\,
\int\frac{d^3\bfk_1\,d^3\bfk_2}{(2\pi)^3}\delta_D(\bfk-\bfk_1-\bfk_2)\,
\gamma_{bcd}(\bfk_1,\bfk_2)\Phi_c(\bfk_1;\eta')\Phi_d(\bfk_2;\eta').
\label{eq:formal_sol}
\end{equation}
%%%%%%%%%%%%%%%%%%%%%%%%%%%%%%%%%%%%%%%%%%%%%%%%%%%%%%%%%%%%%%%%%%%%%%
Here, the quantity $u_a$ is the constant vector which specifies the 
initial condition (see next section), and the quantity $g_{ab}$ denotes 
the linear propagator satisfying the following equation: 
%%%%%%%%%%%%%%%%%%%%%%%%%%%%%%%%%%%%%%%%%%%%%%%%%%%%%%%%%%%%%%%%%%%%%%
\begin{equation}
\left[\delta_{ab}\frac{\partial}{\partial\eta}+\Omega_{ab}(\eta)\right]
g_{bc}(\eta,\eta')=0, 
\label{eq:linear_prop}
\end{equation}
%%%%%%%%%%%%%%%%%%%%%%%%%%%%%%%%%%%%%%%%%%%%%%%%%%%%%%%%%%%%%%%%%%%%%%
with the boundary condition $g_{ab}(\eta,\eta)=\delta_{ab}$. The 
quantity $\delta_0$ is the random density field given at an 
early time $\eta_0$, which is assumed to obey the Gaussian statistic. 
The power spectrum of density field is defined as 
%%%%%%%%%%%%%%%%%%%%%%%%%%%%%%%%%%%%%%%%%%%%%%%%%%%%%%%%%%%%%%%%%%%%%%
\begin{equation}
\langle\delta_0(\bfk)\delta_0(\bfk')\rangle = 
(2\pi)^3\,\delta_D(\bfk+\bfk')\,P_0(k). 
\label{eq:def_P_0}
\end{equation}
%%%%%%%%%%%%%%%%%%%%%%%%%%%%%%%%%%%%%%%%%%%%%%%%%%%%%%%%%%%%%%%%%%%%%%

Eq.~(\ref{eq:vec_fluid_eq})  or (\ref{eq:formal_sol}) is the 
fundamental building block of large-scale structure, 
and the three quantities $\gamma_{abc}$, $g_{ab}$ and $P_0u_au_b$ 
introduced here constitute the basic pieces of standard PT.  
The graphical representation of them is shown in 
Fig.~\ref{fig:diagram_basic} (see also Ref.~\cite{Crocce:2005xy}).

%%%%%%%%%%%%%%%%%%%%%%%%%%%%%%%%%%%%%%%%%%%%%%%%%%%%%%%
%%%%%%%%%%%%%%%%%%%%%%%%%%%%%%%%%%%%%%%%%%%%%%%%%%%%%%%
\section{Improved Perturbation Theory}
\label{sec:CLA}
%%%%%%%%%%%%%%%%%%%%%%%%%%%%%%%%%%%%%%%%%%%%%%%%%%%%%%%
%%%%%%%%%%%%%%%%%%%%%%%%%%%%%%%%%%%%%%%%%%%%%%%%%%%%%%%

%-%-%-%-%-%-%-%-%-%-%-%-%-%-%-%-%-%-%-%-%-%-%-%-%-%-%-%
%-%-%-%-%-%-%-%-%-%-%-%-%-%-%-%-%-%-%-%-%-%-%-%-%-%-%-%
\subsection{Standard PT vs. Renormalized PT}
\label{subsec:SPT_vs_RPT}
%-%-%-%-%-%-%-%-%-%-%-%-%-%-%-%-%-%-%-%-%-%-%-%-%-%-%-%
%-%-%-%-%-%-%-%-%-%-%-%-%-%-%-%-%-%-%-%-%-%-%-%-%-%-%-%

%%%%%%%%%%%%%%%%%%%%%%%%%%%%%%%%%%%%%%%%%%%%%%%%%%%%%%%%%%%
\begin{figure}[t]
\begin{center}
\includegraphics[width=4cm,angle=0]{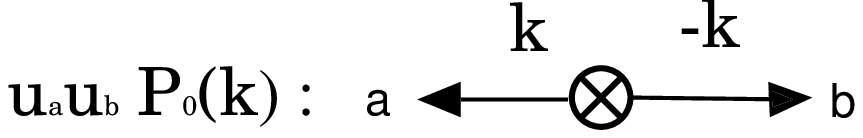}
\hspace*{0.75cm}
\includegraphics[width=4cm,angle=0]{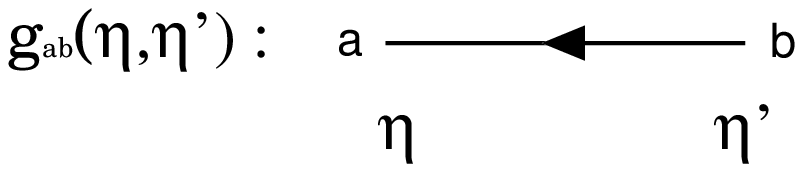}
\hspace*{0.75cm}
\includegraphics[width=5cm,angle=0]{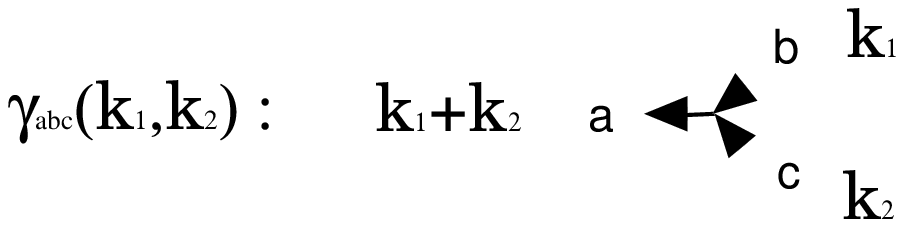}
\end{center}

\vspace*{-0.5cm}

\caption{Diagrammatic notion of the initial power spectrum (left),  
linear propagator (middle), and {\it tree} vertex (right).  The 
linear propagator satisfies the equation (\ref{eq:linear_prop}) 
with boundary condition $g_{ab}(\eta,\eta)=\delta_{ab}$. 
The explicit expression of 
vertex function $\gamma_{abc}$ is given by Eq.~(\ref{eq:def_Gamma}).
\label{fig:diagram_basic}}
\end{figure}
%%%%%%%%%%%%%%%%%%%%%%%%%%%%%%%%%%%%%%%%%%%%%%%%%%%%%%%%%%%

In this paper, we are especially concerned with 
the non-linear evolution of the two-point statistics, defined 
as the ensemble average of $\Phi_a$:  
%%%%%%%%%%%%%%%%%%%%%%%%%%%%%%%%%%%%%%%%%%%%%%%%%%%%%%%%%%%%%%%%%%%%%%
\begin{equation}
\Bigl\langle\Phi_a(\bfk;\eta)\Phi_b(\bfk';\eta')\Bigr\rangle=
(2\pi)^3\,\delta_{\rm D}(\bfk+\bfk')\,P_{ab}(|\bfk|;\eta,\eta')\,;
\quad \eta\geq\eta'.
\label{eq:def_Pk} 
\end{equation}
%%%%%%%%%%%%%%%%%%%%%%%%%%%%%%%%%%%%%%%%%%%%%%%%%%%%%%%%%%%%%%%%%%%%%%
In the above, there are four types of power spectra, 
$P_{11}$, $P_{12}$, $P_{21}$ and $P_{22}$, which respectively correspond to 
the auto- and cross-power spectra, 
$P_{\delta\delta}$, $-P_{\delta\theta}/f$, $-P_{\theta\delta}/f$ and 
$P_{\theta\theta}/f^2$. Note that in general we have 
$P_{12}\ne P_{21}$ unless $\eta=\eta'$.

Consider how to compute the power spectrum based on the 
analytic treatment. 
In the standard treatment of the perturbation theory, we first assume
that the field $\Phi_a$ is a small perturbed quantity and it is expanded as 
%%%%%%%%%%%%%%%%%%%%%%%%%%%%%%%%%%%%%%%%%%%%%%%%%%%%%%%%%%%%%%%%%%%%%%
\begin{equation}
\Phi_a(\bfk;\eta) = \Phi_a^{(1)}(\bfk;\eta)  + \Phi_a^{(2)}(\bfk;\eta)  
+ \Phi_a^{(3)}(\bfk;\eta)+\cdots.
\end{equation}
%%%%%%%%%%%%%%%%%%%%%%%%%%%%%%%%%%%%%%%%%%%%%%%%%%%%%%%%%%%%%%%%%%%%%%
The explicit functional form of the quantity $\Phi_a^{(n)}$ 
is systematically derived through the order-by-order treatment 
of Eq.~(\ref{eq:vec_fluid_eq}). Substituting the above expansion into 
the definition (\ref{eq:def_Pk}) and evaluating 
it perturbatively, the power spectrum $P_{ab}(k;\eta,\eta)$, shortly 
abbreviated as $P_{ab}(k;\eta)$, is schematically expressed as
%%%%%%%%%%%%%%%%%%%%%%%%%%%%%%%%%%%%%%%%%%%%%%%%%%%%%%%%%%%%%%%%%%%%%%
\begin{equation}
P_{ab}(k;\eta) = e^{2\eta} u_au_b\,P_0(k) + 
P^{\rm 1\mbox{-}loop}_{ab}(k;\eta) + 
P^{\rm 2\mbox{-}loop}_{ab}(k;\eta) + \cdots.
\end{equation}
%%%%%%%%%%%%%%%%%%%%%%%%%%%%%%%%%%%%%%%%%%%%%%%%%%%%%%%%%%%%%%%%%%%%%%
where we chose $u_a=(1,1)$, 
which implies that the growing-mode solution is imposed 
at the initial condition\footnote{Strictly speaking, this statement is valid 
only when the universe at an early time is approximately described by the 
Einstein-de Sitter universe }. The function $P_0(k)$ is the linear 
power spectrum given at an early time, obtained from the first-order 
quantity $\Phi_a^{(1)}$ (see Eq.~(\ref{eq:def_P_0} for definition). 
The subsequent terms 
$P^{\rm 1\mbox{-}loop}_{ab}$ and $P^{\rm 2\mbox{-}loop}_{ab}$ represent
the corrections to the linear-order perturbation, 
arising from the higher-order quantities, $\Phi_a^{(2)}$, 
$\Phi_a^{(3)}$, $\cdots$. In terms of the basic pieces of the 
diagrams shown in Fig.~\ref{fig:diagram_basic},  the corrections 
$P_{ab}^{\rm 1\mbox{-}loop}(k)$ and $P_{ab}^{\rm 2\mbox{-}loop}(k)$ 
can be diagrammatically written as the one-loop 
and two-loop diagrams, i.e., connected diagrams 
including one and two closed loops 
(e.g., see Fig.5 in Ref.~\cite{Crocce:2005xy}), and they  
are roughly proportional to 
$P_0\Delta_0^2$ and $P_0\Delta_0^4$, where 
$\Delta_0^2=k^3P_0(k)/(2\pi^2)$. 
The explicit expressions for the power spectra 
together with the solutions of higher-order perturbation 
are summarized in Appendix \ref{app:SPT}.

It should be noted that in the standard PT expansion, 
the positivity of the perturbative corrections is not guaranteed. 
    As we show later, 
    the one- and two-loop contributions change the sign depending on the 
    scale, and the absolute values of their amplitudes become 
    comparable at lower redshift. In this respect, the 
    standard PT has a poor convergence property, and 
    the improvement of PT predictions may not be always guaranteed 
    even including the higher-order corrections.

By contrast, renormalized PT\footnote{In this paper, we intend to 
  make a clear distinction between the terms 'renormalized PT' and 'RPT'. 
  While the renormalized PT indicates the general non-perturbative formalism 
  developed by Ref.~\cite{Crocce:2005xy}, the RPT is meant to imply the 
  practical approximation method for computing the power spectrum 
  based on the renormalized PT, which has been developed 
  by Ref.~\cite{Crocce:2007dt} (see Appendix \ref{app:comparison}). } 
re-organizes 
the naive expansions of the standard PT by introducing the non-perturbative 
statistical quantities \cite{Crocce:2005xy}. In terms of these quantities, 
partial resummation of the naive expansion series is made, and 
the resultant convergence of the expansions is dramatically 
improved. In the renormalized PT, the power spectrum $P_{ab}(k;\eta)$ is 
expressed in the form as 
%%%%%%%%%%%%%%%%%%%%%%%%%%%%%%%%%%%%%%%%%%%%%%%%%%%%%%%%%%%%%%%%%%%%%%
\begin{equation}
P_{ab}(k;\eta) = 
G_{ac}(k|\eta,\eta_0)G_{bd}(k|\eta,\eta_0)P_{cd}(k;\eta_0) + 
P_{ab}^{\rm (MC)}(k;\eta,\eta_0) 
\label{eq:RPT_expansion} 
\end{equation}
%%%%%%%%%%%%%%%%%%%%%%%%%%%%%%%%%%%%%%%%%%%%%%%%%%%%%%%%%%%%%%%%%%%%%%
with $\eta_0$ being the time at which initial condition is imposed. 
Here, $P_{cd}(k;\eta_0)$ is the power spectrum given at an early time 
$\eta_0$. The quantity $G_{ab}$ is one of the non-perturbative 
statistical quantities called non-linear propagator,  
together with the non-linear power spectrum. It is defined by 
%%%%%%%%%%%%%%%%%%%%%%%%%%%%%%%%%%%%%%%%%%%%%%%%%%%%%%%%%%%%%%%%%%%%%%
\begin{equation}
\Bigl\langle\frac{\delta\Phi_a(\bfk;\eta)}{\delta\Phi_b(\bfk';\eta')}
\Bigr\rangle=\delta_{\rm D}(\bfk-\bfk')\,G_{ab}(|\bfk|\,|\eta,\eta') 
\,;
\quad \eta\geq\eta',
\end{equation}
%%%%%%%%%%%%%%%%%%%%%%%%%%%%%%%%%%%%%%%%%%%%%%%%%%%%%%%%%%%%%%%%%%%%%%
where $\delta$ stands for a functional derivative. The propagator $G_{ab}$ 
describes the influence of an infinitesimal disturbance for 
$\Phi_a(\bfk';\eta')$ on $\Phi_a(\bfk;\eta)$, and it coincides with 
the linear propagator $g_{ab}$ in the limit $k\to0$. Note that there is 
another non-perturbative statistical quantity called full 
vertex, which is the non-linear counterpart of the vertex function 
$\gamma_{abc}$ \cite{Crocce:2005xy}.

In the expression (\ref{eq:RPT_expansion}),  
    the term $P_{ab}^{\rm (MC)}$ represents the corrections 
    coming from the loop diagrams. In contrast to the standard PT, 
    the loop diagrams in $P_{ab}^{\rm (MC)}$ are whole 
    {\it irreducible}, as the result of renormalization or re-organization. 
    Further, each of the irreducible diagrams consists 
    of the non-perturbative quantities of non-linear power 
    spectrum, non-linear propagator and full vertex. 
    In this respect, renormalized PT is a fully non-perturbative 
    formulation, and even the expansions truncated at some levels 
    still contain the higher-order effects of non-linear gravitational 
    evolution. This is the basic reason why the convergence properties 
    in the renormalized PT are expected to be improved. 
    As a trade-off, however, a straightforward application of 
    renormalized PT seems difficult because of its non-perturbative 
    formulation. While the term $P_{ab}^{\rm MC}$ collects only the 
    irreducible diagrams, it is expressed as an infinite sum of 
    the loop diagrams, each of which involve the non-linear power spectrum 
    itself. In practice, the approximation or simplification is needed to 
    evaluate the expressions (\ref{eq:RPT_expansion}), which we will 
    discuss in next subsection.

%-%-%-%-%-%-%-%-%-%-%-%-%-%-%-%-%-%-%-%-%-%-%-%-%-%-%-%
%-%-%-%-%-%-%-%-%-%-%-%-%-%-%-%-%-%-%-%-%-%-%-%-%-%-%-%
\subsection{Closure approximation}
%-%-%-%-%-%-%-%-%-%-%-%-%-%-%-%-%-%-%-%-%-%-%-%-%-%-%-%
%-%-%-%-%-%-%-%-%-%-%-%-%-%-%-%-%-%-%-%-%-%-%-%-%-%-%-%

In this subsection, taking a great advantage of the 
formulation of renormalized PT, we discuss how to approximately treat 
Eq.~(\ref{eq:RPT_expansion}) without losing its non-perturbative aspect 
as much as possible.

In the framework of renormalized PT, 
the non-perturbative effects on the power spectrum 
are largely attributed to the non-linear propagator. Thus, it seems 
essential to give a framework to treat both the non-linear propagator 
and power spectrum on an equal footing. 
As it has been pointed out by Ref.~\cite{Crocce:2005xy}, a similar 
    kind of the renormalized 
    expansion to the power spectrum (\ref{eq:RPT_expansion}) 
    can be made for the non-linear propagator:  
%%%%%%%%%%%%%%%%%%%%%%%%%%%%%%%%%%%%%%%%%%%%%%%%%%%%%%%%%%%%%%%%%%%%%%
\begin{equation}
G_{ab}(k|\eta,\eta')=g_{ab}(\eta,\eta') + G_{ab}^{\rm (MC)}(k;\eta,\eta'), 
\label{eq:RPT_expansion2}
\end{equation}
%%%%%%%%%%%%%%%%%%%%%%%%%%%%%%%%%%%%%%%%%%%%%%%%%%%%%%%%%%%%%%%%%%%%%%
where the term $G_{ab}^{\rm (MC)}$ represents the mode-coupling 
correction, which is also made of the infinite sum of irreducible 
loop diagrams.

In order to give a self-consistent treatment for 
both Eqs.~(\ref{eq:RPT_expansion}) and (\ref{eq:RPT_expansion2}),  
a simple but transparent approach is to first (i) adopt the tree-level 
approximation of the full vertex function, and to (ii) apply the 
truncation procedure to the mode-coupling terms. This treatment  
has been frequently used in the statistical theory of turbulence 
in order to deal with the Navier-stokes equation, 
and is called {\it closure approximation} \cite{Taruya:2007xy}. 
In the first approximation (i), the 
full vertex function is simply replaced with the linear-order one, 
i.e., $\gamma_{abc}$ defined in Eq.~(\ref{eq:def_Gamma}). As for the 
truncation (ii), the simplest choice is to keep the one-loop 
renormalized diagram only, and to discard all other contributions.

With this approximation, 
the mode-coupling terms in $P_{ab}$ and $G_{ab}$ are 
simply described by 
$P_{ab}^{\rm(MC)}\simeq P_{ab}^{\rm(MC,1\mbox{-}loop)}$ and 
$G_{ab}^{\rm(MC)}\simeq G_{ab}^{\rm(MC,1\mbox{-}loop)}$. 
The analytical expressions for the one-loop contributions becomes 
\cite{Taruya:2007xy}
%%%%%%%%%%%%%%%%%%%%%%%%%%%%%%%%%%%%%%%%%%%%%%%%%%%%%%%%%%%%%%%%%%%%%%
\begin{eqnarray}
P_{ab}^{\rm(MC,1\mbox{-}loop)}(k;\eta,\eta') &=& 
\int_{\eta_0}^{\eta}d\eta_1 \int_{\eta_0}^{\eta'}d\eta_2\,
G_{ac}(k|\eta,\eta_1)\,G_{bd}(k|\eta',\eta_2)\,\,\Phi_{cd}(k;\eta_2,\eta_1),  
\label{eq:P_MC_CLA}
\\
G_{ab}^{\rm(MC,1\mbox{-}loop)}(k;\eta,\eta') &= &
\int_{\eta_1}^{\eta} d\eta_1 \int_{\eta'}^{\eta_1} d\eta_2
\,g_{ac}(\eta,\eta_1)\,G_{sb}(k|\eta_2,\eta')
\,
\nonumber\\
&&
\times\,\,4\int \frac{d^3\bfq}{(2\pi)^3} 
\gamma_{cpq}(\bfq,\bfk-\bfq)\,P_{pr}(q;\eta_1,\eta_2)
\,G_{ql}(|\bfk-\bfq||\eta_1,\eta_2)\,\gamma_{lrs}(-\bfq,\bfk).
\label{eq:G_MC_CLA}
\end{eqnarray}
%%%%%%%%%%%%%%%%%%%%%%%%%%%%%%%%%%%%%%%%%%%%%%%%%%%%%%%%%%%%%%%%%%%%%%

\begin{figure}[t]
\begin{center}
\includegraphics[width=14cm,angle=0]{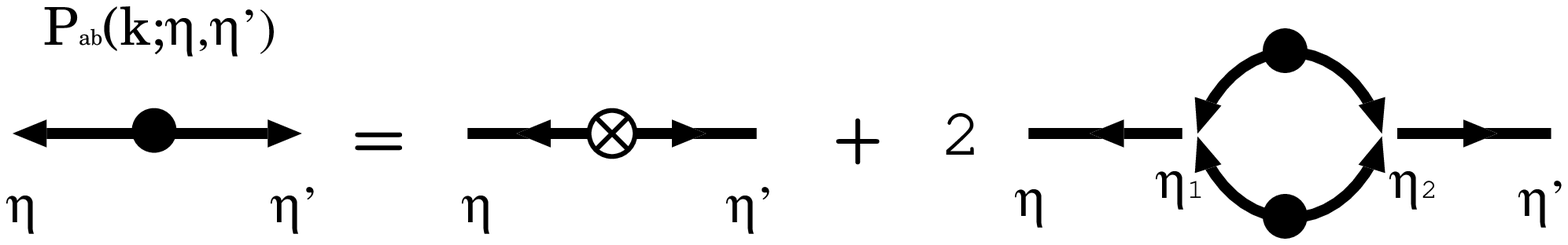}

\vspace*{0.5cm}

\includegraphics[width=14cm,angle=0]{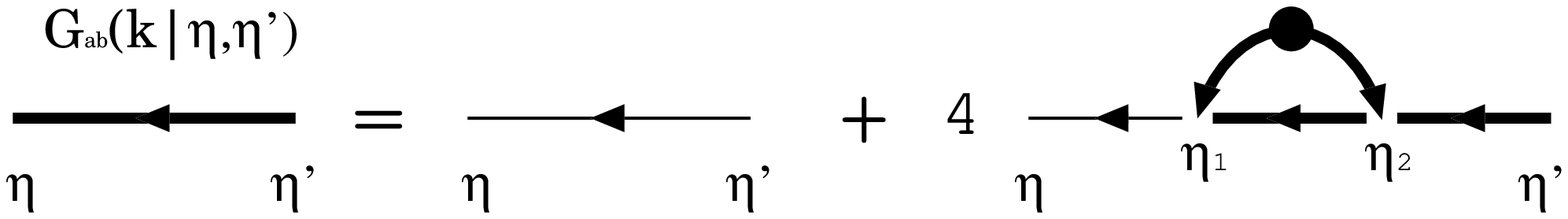}
\end{center}

\vspace*{-0.3cm}

\caption{Diagrammatic representation of the power spectrum and 
  non-linear propagator in closure approximation. The thick lines 
  represent the full-order quantities, while the thin line indicates 
  the linear-order one. The second terms 
  at right-hand side indicate the irreducible one-loop diagrams 
  of the mode-coupling terms, 
  $P_{ab}^{\rm(MC,1\mbox{-}loop)}$ and $G_{ab}^{\rm(MC,1\mbox{-}loop)}$.  
  In the renormalized PT, the mode-coupling term is expressed as 
  an infinite sum of the irreducible loop corrections. 
  Truncating the infinite sum at one-loop order and adopting the 
  tree-level approximation of the full vertex function, we obtain 
  the closed system of power spectrum and propagator, as shown in 
  the figure. 
\label{fig:diagram_CLA}}
\end{figure}

The integrand in $P_{ab}^{\rm(MC,1\mbox{-}loop)}$ 
contain the function $\Phi(k;\eta_1,\eta_2)$,  
which represents the non-linear mode-coupling between different Fourier 
modes, given by 
%%%%%%%%%%%%%%%%%%%%%%%%%%%%%%%%%%%%%%%%%%%%%%%%%%%%%%%%%%%%%%%%%%%%%%
\begin{eqnarray}
&&\Phi_{ab}(k;\eta_1,\eta_2)=2\,\int\frac{d^3\bfq}{(2\pi)^3}\,
\gamma_{ars}(\bfq,\bfk-\bfq)\,\gamma_{bpq}(\bfq,\bfk-\bfq)
\nonumber\\
&&\quad\quad\times\,
\Bigl\{
P_{pr}(q;\eta_1,\eta_2)P_{qs}(|\bfk-\bfq|;\eta_1,\eta_2)\,
\Theta(\eta_1-\eta_2) + 
P_{rp}(q;\eta_2,\eta_1)P_{sq}(|\bfk-\bfq|;\eta_2,\eta_1)\,
\Theta(\eta_2-\eta_1)
\Bigr\}. 
\label{eq:mode_coupling}
\end{eqnarray}
%%%%%%%%%%%%%%%%%%%%%%%%%%%%%%%%%%%%%%%%%%%%%%%%%%%%%%%%%%%%%%%%%%%%%%
Note that the mode-coupling function $\Phi$ 
possesses the following symmetry: 
$\Phi_{ab}(k;\eta_1,\eta_2)=\Phi_{ba}(k;\eta_2,\eta_1)$. 
The corresponding diagrams to the integral expressions for 
power spectrum and non-linear propagator, 
i.e., Eqs.~(\ref{eq:RPT_expansion}) and (\ref{eq:RPT_expansion2}) 
with mode-coupling terms (\ref{eq:P_MC_CLA}) and (\ref{eq:G_MC_CLA}), 
are shown in Fig.~\ref{fig:diagram_CLA}. 
%%%%%%%%%%%%%%%%%%%%%%%%%%%%%%%%%%%%%%%%%%%%%%%%%%%%%%%%%%%
%%%%%%%%%%%%%%%%%%%%%%%%%%%%%%%%%%%%%%%%%%%%%%%%%%%%%%%%%%%

It is worth mentioning 
that the integral equations (\ref{eq:RPT_expansion}) and 
(\ref{eq:RPT_expansion2}) with truncated mode-coupling terms 
(\ref{eq:P_MC_CLA}) and (\ref{eq:G_MC_CLA}) can be 
recast in the form of the integro-differential equations, 
and both the power spectrum and non-linear propagator can be computed 
by solving the evolution equations. 
This forward treatment seems especially suited for the full non-linear 
treatment of closure approximation and would be faster 
than directly treating the integral equations.  
Numerical algorithm to solve evolution equations, 
together with preliminary results, is presented in details 
in Ref.~\cite{Hiramatsu:2009ki} (see also \cite{Koyama:2009me}).

In the present paper, we are especially concerned with 
the evolution of BAOs around $k\lesssim0.4h\,$Mpc$^{-1}$,    
where the non-linearity of gravitational clustering is rather mild, and 
the analytical treatment even involving some approximations 
is still useful. Here, employing the Born approximation,  
we analytically evaluate the integral equations (\ref{eq:RPT_expansion}) 
and (\ref{eq:P_MC_CLA}) \cite{Taruya:2007xy}. 
A fully numerical study on BAOs without Born approximation 
will be discussed in a separate paper.

The Born approximation is the 
iterative approximation scheme in which the leading-order solutions are first 
obtained by replacing the quantities in the non-linear integral terms 
with linear-order ones. The solutions can be improved by 
repeating the iterative substitution of the leading-order solutions 
into the non-linear integral terms. 
Consider the time evolution of the power spectrum started from 
the time $\eta_0$. For a sufficiently small value of $\eta_0$, 
the early-time evolution of power spectrum is well-approximated by the 
linear theory. Assuming the growing-mode initial condition, we have
%%%%%%%%%%%%%%%%%%%%%%%%%%%%%%%%%%%%%%%%%%%%%%%%%%%%%%%%%%%%%%%%%%%%%
\begin{equation}
P_{ab}(k;\eta_0) = e^{2\eta_0}\,u_a\,u_b\,P_0(k)
\label{eq:initial_Pk}
\end{equation}
%%%%%%%%%%%%%%%%%%%%%%%%%%%%%%%%%%%%%%%%%%%%%%%%%%%%%%%%%%%%%%%%%%%%%
with $u_a=(1,\,1)$.
Then, substituting Eq.~(\ref{eq:initial_Pk}) into (\ref{eq:RPT_expansion}),  
the iterative evaluation of the the integral 
equations (\ref{eq:RPT_expansion}) with (\ref{eq:P_MC_CLA}) by the 
Born approximation leads to \cite{Taruya:2007xy}
%%%%%%%%%%%%%%%%%%%%%%%%%%%%%%%%%%%%%%%%%%%%%%%%%%%%%%%%%%%%%%%%%%%%%
\begin{eqnarray}
P_{ab}(k;\eta)=\widetilde{G}_{a}(k|\eta,\eta_0) 
\widetilde{G}_{b}(k|\eta,\eta_0) e^{2\eta_0} P_0(k)+ 
P_{ab}^{\rm(MC1)}(k;\eta)+P_{ab}^{\rm(MC2)}(k;\eta)+\cdots   ,
\label{eq:Pk_Born}
\end{eqnarray}
%%%%%%%%%%%%%%%%%%%%%%%%%%%%%%%%%%%%%%%%%%%%%%%%%%%%%%%%%%%%%%%%%%%%%
where we define $\widetilde{G}_a\equiv G_{a1}+G_{a2}$. 
The terms $P_{ab}^{\rm(MC1)}$ and $P_{ab}^{\rm(MC2)}$ respectively 
represent the leading- and next-to-leading order results of the 
Born approximation to the 
mode-coupling term (\ref{eq:P_MC_CLA}). The explicit expressions become
%%%%%%%%%%%%%%%%%%%%%%%%%%%%%%%%%%%%%%%%%%%%%%%%%%%%%%%%%%%%%%%%%%%%%
\begin{eqnarray}
P_{ab}^{\rm(MC1)}(k;\eta)&=& 2\int\frac{d^3\bfq}{(2\pi)^3}\,
I_{a}(\bfk,\,\bfq;\eta,\,\eta_0)\,
I_{b}(\bfk,\,\bfq;\eta,\,\eta_0)
\,\,e^{4\eta_0} P_0(q)\, P_0(|\bfk-\bfq|), 
\label{eq:CLA_MC1}\\
P_{ab}^{\rm (MC2)}(k;\eta) &=&
8\int\frac{d^3\bfp}{(2\pi)^3}\int\frac{d^3\bfq}{(2\pi)^3} 
J_a(\bfk,\bfp,\bfq;\eta,\,\eta_0)J_b(\bfk,\,\bfp,\,\bfq;\eta,\,\eta_0)
\nonumber\\
&&\quad\quad\quad\quad\quad\times\,\,e^{6\eta_0} \,\,
P_0(|\bfk-\bfp|)P_0(q)P_0(|\bfp-\bfq|).
\label{eq:CLA_MC2}
\end{eqnarray}
%%%%%%%%%%%%%%%%%%%%%%%%%%%%%%%%%%%%%%%%%%%%%%%%%%%%%%%%%%%%%%%%%%%%%
The kernels $I_a$ and $J_a$ are respectively given by 
%%%%%%%%%%%%%%%%%%%%%%%%%%%%%%%%%%%%%%%%%%%%%%%%%%%%%%%%%%%%%%%%%%%%%
\begin{eqnarray}
&&I_{a}(\bfk,\,\bfq;\eta,\,\eta_0)=\int_{\eta_0}^{\eta}d\eta'\,
G_{al}(k|\eta,\eta')\,\gamma_{lrs}(\bfq,\bfk-\bfq)
\,\widetilde{G}_{r}(q|\eta',\eta_0)\,
\widetilde{G}_{s}(|\bfk-\bfq||\eta',\eta_0),
\label{eq:func_I}\\
&&J_{a}(\bfk,\,\bfp,\,\bfq;\eta,\,\eta_0)=
\int_{\eta_0}^{\eta}d\eta_1\,\int_{\eta_0}^{\eta}d\eta_2
G_{al}(k|\eta,\eta_1)\,\gamma_{lrs}(\bfp,\bfk-\bfp)\,G_{rc}(p|\eta_1,\eta_2)
\nonumber\\
&&\quad\quad\quad\quad\quad\quad\quad\quad\quad
\times\,\,\,\gamma_{cpq}(\bfq,\bfp-\bfq)\,
\widetilde{G}_{p}(q|\eta_2,\eta_0) 
\widetilde{G}_{q}(|\bfp-\bfq||\eta_2,\eta_0) 
\widetilde{G}_{s}(|\bfk-\bfp||\eta_1,\eta_0).  
\label{eq:func_J}
\end{eqnarray}
%%%%%%%%%%%%%%%%%%%%%%%%%%%%%%%%%%%%%%%%%%%%%%%%%%%%%%%%%%%%%%%%%%%%%

%%%%%%%%%%%%%%%%%%%%%%%%%%%%%%%%%%%%%%%%%%%%%%%%%%%%%%%%%%%
\begin{figure}[t]
\begin{center}
\includegraphics[width=2.6cm,angle=-90]{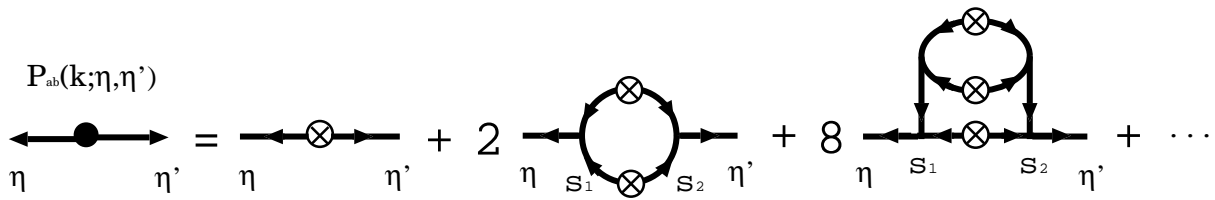}
\end{center}
\caption{Diagrammatic representation for the perturbative treatment of 
  the power spectrum with the Born approximation, i.e., 
  Eq.~(\ref{eq:Pk_Born}). 
\label{fig:Born}}
\end{figure}
%%%%%%%%%%%%%%%%%%%%%%%%%%%%%%%%%%%%%%%%%%%%%%%%%%%%%%%%%%%

The diagram corresponding to the above expressions is shown in 
    Fig.~\ref{fig:Born}. Note that in deriving the expression 
    (\ref{eq:Pk_Born}), we do not expand the propagators $G_{ab}$ and 
    their non-perturbative properties still hold. 
    In order to evaluate Eq.~(\ref{eq:Pk_Born}), we use the analytic 
    solution of $G_{ab}$ derived in Ref.~\cite{Taruya:2007xy}, where the 
    non-linear propagator was constructed approximately by matching the 
    asymptotic behaviors at low- and high-k modes, based on 
    Eqs.~(\ref{eq:RPT_expansion2}) with (\ref{eq:G_MC_CLA}). 
    The resultant analytic solution behaves like 
    $G_{ab}\to g_{ab}\,J_1(2x)/x$ 
    at $k\to\infty$, where the quantity $J_1$ is the Bessel 
    function with its argument $x=k\sigma_{\rm v}(e^\eta-e^{\eta'})$,  
    and the velocity dispersion $\sigma_{\rm v}$ is approximately 
    described by the linear theory, i.e., 
    $\sigma_{\rm v}^2\simeq\sigma_{\rm v,lin}^2=
    \int dq\,P_{\rm lin}(q;z)/(6\pi^2)$.  
    Note that the final results of the power spectrum are 
    a little bit sensitive to the high-$k$ behavior of the propagator, 
    and a naive application of the approximate solution 
    leads to a slight shift in the amplitude of power spectrum. 
    While this is not serious at all for the leading-order 
    calculation, it amounts a percent-level shift  
    when we consider the higher-order correction, $P_{ab}^{\rm (MC2)}$. 
    As discussed by Crocce \& Scoccimarro (2008), 
    one possible reason for this may be a small contribution from 
    the sub-leading corrections in the propagator.  
    In order to remedy the effect of small corrections, 
    we follow the method proposed by Ref.~\cite{Crocce:2007dt}. 
    We define
%%%%%%%%%%%%%%%%%%%%%%%%%%%%%%%%%%%%%%%%%%%%%%%%%%%%%%%%%%%%%%%%%%%%%%
\begin{equation}
\alpha(z)\equiv\left[\frac{\int_0^{k_{\rm max}} dq\,P_{\rm nl}(q;z)}
{\int_0^{k_{\rm max}}dk\,P_{\rm lin}(k;z)}\right]^{1/2},
\end{equation}
%%%%%%%%%%%%%%%%%%%%%%%%%%%%%%%%%%%%%%%%%%%%%%%%%%%%%%%%%%%%%%%%%%%%%%
where $P_{\rm nl}$ means the non-linear matter power spectrum. Then,  
the sub-leading correction can be corrected by simply multiplying the 
factor $\alpha$ by $\sigma_{\rm v}$, i.e., 
$\sigma_{\rm v}\to \alpha(z)\,\sigma_{\rm v}$. Note that 
this treatment is only applied to the propagator in the lowest-order 
term in Eq.~(\ref{eq:Pk_Born}), which most sensitively affects 
the power spectrum amplitude on small scales. For simplicity, we use 
\verb|halofit| \cite{Smith:2002dz} to compute $P_{\rm nl}$ and adopt the 
cutoff wavenumber, $k_{\rm max}=k_{\sigma}$, where $k_{\sigma}$ is the 
non-linear scale defined by Ref.~\cite{Smith:2002dz}.

In the rest of this paper, we present the results for the 
analytic treatment based on the expression (\ref{eq:Pk_Born}). 
In computing the mode-coupling terms $P_{ab}^{\rm(MC1)}$ 
and $P_{ab}^{\rm(MC2)}$, we must first evaluate the functions 
$I_a$ and $J_a$ for a given set of arguments, which involve 
the one- and two-dimensional integrals over time $\eta$. 
We use the Gaussian quadrature for these time integrations. 
As for the momentum integrals in the mode-coupling terms, thanks to 
the symmetry of the functions $I_a$ and $J_a$, the multi-dimensional 
integrals in $P_{ab}^{\rm(MC1)}$ and $P_{ab}^{\rm(MC2)}$ can be reduced 
to the two- and four-dimensional integrals, respectively. We use the 
Gaussian quadratures for the momentum integral in $P_{ab}^{\rm(MC1)}$. 
The four-dimensional momentum integration in 
the mode-coupling term $P_{ab}^{\rm(MC2)}$ is performed with Monte 
Carlo technique of quasi-random sampling using the library, 
\verb|Cuba|\cite{Hahn:2004fe}\footnote{\tt http://www.feynarts.de/cuba/ }.

Finally, we note that the formulation and analytic treatment 
presented here have several distinctions and similarities to 
the other non-perturbative calculations proposed recently. 
In Appendix~\ref{app:comparison}, we compare the present 
work with a subset of these 
treatments, and discuss how the approach developed here is complementary 
to or expands on these studies.

%%%%%%%%%%%%%%%%%%%%%%%%%%%%%%%%%%%%%%%%%%%%%%%%%%%%%%%
%%%%%%%%%%%%%%%%%%%%%%%%%%%%%%%%%%%%%%%%%%%%%%%%%%%%%%%
\section{Improved PT vs. Numerical Simulations}
\label{sec:PT_vs_N-body}
%%%%%%%%%%%%%%%%%%%%%%%%%%%%%%%%%%%%%%%%%%%%%%%%%%%%%%%
%%%%%%%%%%%%%%%%%%%%%%%%%%%%%%%%%%%%%%%%%%%%%%%%%%%%%%%

In this section, particularly focusing on the BAOs, 
we compare the improved PT predictions from the 
analytic treatment of closure approximation with results of N-body 
simulations.  

%-%-%-%-%-%-%-%-%-%-%-%-%-%-%-%-%-%-%-%-%-%-%-%-%-%-%-%
%-%-%-%-%-%-%-%-%-%-%-%-%-%-%-%-%-%-%-%-%-%-%-%-%-%-%-%
\subsection{N-body simulations}
\label{sec:N-body}
%-%-%-%-%-%-%-%-%-%-%-%-%-%-%-%-%-%-%-%-%-%-%-%-%-%-%-%
%-%-%-%-%-%-%-%-%-%-%-%-%-%-%-%-%-%-%-%-%-%-%-%-%-%-%-%

We use a publicly available cosmological N-body code, \verb|Gagdet2| 
\cite{Springel:2005mi}. We ran two sets of simulations, \verb|wmap3| and 
\verb|wmap5|, in which 
we adopt the standard Lambda CDM model with cosmological 
parameters determined from the WMAP3 and WMAP5, respectively 
\cite{Spergel:2006hy,Komatsu:2008hk}. The \verb|wmap3| 
run is basically the same N-body run as described in 
Ref.~\cite{Nishimichi:2008ry}, and a quantitative comparison 
between the leading-order results of improved PT and simulations 
has been previously made. We basically use the results of 
\verb|wmap3| run to check the consistency of the present calculations with 
the previous work. The \verb|wmap3| run is also helpful 
to cross-check the convergence properties in 
the new simulation, \verb|wmap5|, which increase the number of realizations 
to $30$. 
Table \ref{tab:n-body_params} summarizes the parameters used in the
simulations. 
The initial conditions were created with the \verb|2LPT| 
code \cite{Crocce:2006ve} at initial redshift $z_{\rm ini}=31$, 
based on the linear transfer function calculated 
from \verb|CAMB| \cite{Lewis:1999bs}. 
The number of meshes used in the particle-mesh computation is $1,024^3$. 
We adopt a softening length of $0.1h^{-1}$Mpc for tree forces.

We store three output redshifts for \verb|wmap3| run, whereas 
we select four output redshifts for \verb|wmap5| run; 
$z=3$, $1$, and $0$ (\verb|wmap3|) : $z=3$, $2$, $1$, and $0.5$ 
(\verb|wmap5|). Using these outputs, 
we compute the power spectrum and two-point correlation 
function in both real and redshift spaces.

The calculation of the matter power spectrum adopted here is 
basically the same treatment as in Ref.~\cite{Nishimichi:2008ry}. The standard 
method to compute the power spectrum is to square the Fourier transform of the 
density field and to take an average over realizations and Fourier modes. 
This is given by
%%%%%%%%%%%%%%%%%%%%%%%%%%%%%%%%%%%%%%%%%%%%%%%%%%%%%%%%%%%%%%%%%%%%%%
\begin{equation}
\widehat{P}(k_n)=\frac{1}{N_n^{\rm k} N^{\rm run}}
\sum_{m=1}^{N^{\rm run}} \sum_{k_n^{\rm min}<|\bfk|<k_n^{\rm max}}
\,\,\left|\delta^{m{\rm\mbox{-}th}}(\bfk)\right|^2\,\,;\quad
k_n\equiv \frac{1}{N_n^{\rm k}}\sum_{k_n^{\rm min}<|\bfk|<r_k^{\rm max}}
\left|\bfk\right|,
\label{eq:estimator_pk}
\end{equation}
%%%%%%%%%%%%%%%%%%%%%%%%%%%%%%%%%%%%%%%%%%%%%%%%%%%%%%%%%%%%%%%%%%%%%%
where $N_n^{\rm k}$ and $N^{\rm run}$ are the number of Fourier modes 
in the $n$-th wavenumber bin and the number of realizations, and 
$k_n^{\rm min}$ and $k_n^{\rm max}$ are the minimum and the maximum 
wavenumber of the $n$-th bin, respectively. The quantity 
$\delta^{m{\rm\mbox{-}th}}(\bfk)$ means the density field in Fourier space 
obtained from the $m$-th realization data. 
We use the Cloud-in-Cells interpolation for the density assignment of 
particles onto a $1,024^3$ mesh, and correct the 
window function. Note that the power spectra measured from the 
standard treatment above suffer from the effect of finite-mode sampling 
discussed by
Ref.~\cite{Takahashi:2008wn}. The resultant power spectrum deviates from 
the prediction for the ideal ensemble average, and exhibits the 
anomalous growth of power spectrum amplitude on large scales. 
In order to reduce the effect of finite mode sampling at 
$k\lesssim0.1h$Mpc$^{-1}$, 
we multiply the measured power spectrum by the ratio, 
$\widehat{P}^{\rm PT}(k)/P_{\rm lin}(k)$, where the quantity 
$\widehat{P}^{\rm PT}(k)$ is calculated from the perturbation theory up to the 
third-order in density field, and $P_{\rm lin}(k)$ is the input linear 
power spectrum extrapolated to a given output redshift. Note that 
in computing $\widehat{P}^{\rm PT}(k)$, we use the Gaussian-sampled 
density field 
used to generate the initial condition of each N-body run. With this treatment, 
the individual random nature of each N-body run is weakened, 
and the errors associated with anomalous growth is reduced\footnote{In Ref.~
\cite{Nishimichi:2008ry}, the correction to the effect of finite-mode 
sampling has been applied to the real-space power spectra. 
Here, we extend it to compute the redshift-space power spectrum 
by simply replacing the ratio $\widehat{P}^{\rm PT}(k)/P_{\rm lin}(k)$ with 
that in redshift space. To be precise, we compute the multipole moments of
the redshift-space power spectrum, and the ratio, 
$\widehat{P}_\ell^{\rm(S),PT}(k)/P_{\ell,{\rm lin}}^{\rm(S)}(k)$, is 
multiplied for each multipole spectrum (see Sec.~\ref{subsubsec:pk_in_red}).}. 

For the estimation of two-point correlation function, we adopt the 
grid-based calculation using the Fast Fourier Transformation (FFT). 
In this treatment, similar to the 
power spectrum analysis, we first compute the square of the density 
field on each grid of Fourier space. Then, applying the inverse Fourier 
transformation, we take the average over realization and distance, 
and obtain the two-point correlation function. 
Schematically, this is expressed as  
%%%%%%%%%%%%%%%%%%%%%%%%%%%%%%%%%%%%%%%%%%%%%%%%%%%%%%%%%%%%%%%%%%%%%%
\begin{equation}
\widehat{\xi}(r_n) = \frac{1}{N_n^{\rm r} N^{\rm run}}
\sum_{m=1}^{N^{\rm run}} \sum_{r_n^{\rm min}<|\bfr|<r_n^{\rm max}}
\,\,
\widehat{\rm FFT}^{-1}\Bigl[|\delta^{ m{\rm\mbox{-}th}}(\bfk)|^2;\bfr\Bigr]
,\label{eq:estimator_xi}
\end{equation}
%%%%%%%%%%%%%%%%%%%%%%%%%%%%%%%%%%%%%%%%%%%%%%%%%%%%%%%%%%%%%%%%%%%%%%
where the operation $\widehat{\rm FFT}^{-1}$ stands for the inverse FFT of 
the squared density field on each grid. Note here that $r_n$ is simply chosen 
at the center of the $n$-th radial bin, i.e., $r_n=(r_{\rm min}+r_{\rm max})/2$. 
 
Eq.~(\ref{eq:estimator_xi}) usually suffers from the 
ambiguity of the zero-point normalization in the amplitude of 
two-point correlation function, because of the lack of the low-$k$ powers 
due to the finite boxsize of the simulations.  
With the $1,024^3$ grids and the boxsize of $L_{\rm box}=1h^{-1}$Gpc, 
however, we can safely evaluate the two-point correlation function around 
the baryon acoustic peak. Comparison between 
different computational methods, together with 
convergence check of this method, 
is presented in Appendix \ref{app:computing_TPCF}.

Finally, similar to the estimation of power spectrum,  
the finite-mode sampling also affects the calculation of 
the two-point correlation function. We thus correct it by 
subtracting and adding the extrapolated linear density field as, 
$\widehat{\xi}(r)-\widehat{\xi}_{\rm lin}(r)+\xi_{\rm lin}(r)$,   
where $\widehat{\xi}_{\rm lin}$ is the correlation function estimated 
from the Gaussian density field, and $\xi_{\rm lin}$ is the 
linear theory prediction of two-point correlation function. 

%%%%%%%%%%%%%%%%%%%%%%%%%%%  TABLE  %%%%%%%%%%%%%%%%%%%%%%%%%%%%%%%
\begin{table}[t]
\caption{\label{tab:n-body_params} Parameters of N-body simulations}
\begin{ruledtabular}
\begin{tabular}{lcccc|cccccc}
Name & $L_{\rm box}$ & \# of particles & $z_{\rm ini}$ 
& \# of runs & 
$\Omega_{\rm m}$ & $\Omega_{\Lambda}$ & $\Omega_{\rm b}/\Omega_{\rm m}$ & $h$ & $n_s$ & $\sigma_8$ \\
\hline
\verb|wmap3| & $1000h^{-1}$Mpc & $512^3$ & $31$ & $4$ & 0.234 &  0.766 & 0.175 & 0.734 & 0.961 
& 0.76\\
\verb|wmap5| & $1000h^{-1}$Mpc & $512^3$ & $31$ & $30$ & 0.279 &  0.721 & 0.165 & 0.701 & 0.96 
& 0.817\\
\end{tabular}
\end{ruledtabular}
\end{table}
%%%%%%%%%%%%%%%%%%%%%%%%%%%  TABLE  %%%%%%%%%%%%%%%%%%%%%%%%%%%%%%%

%-%-%-%-%-%-%-%-%-%-%-%-%-%-%-%-%-%-%-%-%-%-%-%-%-%-%-%
%-%-%-%-%-%-%-%-%-%-%-%-%-%-%-%-%-%-%-%-%-%-%-%-%-%-%-%
\subsection{Results in real space}
\label{sec:result_real}
%-%-%-%-%-%-%-%-%-%-%-%-%-%-%-%-%-%-%-%-%-%-%-%-%-%-%-%
%-%-%-%-%-%-%-%-%-%-%-%-%-%-%-%-%-%-%-%-%-%-%-%-%-%-%-%

%%%%%%%%%%%%%%%%%%%%%%%%%%%%%%%%%%%%%%%%%%%%%%%%%%%%%%%%%%%
\begin{figure}[t]
\begin{center}
\includegraphics[width=9.5cm,angle=0]{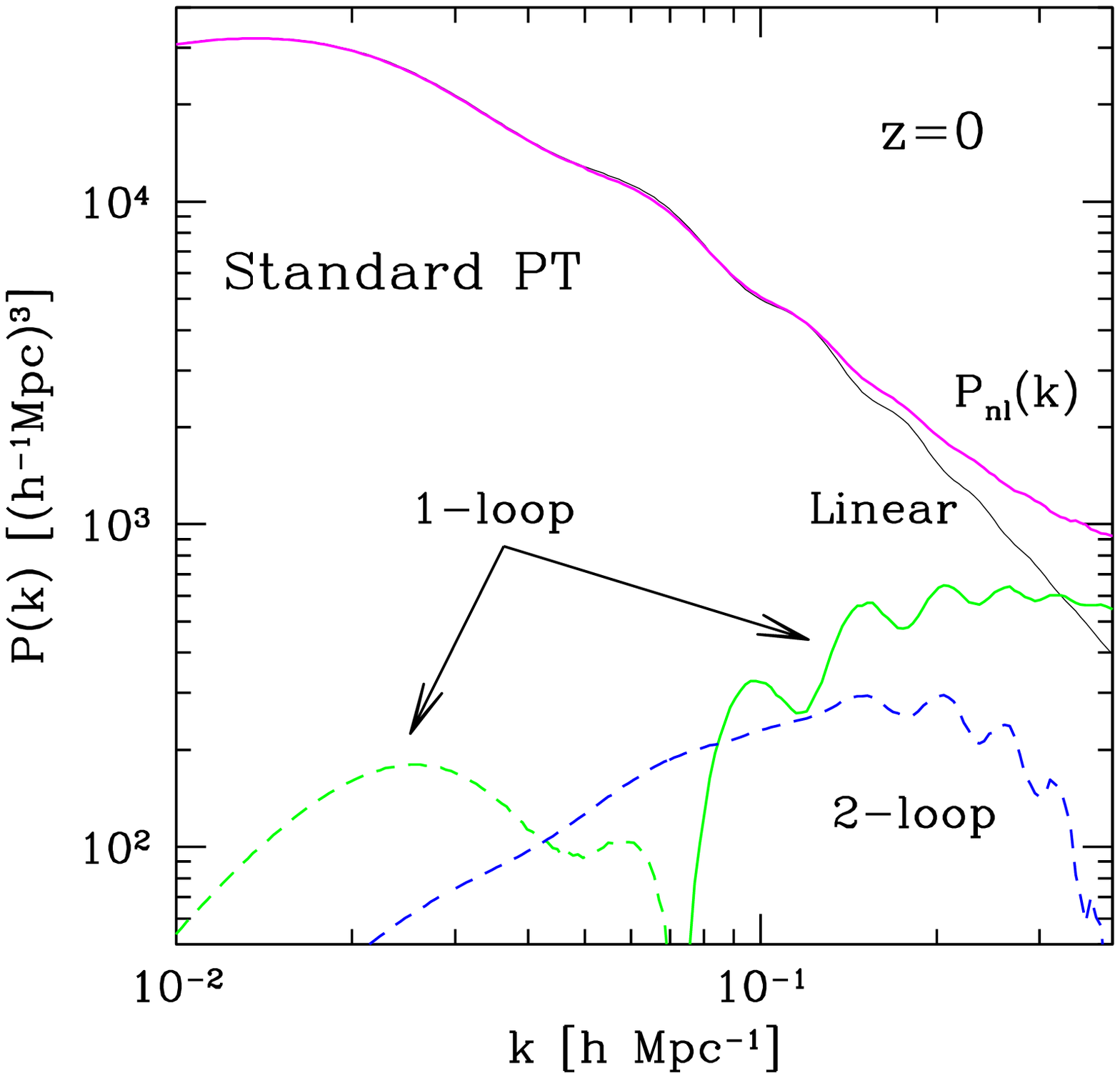}
\hspace*{-1.5cm}
\includegraphics[width=9.5cm,angle=0]{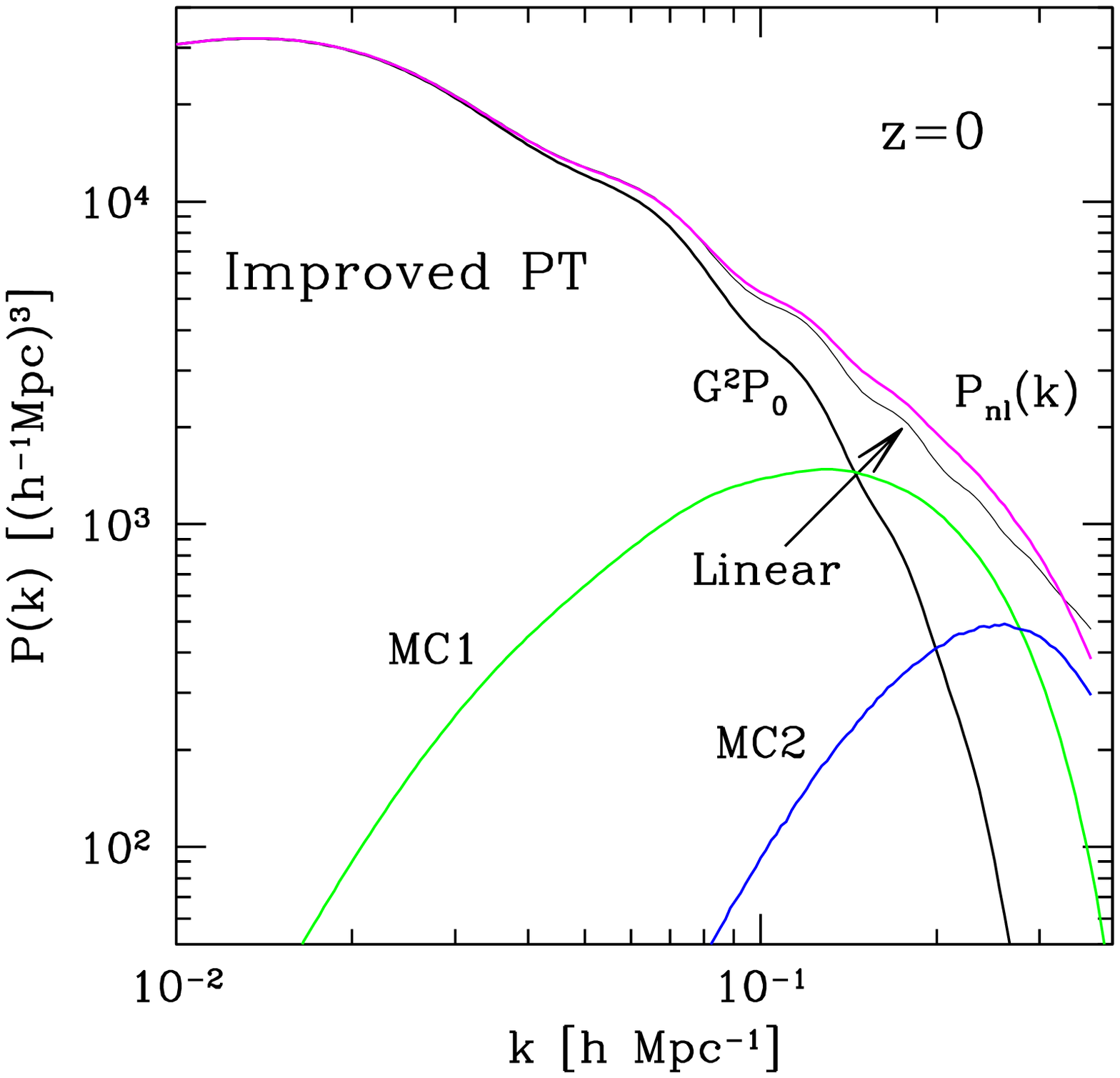}
\end{center}

\vspace*{-1.5cm}

\caption{Convergence properties of standard PT (left) and 
  improved PT (right) expansions in the matter power spectrum. 
  In each panel, the higher-order contributions to 
  the total power spectrum labeled as $P_{\rm nl}$ is separately plotted. 
  In left panel, one-loop and two-loop corrections in the standard PT, 
  $P_{11}^{\rm 1\mbox{-}loop}$ and $P_{11}^{\rm 2\mbox{-}loop}$, are plotted, while 
  in right panel, the mode-coupling corrections $P_{11}^{\rm(MC1)}$ and 
  $P_{11}^{\rm(MC2)}$ in the improved PT 
  given at Eqs.~(\ref{eq:CLA_MC1}) and (\ref{eq:CLA_MC2}) 
  respectively shown (labeled as MC1 and MC2), 
  together with the first term in Eq.~(\ref{eq:Pk_Born}) 
  (labeled as G$^2$P$_0$). Note that dashed lines indicate the negative values. 
\label{fig:convergence}}
\end{figure}
%%%%%%%%%%%%%%%%%%%%%%%%%%%%%%%%%%%%%%%%%%%%%%%%%%%%%%%%%%%

%-%-%-%-%-%-%-%-%-%-%-%-%-%-%-%-%-%-%-%-%-%-%-%-%-%-%-%
\subsubsection{Power spectrum}
\label{subsubsec:real_pk}
%-%-%-%-%-%-%-%-%-%-%-%-%-%-%-%-%-%-%-%-%-%-%-%-%-%-%-%

Before addressing a quantitative comparison between 
N-body simulation and improved PT, we first discuss
the convergence properties of the improved PT, and consider
how well the calculation based on the improved PT does improve the 
prediction compared to the standard PT. 

Fig.~\ref{fig:convergence} plots the overall behaviors of the 
non-linear power spectrum of density fluctuation, $P(k;z)\equiv P_{11}(k;z)$,  
given at $z=0$, adopting the \verb|wmap3| cosmological parameters.    
In left panel, the results of standard PT are shown, and 
the contributions to the total power spectrum up to the two-loop diagrams  
are separately plotted. On the other hand, 
right panel shows the results of improved PT. We 
plot the contributions up to the second-order Born approximation labeled 
as MC1 and MC2. 

In Fig.~\ref{fig:convergence},  
there are clear distinctions between standard and improved PTs. 
While the loop corrections in standard PT 
change their signs depending on the scales and exhibit an oscillatory feature, 
the corrections coming from the Born approximation in the improved PT 
are all positive and mostly the smooth function of $k$. Further, 
the higher-order corrections in the improved PT have a remarkable 
scale-dependent property compared to those in the standard PT; 
their contributions are well-localized around some characteristic 
wavenumbers, and they are shifted to the higher $k$ modes as increasing 
the order of PT. These trends clearly indicate that the improved PT 
with closure approximation has a better convergence property.  
Qualitative behaviors of the higher-order corrections quite resemble 
the predictions of RPT by Crocce \& Scoccimarro (2008)
\cite{Crocce:2007dt}.  

\begin{figure}[t]
\begin{center}
\includegraphics[width=8cm,angle=0]{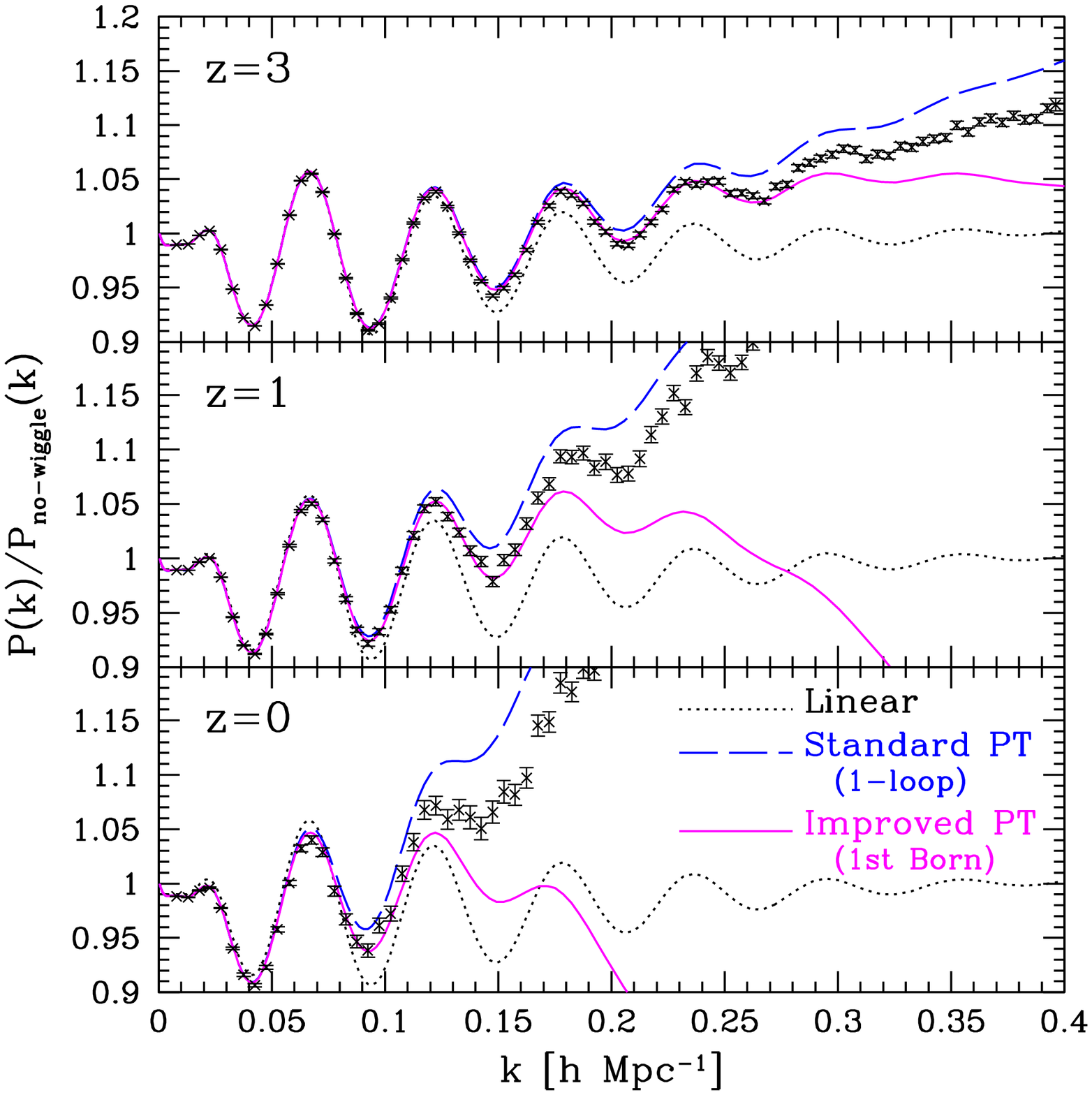}
\includegraphics[width=8cm,angle=0]{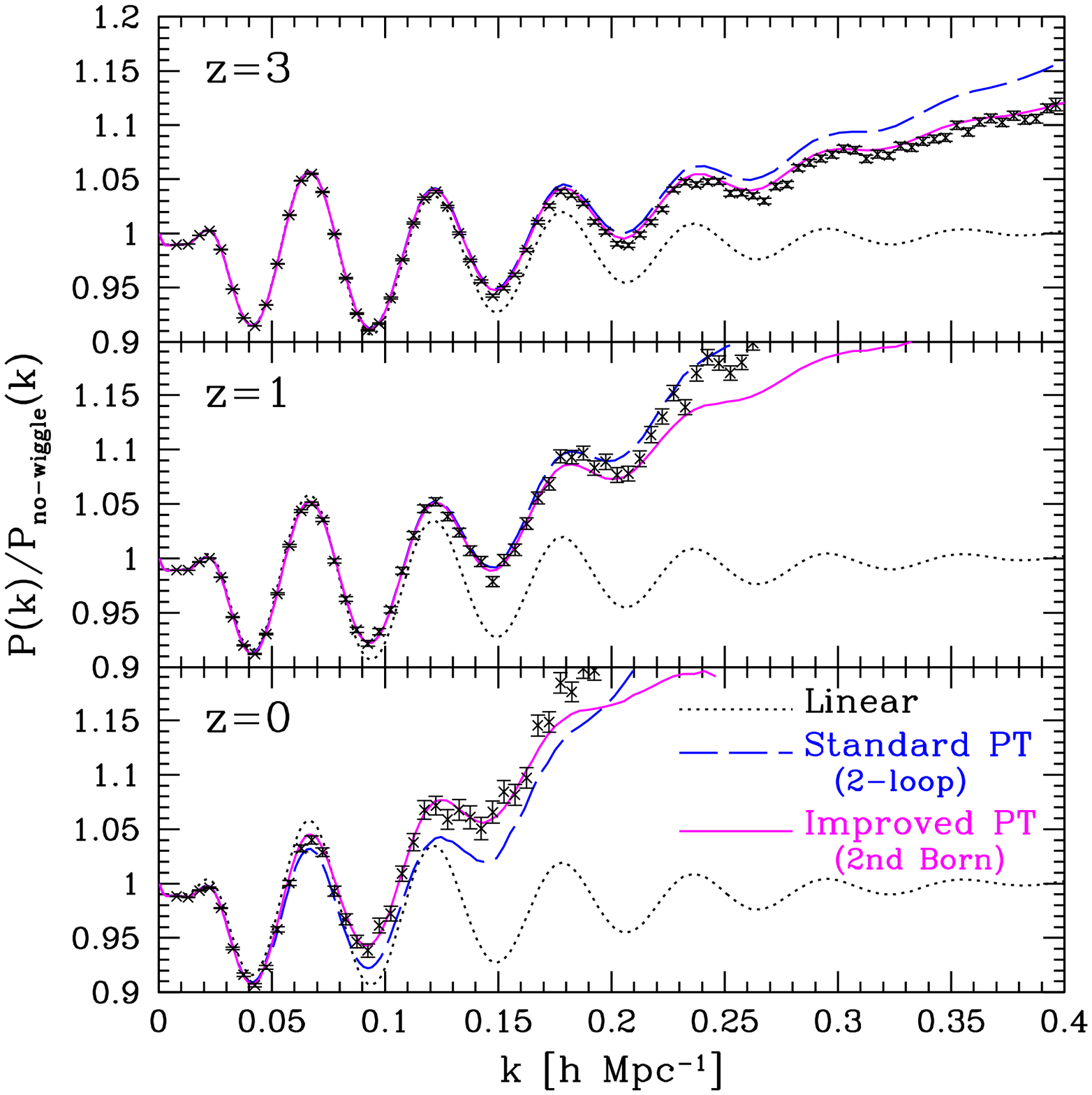}
\end{center}

\vspace*{-0.3cm}

\caption{Ratios of power spectrum to smoothed reference spectrum, 
 $P(k)/P_{\rm no\mbox{-}wiggle}(k)$, given at redshifts $z=3$(top), 
 $1$(middle) and $0$(bottom). Cosmological parameters used in 
 the {\tt wmap3} simulations are adopted to compute the power spectrum from 
 standard PT and improved PT, and the results are compared with N-body 
 simulations (symbols with error-bars). The reference spectrum 
 $P_{\rm no\mbox{-}wiggle}(k)$ is calculated from the no-wiggle formula of 
 the linear transfer function in Ref.~\cite{Eisenstein:1997ik}. 
 In each panel, dotted, 
 dashed and solid lines represent the linear, standard PT and improved PT 
 results, respectively. 
 In left panel, leading-order results of standard PT and improved PT 
 are shown, while in right panel,  
 the results including the higher-order corrections are plotted.
\label{fig:ratio_pk_real}}
\end{figure}

Now, let us focus on the behavior of BAOs, and discuss how the 
convergence properties seen in Fig.~\ref{fig:convergence} affect the 
predictions of BAO features. In Fig.~\ref{fig:ratio_pk_real}, 
adopting the \verb|wmap3| cosmological parameters, 
we plot the ratio, $P(k)/P_{\rm no\mbox{-}wiggle}(k)$, where the function 
$P_{\rm no\mbox{-}wiggle}(k)$ is the linear power spectrum 
from the smooth transfer function neglecting the BAO feature in 
Ref.~\cite{Eisenstein:1997ik}. In left panel, 
N-body simulations are compared with the leading-order results of 
PT predictions, i.e., standard PT including the one-loop 
correction (dashed), and improved PT with first-order Born correction (solid). 
Apart from the wiggle structure, the amplitude of standard PT 
prediction monotonically increases with wavenumber $k$, and tends to  
overestimate the results of N-body simulations. On the other hand, 
the amplitude of improved PT prediction rapidly falls off at a certain 
wavenumber, and the deviation from N-body results becomes significant. 
However, a closer look at the behavior on large scales reveals that 
improved PT prediction gives a better agreement with simulation. 
The results are indeed consistent with the previous findings in
Ref.~\cite{Nishimichi:2008ry}. The situation 
becomes more impressive when we add the next-to-leading order corrections.    
As shown in right panel, the improved PT gets the power on smaller scales, 
and reproduces the N-body results in a wider range of wavenumber. By contrast, 
the prediction of standard PT depicted as dashed lines 
seems a little bit subtle. Compared to the one-loop results,  
the amplitudes of the standard PT prediction including the two-loop correction 
are slightly reduced, and the agreement with N-body simulation 
seems apparently improved a bit at higher redshift. At lower redshift $z=0$, 
however, the correction coming from the two-loop order becomes significant,  
and the prediction eventually underestimates the simulation. The reason for 
these behaviors basically comes from the competition between positive 
and negative contributions of the one-loop and two-loop corrections, 
respectively (see left panel of 
Fig~. \ref{fig:convergence}). These are consistent with those findings in
Ref.~\cite{Carlson:2009it} (see Fig.~1 of their paper).

%%%%%%%%%%%%%%%%%%%%%%%%%%%%%%%%%%%%%%%%%%%%%%%%%%%%%%%%%%%
%%%%%%%%%%%%%%%%%%%%%%%%%%%%%%%%%%%%%%%%%%%%%%%%%%%%%%%%%%%
\begin{figure}[h]
\begin{center}
\includegraphics[width=8cm,angle=0]{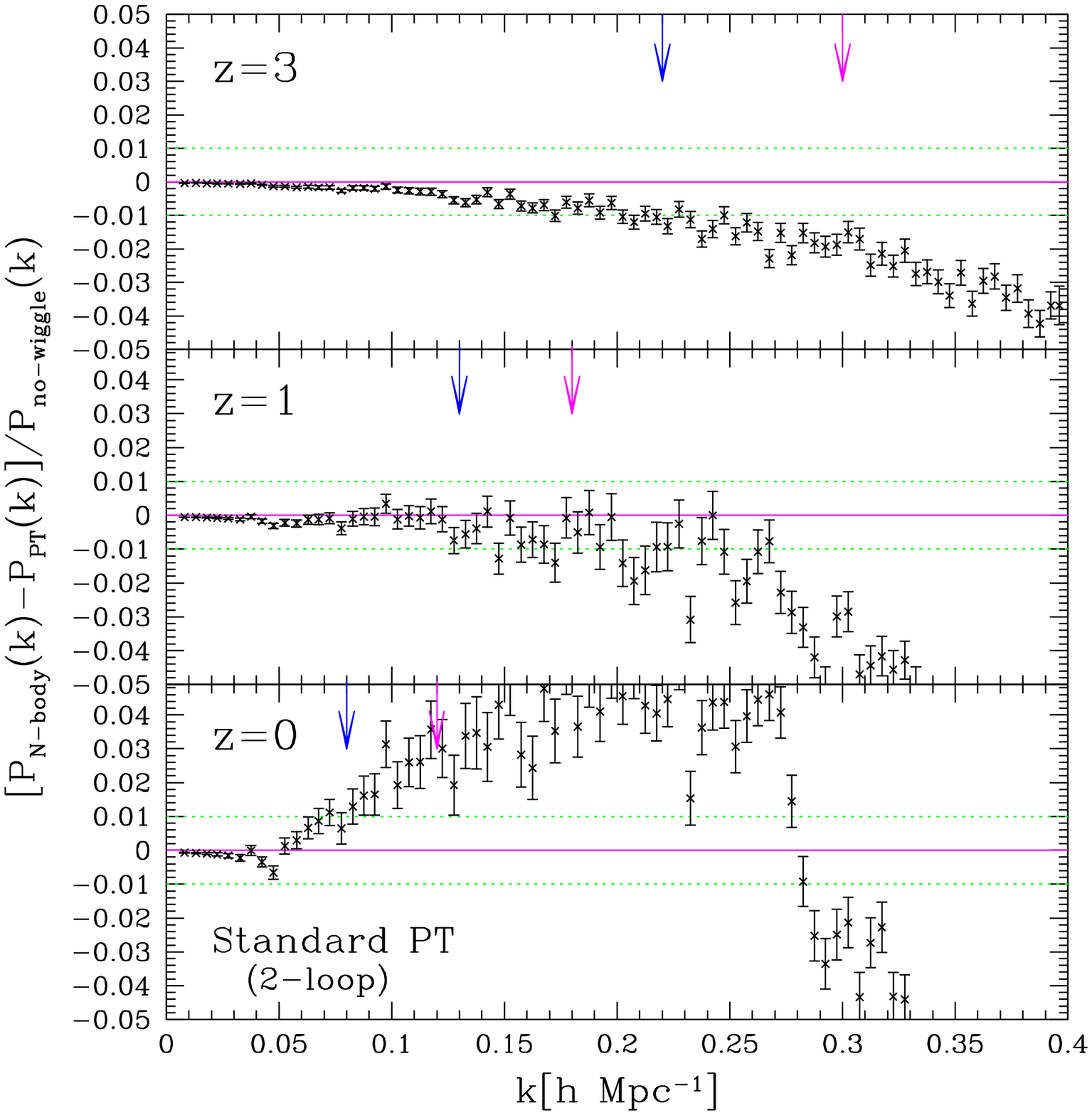}
\includegraphics[width=8cm,angle=0]{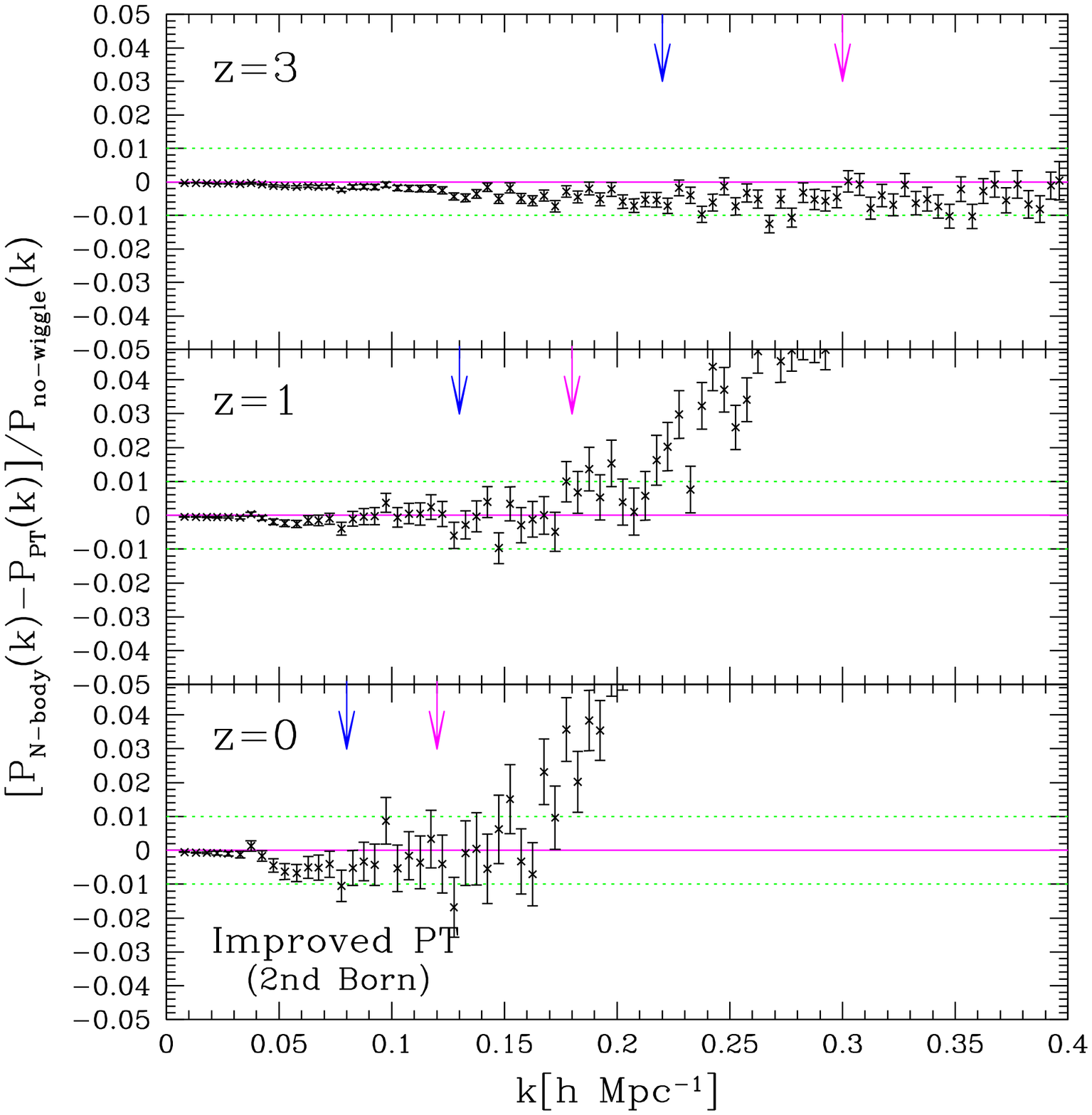}
\end{center}

\vspace*{-0.3cm}

\caption{Difference between N-body and PT results divided by the reference 
  spectrum, $[P_{\rm N\mbox{-}body}(k)-P_{\rm PT}(k)]/P_{\rm no\mbox{-}wiggle}(k)$. 
  Left panel shows the results for standard PT up to the two-loop order. 
  Right panel presents the case of improved PT including the corrections 
  up to the second-order Born approximation of the mode-coupling term. 
  In both panels, vertical arrows represent the wavenumbers $k_{1\%}$ of 
  standard and improved PT (from left to right), below 
  which the leading-order PT predictions 
  reproduce the N-body simulations well within $1\%$ accuracy (see text in 
  details). 
\label{fig:ratio_pk_real2}}
\end{figure}
%%%%%%%%%%%%%%%%%%%%%%%%%%%%%%%%%%%%%%%%%%%%%%%%%%%%%%%%%%%

In Fig.~\ref{fig:ratio_pk_real2}, 
to clarify the range of agreement in more quantitative ways, 
we plot the fractional difference divided by the smoothed reference spectra, 
$[P_{\rm N\mbox{-}body}(k)-P_{\rm PT}(k)]/P_{\rm no\mbox{-}wiggle}$, where  
the quantity $P_{\rm PT}(k)$ implies the standard and improved PT predictions
in left and right panels, respectively. Here, the vertical arrows 
represent the maximum wavenumber $k_{1\%}$, below which 
the leading-order predictions of standard or improved PT 
reproduce the N-body results quite well within the $1\%$ accuracy.  
According to Nishimichi et al. \cite{Nishimichi:2008ry}, 
this has been determined by the detailed comparison between models and 
simulations, and is empirically characterized by solving the following equation:
%%%%%%%%%%%%%%%%%%%%%%%%%%%%%%%%%%%%%%%%%%%%%%%%%%%%%%%%%%%
\begin{equation}
\frac{k_{1\%}^2}{6\pi^2}\int_0^{k_{1\%}} dq P_{\rm lin}(q;z) = C
\label{eq:k_criterion}
\end{equation}
%%%%%%%%%%%%%%%%%%%%%%%%%%%%%%%%%%%%%%%%%%%%%%%%%%%%%%%%%%%
with $C=0.18$ for the one-loop standard PT, and $C=0.35$ for the 
improved PT up to the first-order Born correction. 

Comparing these convergence regimes of the leading-order calculation with 
  results of fractional differences, Fig.~\ref{fig:ratio_pk_real2} 
  shows that the inclusion of higher-order terms does not always 
  improve the prediction in the standard PT treatment. By contrast, 
  the improved PT calculation does improve the predictions, and the 
  range of agreement between N-body simulations and the predictions becomes 
  wider.

\begin{figure}[t]
\begin{center}
\includegraphics[width=8cm,angle=0]{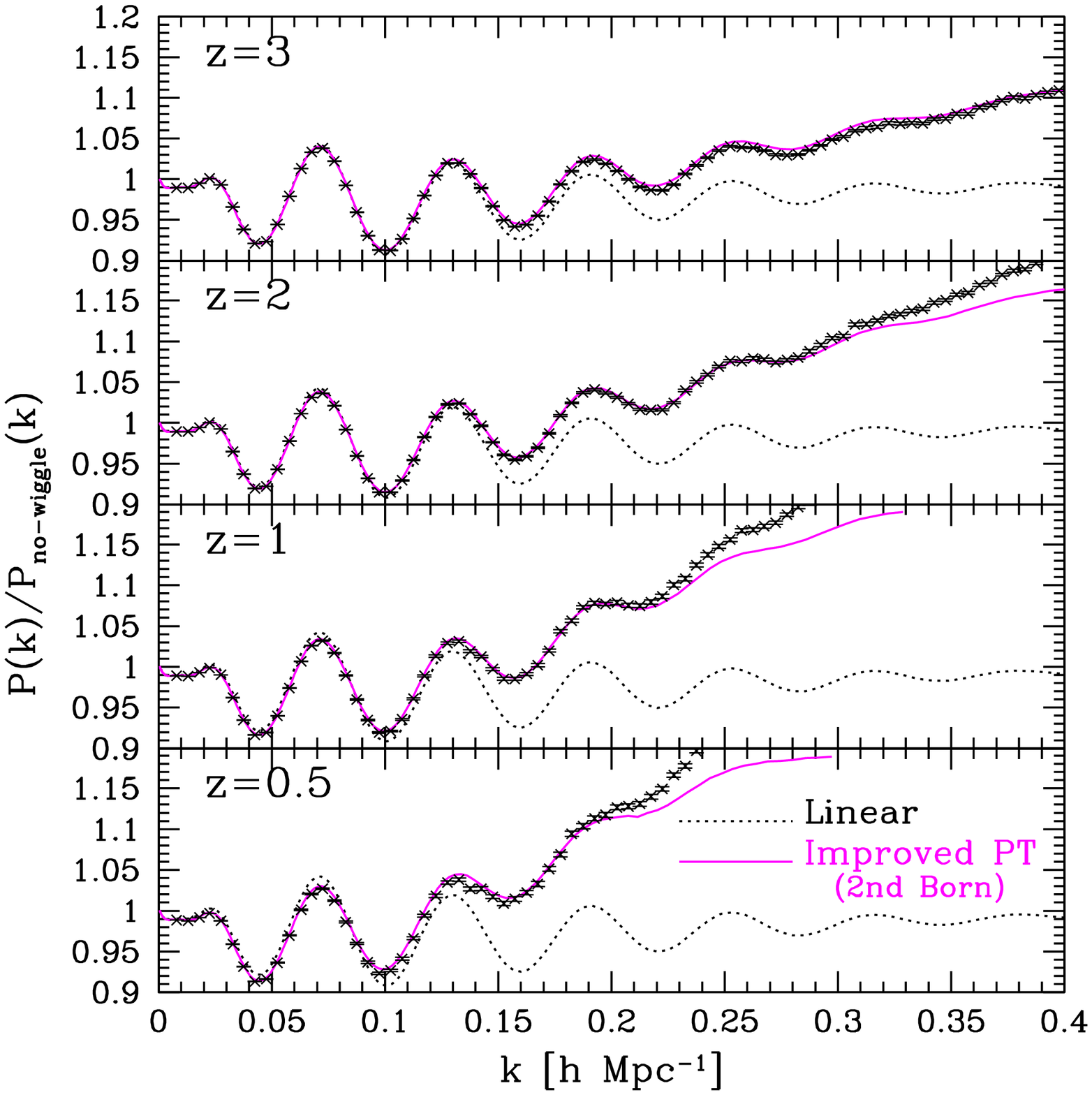}
\includegraphics[width=8cm,angle=0]{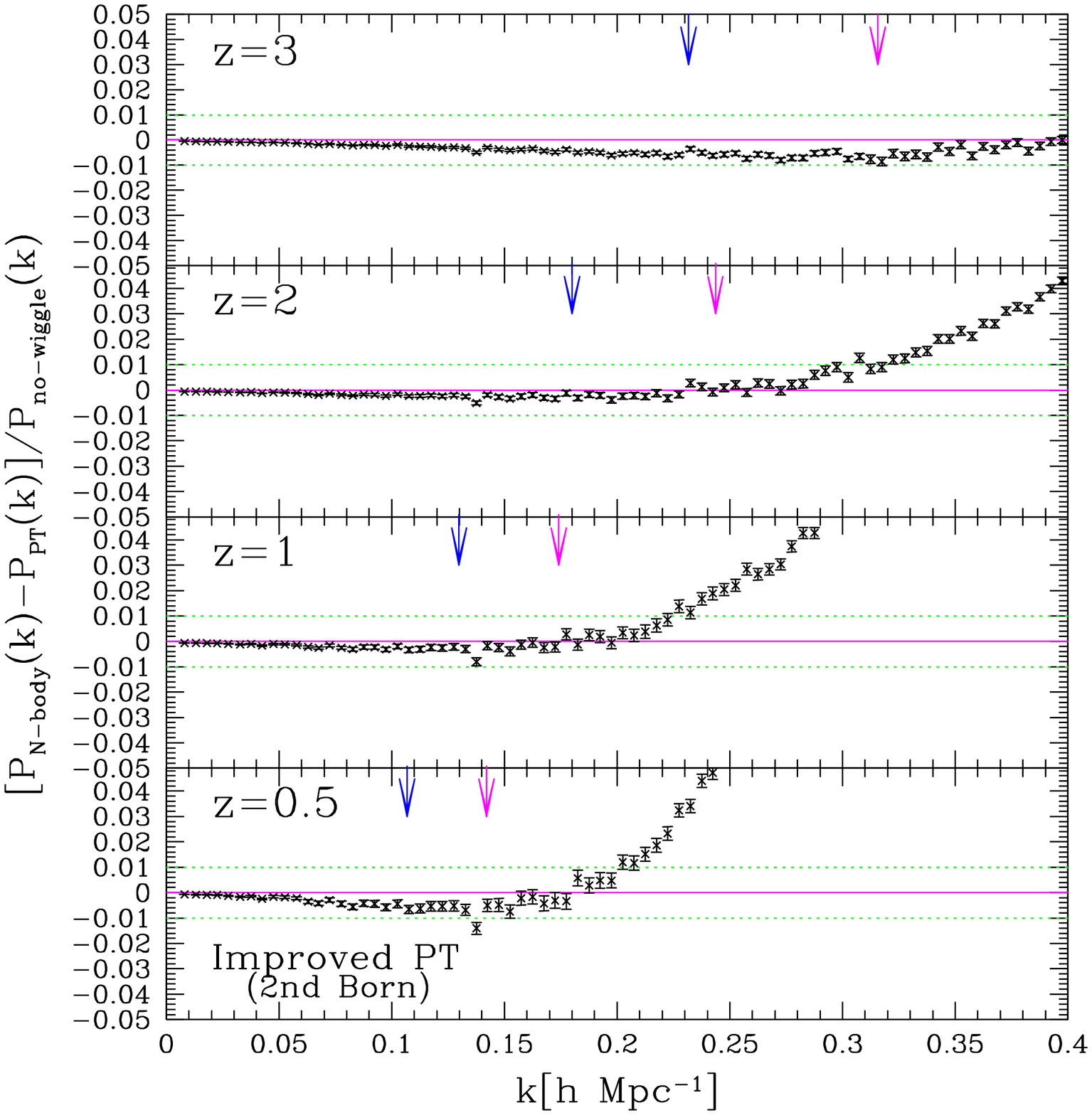}
\end{center}

\vspace*{-0.3cm}

\caption{Comparison between N-body results and improved PT 
predictions in the case adopting {\tt wmap5} cosmological parameters. 
From top to bottom, the results at $z=3$, $2$, $1$ and $0.5$ are shown. 
The improved PT predictions plotted here include the corrections up to the 
second-order Born approximation of the mode-coupling term, $P^{\rm MC2}$. 
Left: ratio of power spectrum to the smoothed reference spectra, 
$P(k)/P_{\rm no\mbox{-}wiggle}(k)$. Solid and dotted lines are improved PT 
and linear theory predictions, respectively. 
  Right: difference between N-body and improved PT results normalized by 
  the no-wiggle formula, 
  $[P_{\rm N\mbox{-}body}(k)-P_{\rm PT}(k)]/P_{\rm no\mbox{-}wiggle}(k)$. 
  In each panel, vertical arrows represent the wavenumber $k_{1\%}$ for 
  the leading-order predictions of standard and improved PT (from left to 
  right).  
\label{fig:ratio_pk2_alpha}}
\end{figure}

In Fig.~\ref{fig:ratio_pk2_alpha}, we plot the results for the \verb|wmap5| 
simulations, which have relatively large value of $\sigma_8$ compared 
to the wmap3 run (see Table \ref{tab:n-body_params}). 
Left and right panels respectively plot the ratio of power spectrum 
amplitude and the fractional difference between N-body results and improved 
PT predictions. With the $30$ runs of N-body simulations, 
the errors in the power spectrum amplitude are greatly reduced, and it 
is clearly shown that  the predictions of 
improved PT including the higher-order corrections almost coincide with the 
N-body results beyond the convergence regime of the 
leading-order calculations (indicated by vertical arrows), 
and achieve a sub-percent accuracy. 
From this plot, 
the maximum wavenumber $k_{1\%}$ at each redshift can be estimated by comparing 
the predictions with N-body results as  
$k_{1\%}=0.20h$Mpc$^{-1}$(z=0.5), $0.23h$Mpc$^{-1}$(z=1), 
$0.33h$Mpc$^{-1}$(z=2) and $0.47h$Mpc$^{-1}$(z=3). 
These values roughly match those determined from 
the criterion (\ref{eq:k_criterion}) with the constant $C=0.70$. 

Although we did not store the $z=0$ data of \verb|wmap5| run to compare with 
analytic prediction, 
Eq. (\ref{eq:k_criterion}) using this constant value implies 
that the maximum wavenumber for improved PT becomes 
$k_{1\%}=0.15h$Mpc$^{-1}$, 
which contrasts with the one for the one-loop prediction of standard PT, 
$k_{1\%}=0.09h$Mpc$^{-1}$. Thus, the improved PT including up 
to the second-order Born approximation is expected to be still accurate 
at $z=0$, and it can cover the major part of the BAOs. A more 
detailed comparison at low redshift including other analytic 
prescriptions can be found in Ref.~\cite{Carlson:2009it}.

%%%%%%%%%%%%%%%%%%%%%%%%%%%%%%%%%%%%%%%%%%%%%%%%%%%%%%%%%%%
%%%%%%%%%%%%%%%%%%%%%%%%%%%%%%%%%%%%%%%%%%%%%%%%%%%%%%%%%%%
\begin{figure}[h]
\begin{center}
\includegraphics[width=11cm,angle=0]{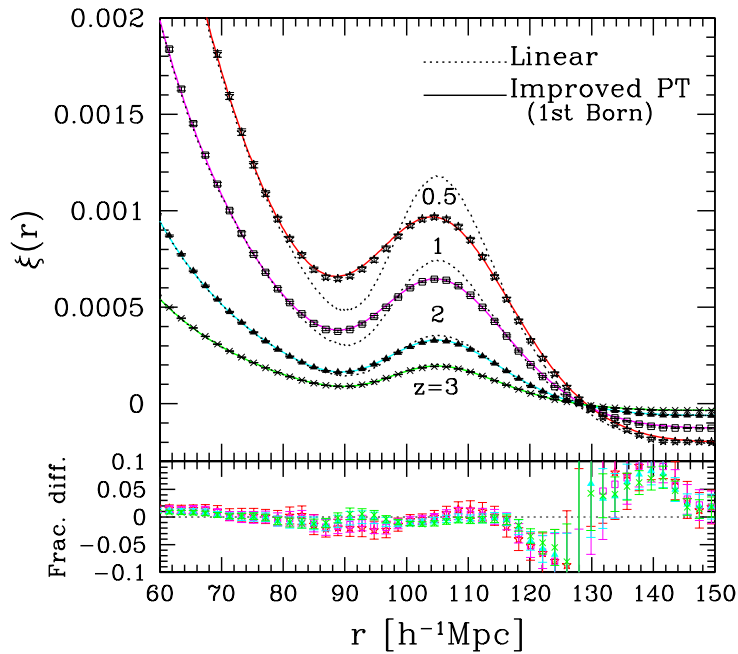}

\vspace*{-1.2cm}

\end{center}
\caption{Top: Two-point correlation functions in real space adopting 
the {\tt wmap5} cosmological parameters. The solid lines represent 
the leading-order predictions of improved PT, while the dotted 
lines show the linear theory results. 
Bottom: Fractional differences between 
N-body and improved PT results, 
$[\xi_{\rm N\mbox{-}body}(r)-\xi_{\rm PT}(r)]/\xi_{\rm PT}(r)$. 
In both panels, 
the symbols with error-bars indicate the N-body results averaged over 
the $30$ realizations in which the effect of finite-mode sampling is 
corrected: $z=0.5$ (open stars), $1$ (open squares), $2$ (filled triangles), 
and $3$ (crosses). } 
\label{fig:xi_real}
\end{figure}
%%%%%%%%%%%%%%%%%%%%%%%%%%%%%%%%%%%%%%%%%%%%%%%%%%%%%%%%%%%

%-%-%-%-%-%-%-%-%-%-%-%-%-%-%-%-%-%-%-%-%-%-%-%-%-%-%-%
\subsubsection{Correlation function}
%-%-%-%-%-%-%-%-%-%-%-%-%-%-%-%-%-%-%-%-%-%-%-%-%-%-%-%

Having confirmed the excellent properties of the improved PT, 
  we turn to focus on the baryon acoustic peak in the two-point correlation 
  function. The two-point correlation function can be computed from the 
  power spectrum as 
%%%%%%%%%%%%%%%%%%%%%%%%%%%%%%%%%%%%%%%%%%%%%%%%%%%%%%%%%%%%%%%%%%%%%%
\begin{equation}
\xi(r) = \int \frac{dk\,k^2}{2\pi^2}\,P_{11}(k) \frac{\sin(k\,r)}{k\,r}.
\label{eq:def_xi} 
\end{equation}
%%%%%%%%%%%%%%%%%%%%%%%%%%%%%%%%%%%%%%%%%%%%%%%%%%%%%%%%%%%%%%%%%%%%%%

Top panel of Fig.~\ref{fig:xi_real} shows the two-point correlation 
functions around the baryon acoustic peak at different 
redshifts $z=0.5, 1, 2$ and $3$ (from top to bottom)  
in the case adopting the $\verb|wmap5|$ cosmological parameters. 
Also, lower panel plots the fractional differences between 
N-body and improved PT results, i.e., 
$[\xi_{\rm N\mbox{-}body}(r)-\xi_{\rm PT}(r)]/\xi_{\rm PT}(r)$.

 After the correction of finite-mode sampling, 
the error-bars in N-body simulations are greatly reduced, and 
the deviation of the N-body results from linear theory predictions 
(depicted as dotted lines) is clearly seen. As decreasing the redshift, 
the baryon acoustic peaks become smeared and the 
position of the peak are slightly shifted to a smaller scale. 
These trends can be accurately described by 
the leading-order calculation of improved PT, and  
the agreement between N-body results and the predictions is excellent. 
The fractional error in amplitude is well within a few percent, 
except for a large separation beyond the location of baryon acoustic 
peak, where the accuracy of N-body results tends to be worsen due to the 
limited simulation boxsize.
Note that the corrections coming from the higher-order Born approximation 
do not alter the behaviors at $r>30h^{-1}$Mpc, and their amplitudes 
are negligibly small compared to the error-bars of N-body simulations. 
Thus the leading-order 
prediction seems robust for describing the baryon acoustic peak. 

It has been recently suggested by several authors that the smearing 
effect on baryon acoustic peak is mostly attributed to the random motion 
of mass distribution \cite{Eisenstein:2006nj}, 
and it is approximately described by the convolution of the Gaussian 
smoothing function (e.g., \cite{Matsubara:2007wj, Smith:2007gi}). 
In the language of improved PT, this effect corresponds to 
the disappearance of the memory of initial condition, which is 
encoded in the non-linear propagator. 
Strictly speaking, 
the asymptotic behavior of the non-linear propagator is not 
a Gaussian form in closure approximation, although 
the damping behavior manifestly exhibits in the approximate solution of 
non-linear propagator. Hence, the prediction for the 
two-point correlation function seems robust against 
the high-$k$ behavior of the non-linear propagator. 

Finally, it should be noted that the standard PT prediction fails 
to converge the integral in Eq.~(\ref{eq:def_xi}), because of the 
high-$k$ behavior of the power spectrum. 
This is true even when including the higher-order 
correction of two-loop order. Thus, the successful results of improved PT 
prediction may be regarded as an outcome of non-perturbative property.

%-%-%-%-%-%-%-%-%-%-%-%-%-%-%-%-%-%-%-%-%-%-%-%-%-%-%-%
%-%-%-%-%-%-%-%-%-%-%-%-%-%-%-%-%-%-%-%-%-%-%-%-%-%-%-%
\subsection{Results in redshift space}
\label{sec:result_redshift}
%-%-%-%-%-%-%-%-%-%-%-%-%-%-%-%-%-%-%-%-%-%-%-%-%-%-%-%
%-%-%-%-%-%-%-%-%-%-%-%-%-%-%-%-%-%-%-%-%-%-%-%-%-%-%-%

In practical observation with galaxy redshift surveys, 
the observed galaxy distribution is inevitably distorted due to 
the peculiar velocity of each galaxy. The so-called redshift-space 
distortion is known to alter the shape of the power spectrum 
in two different ways (e.g., \cite{Hamilton:1997zq}). One is the 
apparent enhancement of the clustering signal called Kaiser effect 
\cite{Kaiser:1987qv}, 
which originates from the bulk motion of mass distribution falling into 
the massive halos. Another important effect is the finger-of-God (FoG) 
effect, which effectively suppresses the power spectrum amplitude on small 
scales by the virialized random motion of the mass residing at a halos.

Although a rigorous non-perturbative treatment of 
the redshift-space distortion is difficult, these two effects 
has been phenomenologically modeled as (e.g., 
\cite{Peacock:1993xg,Cole:1993kh,Park:1994fa,Ballinger:1996cd})
%%%%%%%%%%%%%%%%%%%%%%%%%%%%%%%%%%%%%%%%%%%%%%%%%%%%%%%%%%%%%%%%%%%%%%
\begin{equation}
P^{\rm(S)}(k,\mu)=\left(1+\mu^2\,f\right)^2 P_{11}(k)\,D_{\rm FoG}(k\,\mu),
\label{eq:redshift_model_old}
\end{equation}
%%%%%%%%%%%%%%%%%%%%%%%%%%%%%%%%%%%%%%%%%%%%%%%%%%%%%%%%%%%%%%%%%%%%%%
where $\mu$ is the cosine of the angle between the line-of-sight direction 
and the Fourier mode $\bfk$, and $f$ is the logarithmic derivative of linear 
growth factor, defined as $f\equiv d\ln D/d\ln a$. 
The function $D_{\rm FoG}$ represents the 
damping function which mimics the FoG effect, and it asymptotically approaches 
unity in the $k\to0$ limit, where the linear-theory formula by Kaiser 
is recovered.

Recently, Scoccimarro \cite{Scoccimarro:2004tg} proposed an improved 
version of the model (\ref{eq:redshift_model_old}) to properly take account of 
the non-linear evolution of density and velocity fields on the Kaiser effect 
(see also \cite{Percival:2008sh,:2008yg}).  
This is expressed as
%%%%%%%%%%%%%%%%%%%%%%%%%%%%%%%%%%%%%%%%%%%%%%%%%%%%%%%%%%%%%%%%%%%%%%
\begin{equation}
P^{\rm(S)}(k,\mu)=\left[P_{11}(k)+
2f\mu^2\,P_{12}(k)+f^2\mu^4P_{22}(k)\right]
\exp\{-(f\mu k\,\sigma_{\rm v})^2\}. 
\label{eq:redshift_model}
\end{equation}
%%%%%%%%%%%%%%%%%%%%%%%%%%%%%%%%%%%%%%%%%%%%%%%%%%%%%%%%%%%%%%%%%%%%%%
Here, the quantity $\sigma_{\rm v}$ is the one-dimensional velocity 
dispersion given by 
%%%%%%%%%%%%%%%%%%%%%%%%%%%%%%%%%%%%%%%%%%%%%%%%%%%%%%%%%%%%%%%%%%%%%%
\begin{equation}
\sigma_{\rm v}^2=\frac{1}{3}\int\frac{d^3\bfq}{(2\pi)^3}\,
\frac{P_{22}(q)}{q^2}.
\end{equation}
%%%%%%%%%%%%%%%%%%%%%%%%%%%%%%%%%%%%%%%%%%%%%%%%%%%%%%%%%%%%%%%%%%%%%%

 In what follows, we adopt the model (\ref{eq:redshift_model}) to 
calculate the redshift-space power spectrum. Although this model is 
still phenomenological and may not be regarded as the best one, 
a comparison between the model predictions and 
N-body simulations shows that the prediction based on the model 
(\ref{eq:redshift_model}) gives a better result. 
Taking Eq.~(\ref{eq:redshift_model}) as a canonical model of 
the redshift-space distortion, we will investigate 
the extent to which the model (\ref{eq:redshift_model}) 
faithfully reproduces the N-body results well, and discuss how to improve 
the model prescription.  

%-%-%-%-%-%-%-%-%-%-%-%-%-%-%-%-%-%-%-%-%-%-%-%-%-%-%-%
\subsubsection{Power spectrum}
\label{subsubsec:pk_in_red}
%-%-%-%-%-%-%-%-%-%-%-%-%-%-%-%-%-%-%-%-%-%-%-%-%-%-%-%

%%%%%%%%%%%%%%%%%%%%%%%%%%%%%%%%%%%%%%%%%%%%%%%%%%%%%%%%%%%
\begin{figure}[t]
\begin{center}
\includegraphics[width=8cm,angle=0]{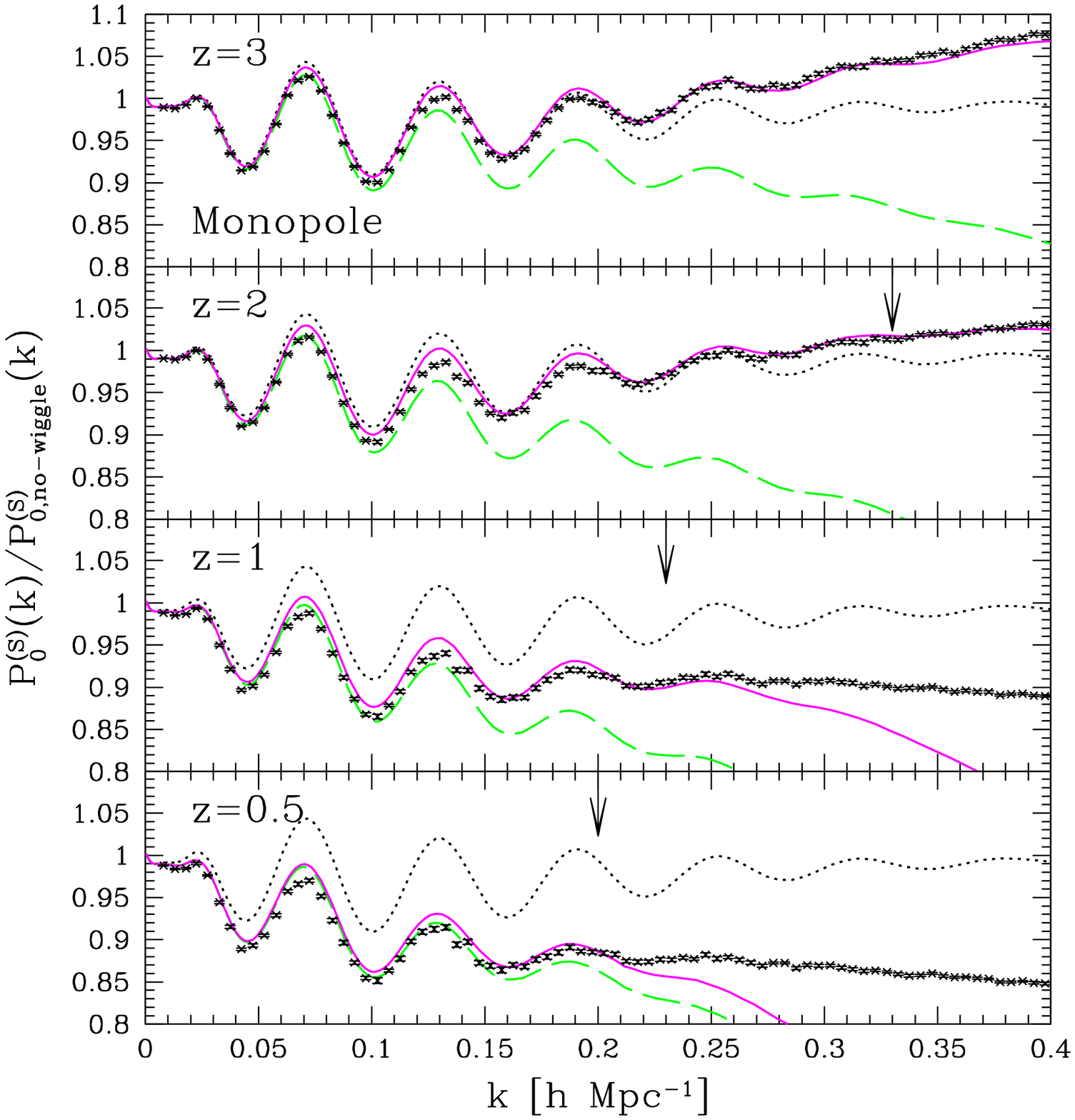}
\includegraphics[width=8cm,angle=0]{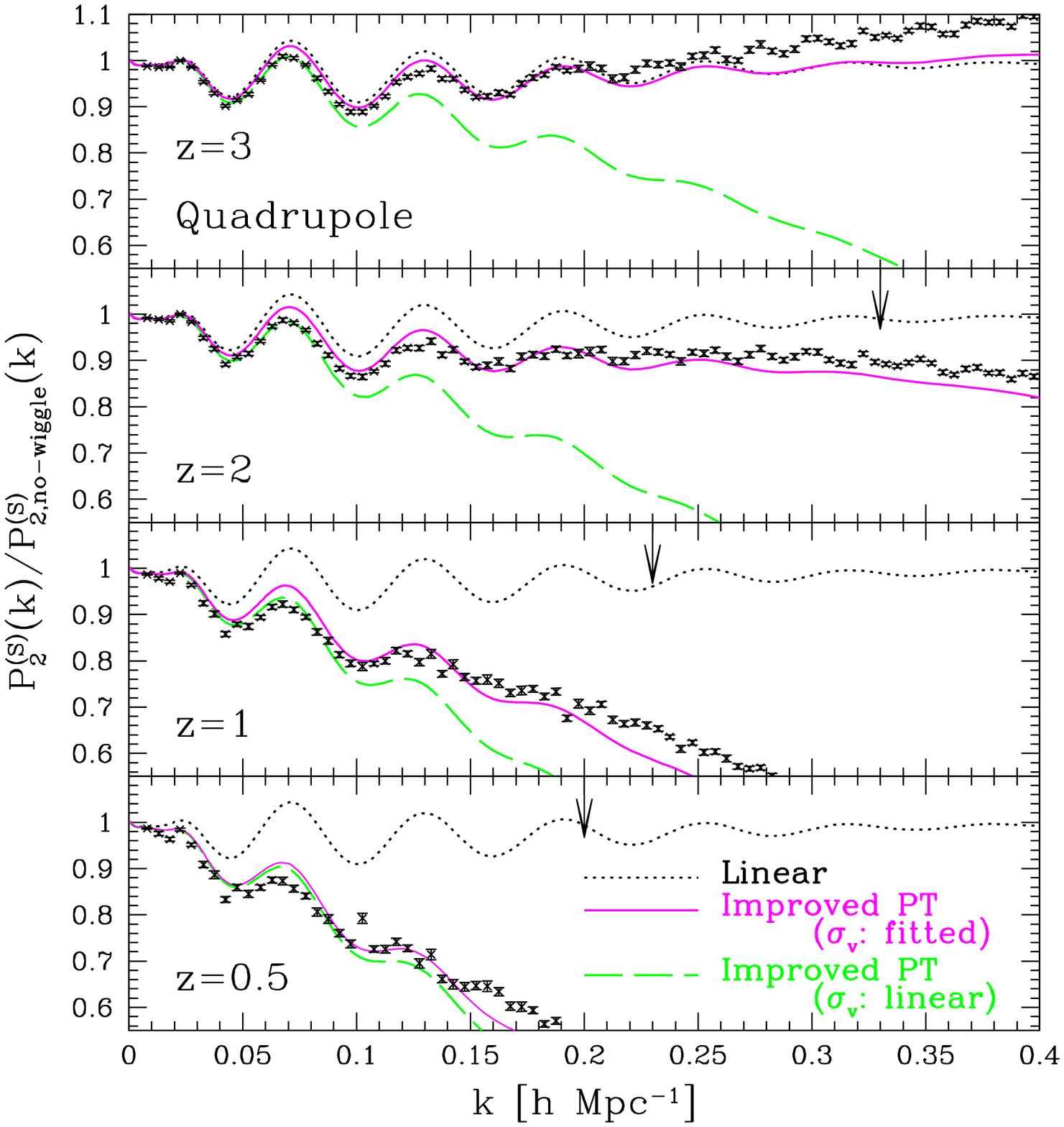}
\end{center}

\vspace*{-0.3cm}

\caption{Ratio of power spectra to smoothed reference spectra in 
  redshift space, 
  $P_\ell^{\rm (S)}(k)/P_{\ell,{\rm no\mbox{-}wiggle}}^{\rm (S)}(k)$, from the 
  {\tt wmap5} simulations. The reference spectrum 
  $P_{\ell,{\rm no\mbox{-}wiggle}}^{\rm (S)}$ is calculated 
  from the no-wiggle approximation of the linear transfer function, and 
  the linear theory of the Kaiser effect is taken into account. 
  Left panel shows the monopole power spectra ($\ell=0$), and 
  the right panel shows the quadrupole spectra ($\ell=2$). Solid and dashed
  lines represent the results from the improved PT adopting the model 
  of redshift-space distortion (\ref{eq:redshift_model}). 
  To plot the results, the linear theory 
  was used to compute $\sigma_{\rm v}$ in dashed lines, 
  while in solid lines, $\sigma_{\rm v}$ was determined 
  by fitting the predictions to the N-body simulations.  In each 
  panel, vertical arrow indicates the maximum wavenumber $k_{1\%}$ 
  for improved PT prediction including up to the second-order Born 
  approximation, which has been estimated from 
  Fig.~\ref{fig:ratio_pk2_alpha} (see Sec.~\ref{subsubsec:real_pk} for 
  definition of $k_{1\%}$). 
} 
\label{fig:ratio_pk2_red}
\end{figure}
%%%%%%%%%%%%%%%%%%%%%%%%%%%%%%%%%%%%%%%%%%%%%%%%%%%%%%%%%%%
%%%%%%%%%%%%%%%%%%%%%%%%%%%%%%%%%%%%%%%%%%%%%%%%%%%%%%%%%%%
\begin{figure}[h]
\begin{center}
\includegraphics[width=9cm,angle=0]{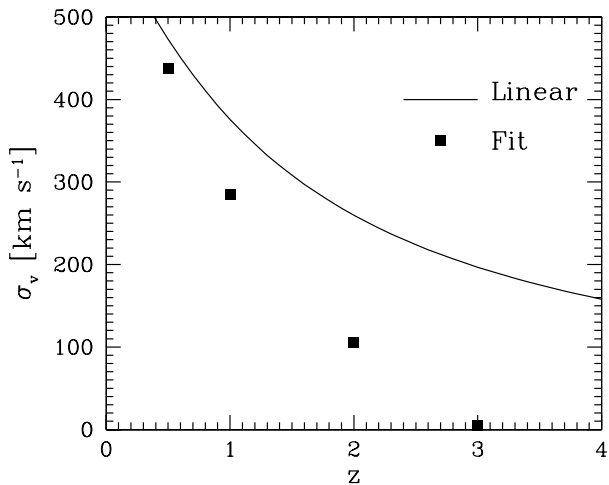}
\end{center}

\vspace*{-1.8cm}

\caption{Redshift evolution of velocity dispersion $\sigma_{\rm v}$. 
  While the solid lines represent the linear theory prediction, 
  the open squares indicate the results obtained by fitting the 
  model (\ref{eq:redshift_model}) to the monopole and quadrupole 
  spectra of N-body simulations (see Fig.~\ref{fig:ratio_pk2_red}). 
\label{fig:sigma_v}}
\end{figure}
%%%%%%%%%%%%%%%%%%%%%%%%%%%%%%%%%%%%%%%%%%%%%%%%%%%%%%%%%%%
%%%%%%%%%%%%%%%%%%%%%%%%%%%%%%%%%%%%%%%%%%%%%%%%%%%%%%%%%%%
\begin{figure}[h]
\begin{center}
\includegraphics[width=8cm,angle=0]{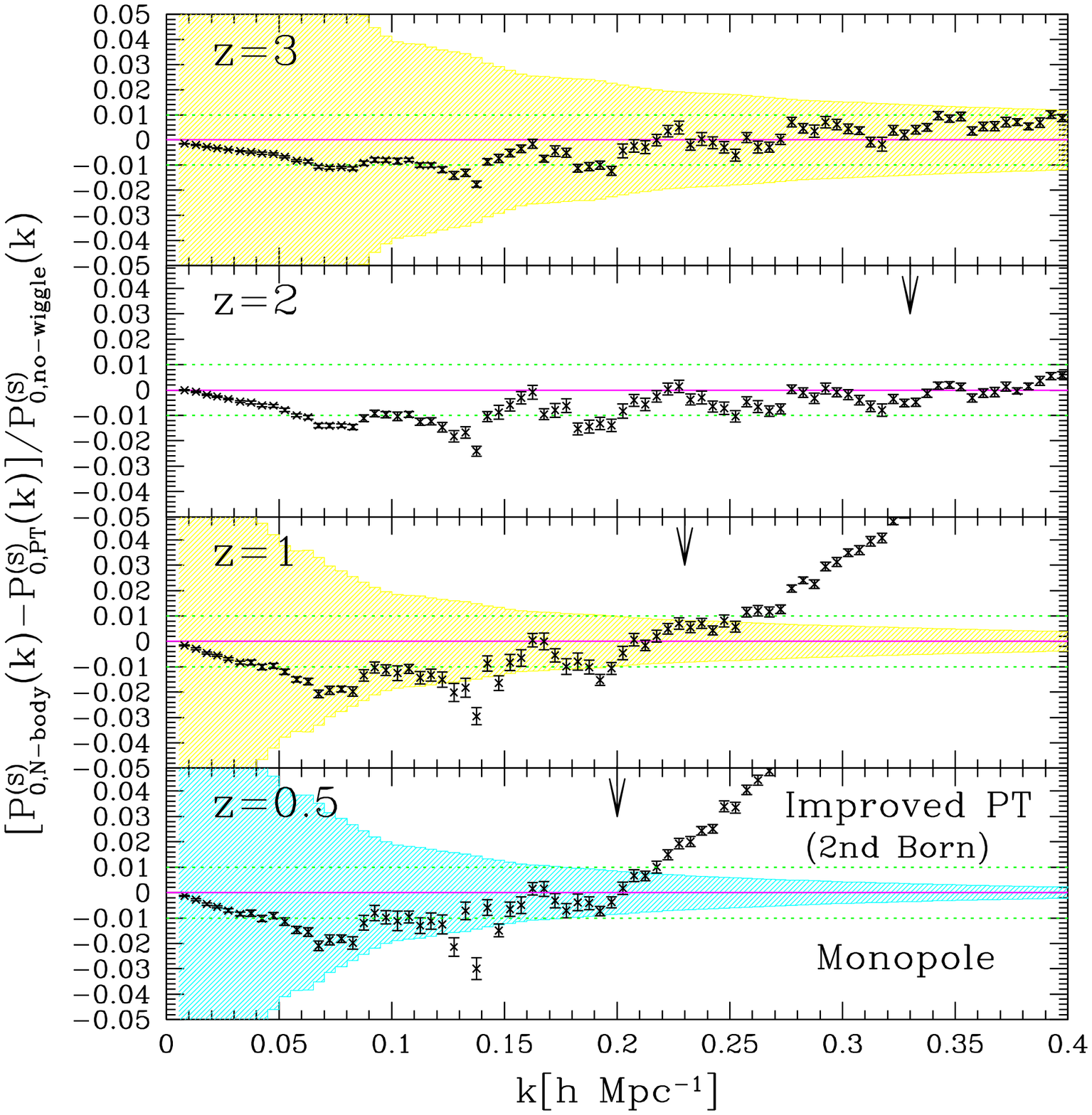}
\includegraphics[width=8cm,angle=0]{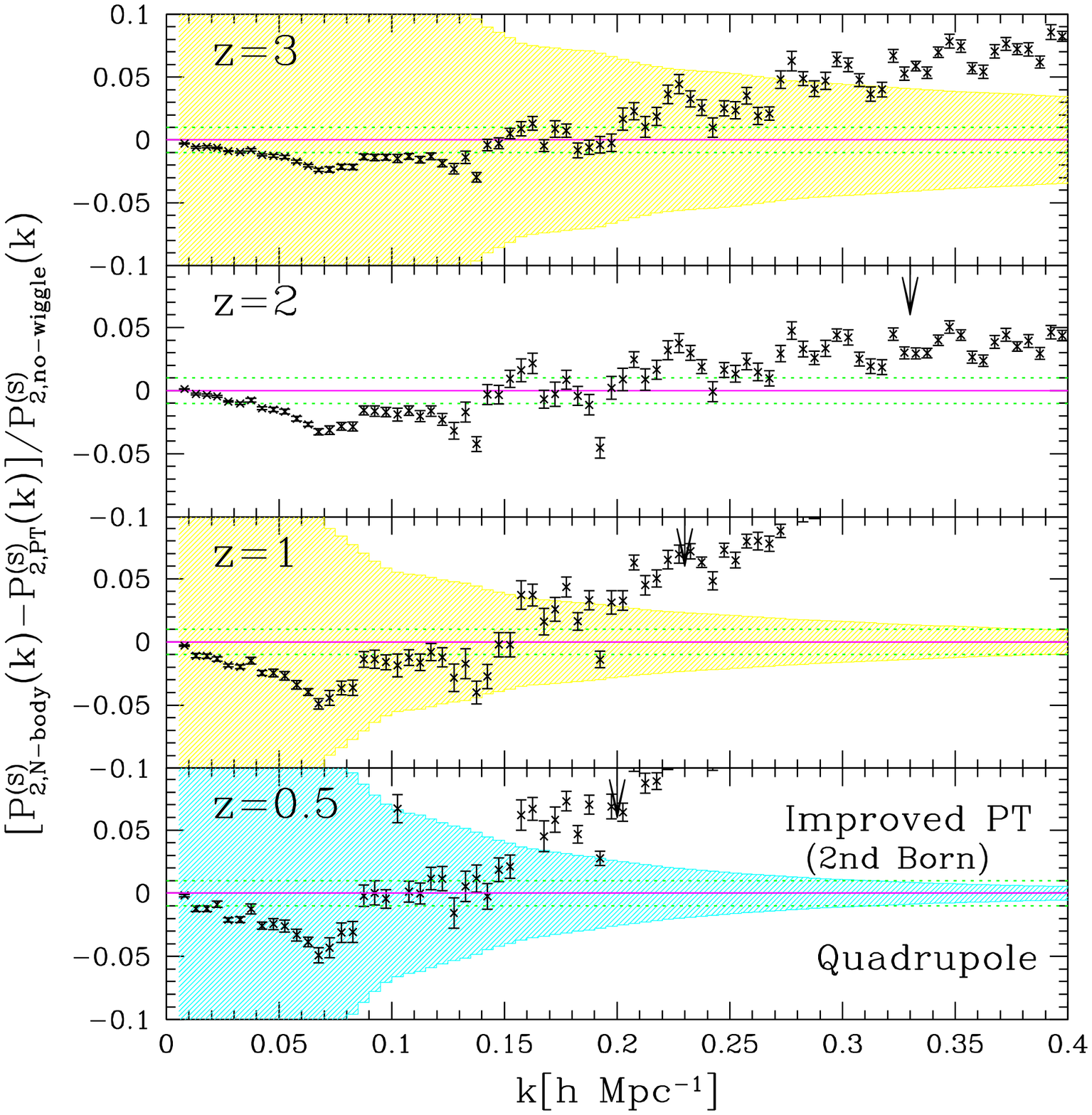}
\end{center}

\vspace*{-0.5cm}

\caption{Difference between N-body and PT results divided by the 
  reference spectrum in redshift space, i.e., 
  $[P_{\ell,{\rm N\mbox{-}body}}^{\rm(S)}(k)-P_{\ell,{\rm PT}}^{\rm(S)}(k)]/
  P_{\ell,{\rm no\mbox{-}wiggle}}^{\rm(S)}(k)$. 
 The left and right panels respectively represent the results 
  from monopole and quadrupole power spectra. 
  Note that the improved PT predictions are computed based on the 
  model (\ref{eq:redshift_model}) adopting the fitted value of 
  $\sigma_{\rm v}$. For comparison, the statistical errors limited by 
  the cosmic variance of the survey volumes roughly corresponding to 
  those of WFMOS-like survey\cite{Bassett:2005kn} 
  and BOSS\cite{Schlegel:2009hj} are shown as shaded regions 
  in panels of $z=3$, $z=1$ and $z=0.5$, assuming respectively
  the survey volumes of $V=1 h^{-3}\mbox{Gpc}^3$, 
  $4 h^{-3}\mbox{Gpc}^3$ and $4.5 h^{-3}\mbox{Gpc}^3$. 
  Note that in each panel, vertical arrow indicates the maximum wavenumber 
  $k_{1\%}$ determined from Fig.~\ref{fig:ratio_pk2_alpha} by comparison 
  between N-body and improved PT results. 
\label{fig:ratio_pk2_error_red}}
\end{figure}
%%%%%%%%%%%%%%%%%%%%%%%%%%%%%%%%%%%%%%%%%%%%%%%%%%%%%%%%%%%

For a quantitative comparison of model prediction with N-body simulation, 
we compute the multipole moments of the two-dimensional power 
spectrum  $P^{\rm(S)}(k,\mu)$: 
%%%%%%%%%%%%%%%%%%%%%%%%%%%%%%%%%%%%%%%%%%%%%%%%%%%%%%%%%%%%%%%%%%%%%%
\begin{equation}
P_{\ell}^{\rm(S)}(k)=\frac{2\ell+1}{2}\int_{-1}^{1}d\mu\, P^{\rm(S)}(k,\mu)\,
\mathcal{P}_\ell(\mu), 
\end{equation}
%%%%%%%%%%%%%%%%%%%%%%%%%%%%%%%%%%%%%%%%%%%%%%%%%%%%%%%%%%%%%%%%%%%%%%
with $\mathcal{P}_\ell$ being the Legendre polynomials. 

Substituting the model (\ref{eq:redshift_model}) into the above, 
the monopole, quadrupole and hexadecapole contribution to the 
redshift-space power spectrum are analytically expressed as 
%%%%%%%%%%%%%%%%%%%%%%%%%%%%%%%%%%%%%%%%%%%%%%%%%%%%%%%%%%%%%%%%%%%%%%
\begin{eqnarray}
P_0^{\rm(S)}(k)&=& p_0(k),
\label{eq:monopole_pk}
\\
P_2^{\rm(S)}(k)&=& \frac{5}{2}\left\{3p_1(k)-p_0(k)\right\},
\label{eq:quadrupole_pk}
\\
P_4^{\rm(S)}(k)&=&\frac{9}{8}\left\{35p_2(k)-30p_1(k)+3p_0(k)\right\}.
\label{eq:hexadecapole_pk}
\end{eqnarray}
%%%%%%%%%%%%%%%%%%%%%%%%%%%%%%%%%%%%%%%%%%%%%%%%%%%%%%%%%%%%%%%%%%%%%%
where the function $p_n(k)$ is defined by 
%%%%%%%%%%%%%%%%%%%%%%%%%%%%%%%%%%%%%%%%%%%%%%%%%%%%%%%%%%%%%%%%%%%%%%
\begin{equation}
p_n(k)=\frac{1}{2}\left[
 \frac{\gamma(n+1/2,\kappa)}{\kappa^{n+1/2}}P_{11}(k)+
 2\,\frac{\gamma(n+3/2,\kappa)}{\kappa^{n+3/2}}\,f\,P_{12}(k)+
 \frac{\gamma(n+5/2,\kappa)}{\kappa^{n+5/2}}\,f^2\,P_{22}(k)\right].
\end{equation}
%%%%%%%%%%%%%%%%%%%%%%%%%%%%%%%%%%%%%%%%%%%%%%%%%%%%%%%%%%%%%%%%%%%%%%
The quantity $\gamma(n,\kappa)$ is the incomplete gamma function 
of the first kind 
%%%%%%%%%%%%%%%%%%%%%%%%%%%%%%%%%%%%%%%%%%%%%%%%%%%%%%%%%%%%%%%%%%%%%%
\begin{equation}
\gamma\left(n,\kappa\right)=\int_0^\kappa dt \,\,t^{n-1}e^{-t}
\label{eq:def_incomplete_gamma}
\end{equation}
%%%%%%%%%%%%%%%%%%%%%%%%%%%%%%%%%%%%%%%%%%%%%%%%%%%%%%%%%%%%%%%%%%%%%%
with its argument $\kappa=(k\,f\,\sigma_{\rm v})^2$.

Fig.~\ref{fig:ratio_pk2_red} shows the monopole (left) and quadrupole 
(right) moments  
of the redshift-space power spectra at different redshifts, obtained from 
the \verb|wmap5| simulations. We do not plot here the hexadecapole 
contributions,   
because the power spectrum estimated from the N-body simulations is 
still noisy even with the 30 realizations. 
In each panel of Fig.~\ref{fig:ratio_pk2_red}, the dashed lines indicate 
the improved PT predictions based on the model (\ref{eq:redshift_model}), 
where the corrections up to the second-order Born approximations are included. 
Note that the velocity dispersion $\sigma_{\rm v}$ is computed from the 
linear theory. Clearly, the predictions all underestimate the N-body results, 
and the agreement between predictions and N-body simulations is restricted to 
a quite narrow range on large scales. As a reference, we also show 
the maximum wavenumber $k_{1\%}$ of the improved PT prediction 
(vertical arrows), in which we include the corrections up to the 
second-order Born approximation in real space (see 
Fig.~\ref{fig:ratio_pk2_alpha}).

The reason why the prediction generically underestimates the N-body 
simulations would be partly attributed to the calculation of the 
velocity dispersion $\sigma_{\rm v}$ 
using the linear theory. It has been advocated by several authors that 
the suppression of power spectrum by FoG effect is originated from the 
non-linear structure of virialized halos,  
and thereby the linear theory estimation of $\sigma_{\rm v}$ may be 
inappropriate. In this respect, we admittedly regard $\sigma_{\rm v}$ as 
an uncontrollable parameter, which should be determined by fitting the 
predictions to N-body results.

The solid lines in each panel of Fig.~\ref{fig:ratio_pk2_red} show the 
results of redshift-space spectra adopting the fitted values of 
$\sigma_{\rm v}$. In estimating $\sigma_{\rm v}$, both the monopole and 
quadrupole spectra were 
fitted to the N-body results in the range of $0\leq k \leq k_{\rm 1\%}$. 
Fig.~\ref{fig:sigma_v} summarizes the fitted results of $\sigma_{\rm v}$, 
which significantly deviate from the linear theory prediction at higher 
redshifts.

Then, apparently, overall agreement between prediction and 
simulation becomes fairly improved, although as a trade-off, 
small discrepancy manifests at low-k mode, where the 
N-body results rather agree well with the prediction 
adopting $\sigma_{\rm v}$ calculated from linear theory. 
In Fig.~\ref{fig:ratio_pk2_error_red},  left and right panels respectively 
plot the fractional differences of the monopole and quadrupole moments 
between the model predictions and  N-body simulations. 
Except for the narrow range of low-k modes,  
a percent-level agreement is almost achieved 
for the monopole power spectrum. This is true    
at least within the convergence regime calibrated in real space (see 
vertical arrows in Fig.~\ref{fig:ratio_pk2_error_red}). 
However, the fractional error of the quadrupole power spectrum 
still exhibits a little bit large discrepancy, 
signaling the fact that the model (\ref{eq:redshift_model}) 
misses something important for higher-multipole moment of 
redshift-space distortion.

To see the significance of this deviation in practice, 
in Fig.~\ref{fig:ratio_pk2_error_red}, 
the expected $1\mbox{-}\sigma$ errors limited by the cosmic variance, 
$\Delta P_\ell^{\rm(S)}(k)$, are shown, depicted as the shaded region. 
Here, we specifically consider the ground-based BAO surveys like WFMOS 
survey\cite{Bassett:2005kn} and BOSS\cite{Schlegel:2009hj},  
assuming the survey volumes of $V=1\,h^{-3}\mbox{Gpc}^3$ at $z=3$ and 
$4\,h^{-3}\mbox{Gpc}^3$ at $z=1$ for WFMOS survey, and 
$V=4.5\,h^{-3}\mbox{Gpc}^3$ at $z=0.5$ for BOSS\footnote{Strictly speaking, 
BOSS project is a part of Sloan Digital Sky Survey III, aiming at precisely 
measuring the cosmological distance and expansion rate at $z=0.35$, $0.6$ and 
$z=2.5$. Here, we only consider the low-$z$ measurement with survey depth 
$0.2\lesssim z \lesssim0.8$. }. 
Based on the approximation that the 
density field is well-described by a Gaussian random field, 
the cosmic-variance error $\Delta P_\ell^{\rm(S)}(k)$ can be 
estimated as
%%%%%%%%%%%%%%%%%%%%%%%%%%%%%%%%%%%%%%%%%%%%%%%%%%%%%%%%%%%%%%%%%%%%%%
\begin{equation}
[\Delta P_\ell^{\rm(S)}(k)]^2=\frac{2}{N_k}\,\sigma_{P,\ell}^2(k),
\label{eq:Var_Pk}
\end{equation}
%%%%%%%%%%%%%%%%%%%%%%%%%%%%%%%%%%%%%%%%%%%%%%%%%%%%%%%%%%%%%%%%%%%%%%
where the quantity $N_k$ is the number of Fourier modes within a given 
bin at $k$, and is given by 
$N_k=4\pi\,k^2\Delta k/(2\pi/L_{\rm box})^3/2=V\,k^2\,\Delta k/(2\pi)^2$. 
The function $\sigma_{P,\ell}$ is 
%%%%%%%%%%%%%%%%%%%%%%%%%%%%%%%%%%%%%%%%%%%%%%%%%%%%%%%%%%%%%%%%%%%%%%
\begin{equation}
\sigma_{P,\ell}^2(k)=\frac{(2\ell+1)^2}{2}
\int_{-1}^{1} d\mu\,\left\{P^{\rm(S)}(k,\mu)\,
\mathcal{P}_\ell(\mu)\right\}^2.
\label{eq:sigma_P_ell}
\end{equation}
%%%%%%%%%%%%%%%%%%%%%%%%%%%%%%%%%%%%%%%%%%%%%%%%%%%%%%%%%%%%%%%%%%%%%%
The expression (\ref{eq:Var_Pk}) with (\ref{eq:sigma_P_ell}) 
is a generalization of 
the cosmic-variance error in real space 
(e.g., \cite{Bernardeau:2001qr,Meiksin:1998mu,Scoccimarro:1999kp,
Takahashi:2009bq}) to the multipole moments 
in redshift space. Note that the error $\Delta P_\ell^{\rm(S)}(k)$ 
depends on the bin width $\Delta k$, for which we simply adopt 
the same bin size as used in the power spectrum analysis of N-body 
data. The analytic estimate of $\Delta P_\ell^{\rm(S)}$ based on 
Eq.~(\ref{eq:Var_Pk}) is roughly consistent with the statistical 
errors estimated from the N-body data of 30 realizations.

Comparison between the cosmic-variance errors and fractional 
differences shows that the discrepancy seen in the quadrupole 
power spectrum is definitely large, and it eventually exceeds the 
statistical error at large $k$ modes. Since this has happened  
inside the valid range of the improved PT calibrated in real space 
(indicated as vertical arrows),  we conclude that 
the current model prediction with (\ref{eq:redshift_model}) 
is insufficient to describe the higher-multipole moments of BAOs, and   
a more elaborate work on the models of redshift-space distortion is 
needed for upcoming BAO measurement.

%-%-%-%-%-%-%-%-%-%-%-%-%-%-%-%-%-%-%-%-%-%-%-%-%-%-%-%
\subsubsection{Correlation function}
%-%-%-%-%-%-%-%-%-%-%-%-%-%-%-%-%-%-%-%-%-%-%-%-%-%-%-%

Finally, we discuss the correlation functions in redshift space. 
Similar to the power spectrum, we apply the multipole expansion to  
the anisotropic two-point correlation function as 
%%%%%%%%%%%%%%%%%%%%%%%%%%%%%%%%%%%%%%%%%%%%%%%%%%%%%%%%%%%%%%%%%%%%%%
\begin{equation}
\xi^{\rm(S)}(s_\parallel,s_\perp)=
\int \frac{d^3\bfk}{(2\pi)^3}\,P^{\rm(S)}(k,\mu)e^{i\bfk\cdot\bfs}=
\sum_{\ell:{\rm even}}\xi_{\ell}^{\rm(S)}(s)\mathcal{P}_{\ell}(\nu)
\end{equation}
%%%%%%%%%%%%%%%%%%%%%%%%%%%%%%%%%%%%%%%%%%%%%%%%%%%%%%%%%%%%%%%%%%%%%%
with $\nu=s_\parallel/s$. The multipole moment of the correlation function, 
$\xi_{\ell}^{\rm(S)}$, is directly related to the Fourier counterpart, 
$P^{\rm(S)}_\ell$ through 
%%%%%%%%%%%%%%%%%%%%%%%%%%%%%%%%%%%%%%%%%%%%%%%%%%%%%%%%%%%%%%%%%%%%%%
\begin{equation}
\xi_{\ell}^{\rm(S)}(s)=i^{\ell} \int\frac{dkk^2}{2\pi^2}\,
P_\ell^{\rm(S)}(k)j_\ell(ks). 
\end{equation}
%%%%%%%%%%%%%%%%%%%%%%%%%%%%%%%%%%%%%%%%%%%%%%%%%%%%%%%%%%%%%%%%%%%%%%

\begin{figure}[t]
\begin{center}
\includegraphics[width=6.5cm,angle=0]{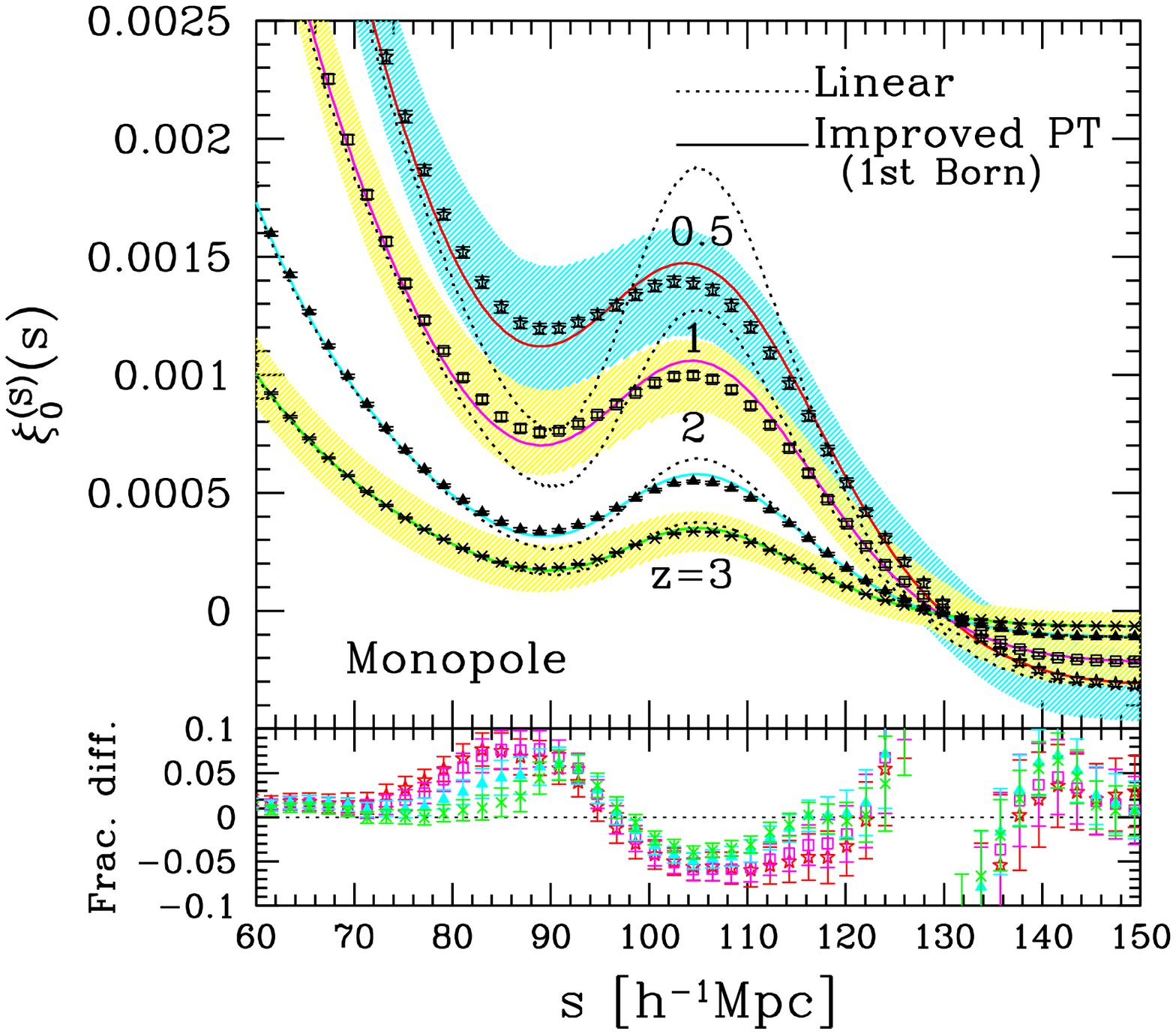}
\hspace*{-1.0cm}
\includegraphics[width=6.5cm,angle=0]{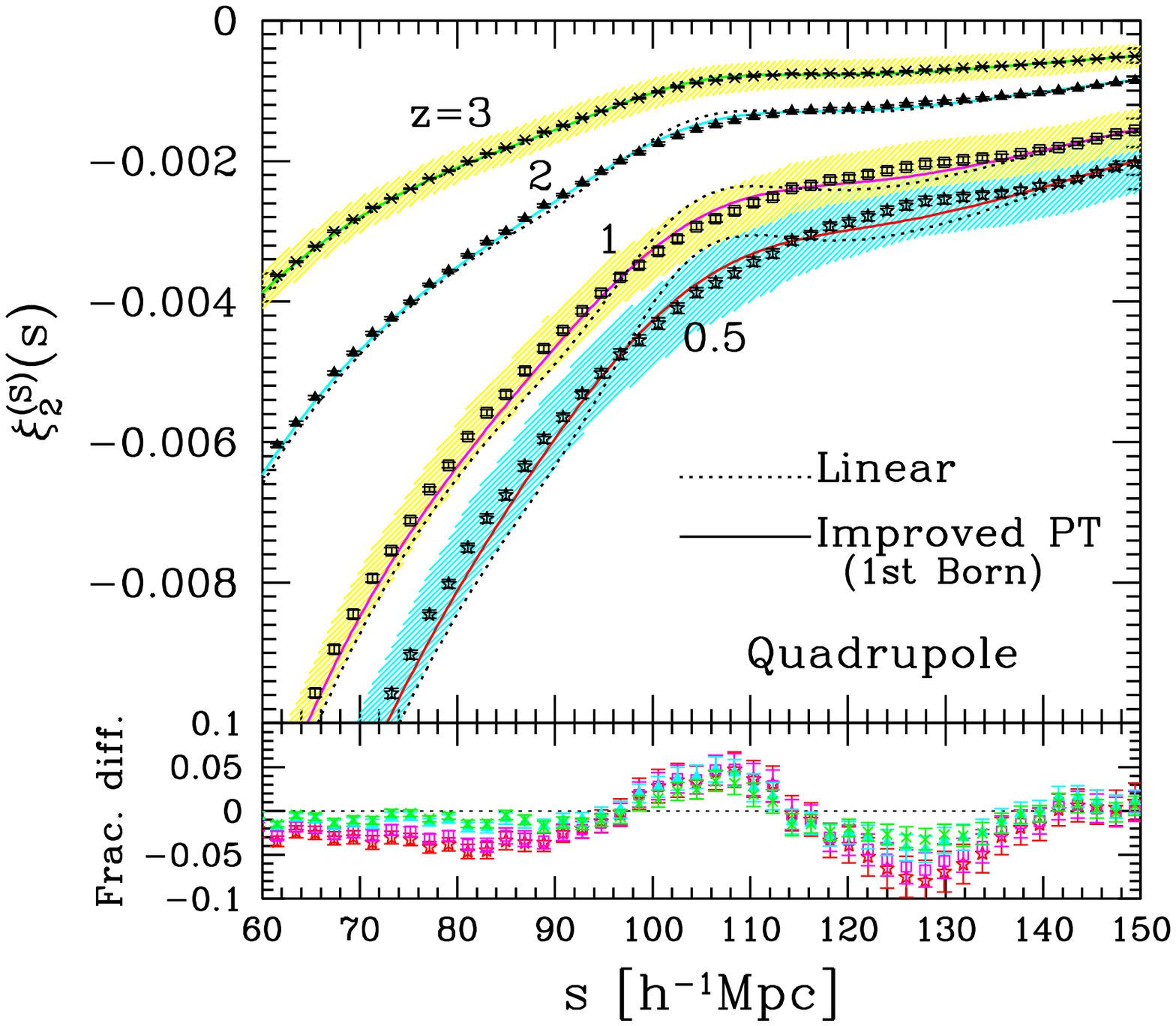}
\hspace*{-1.0cm}
\includegraphics[width=6.5cm,angle=0]{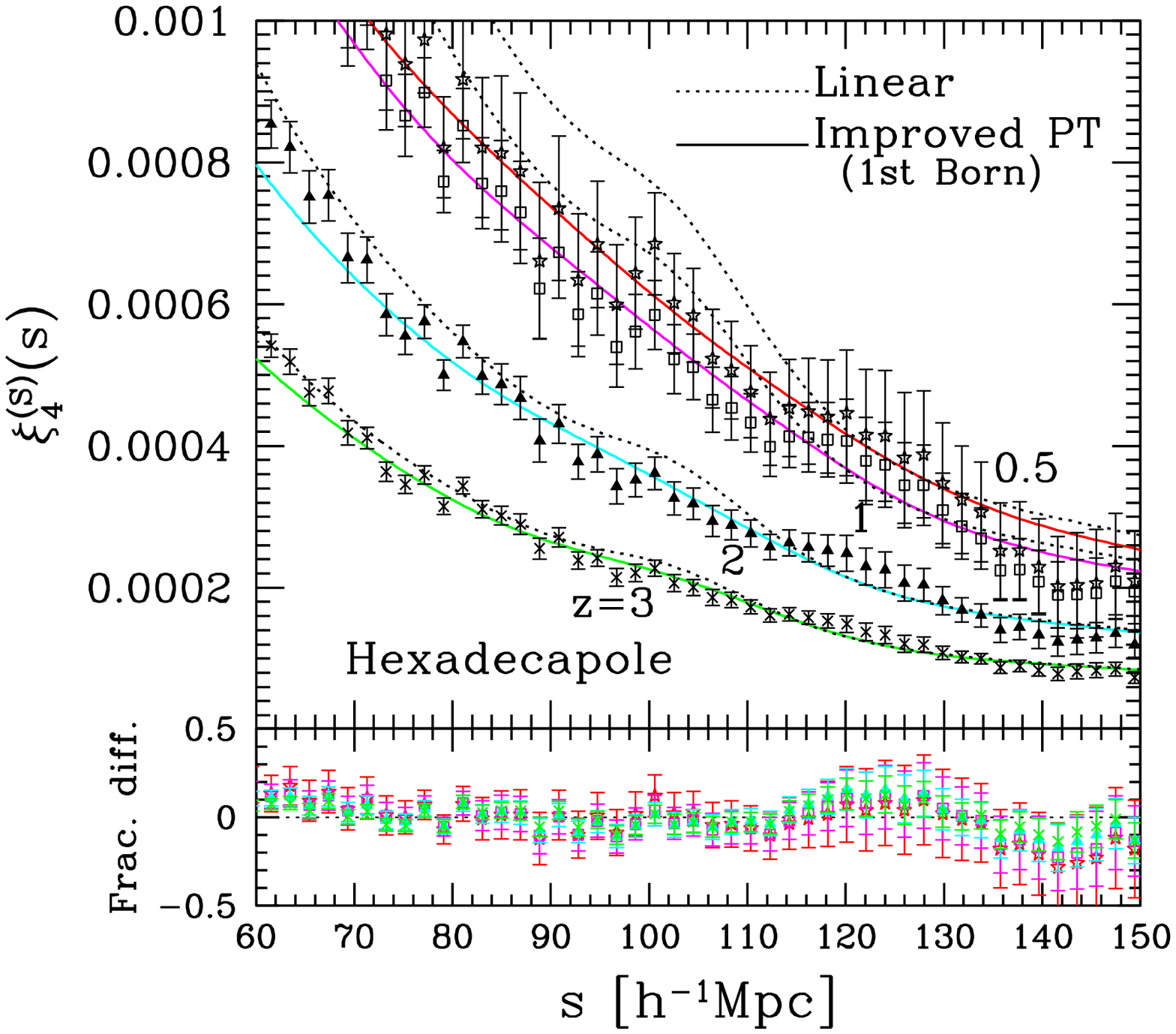}
\end{center}

\vspace*{-0.8cm}

\caption{Top: Correlation function in redshift space. 
  Left, middle and right panel respectively 
  shows the monopole, quadrupole and hexadecapole contributions 
  to the anisotropic correlation function
  $\xi^{\rm(S)}$. The solid and dotted lines 
  are the predictions from the improved PT based on the model 
  (\ref{eq:redshift_model}) and linear theory, respectively. 
  Note that only the leading-order 
  Born approximation to the the mode coupling term is included in 
  the improved PT.  $z=0.5$ (red); $z=1$ (magenta); $z=2$ (cyan); 
  $z=3$ (green). For comparison, the statistical errors limited by 
  the cosmic variance of the survey volumes $V=1 h^{-3}\mbox{Gpc}^3$,  
  $4 h^{-3}\mbox{Gpc}^3$ and $4.5 h^{-3}\mbox{Gpc}^3$ 
  are estimated from Eq.~(\ref{eq:delta_xi}),  
  and are depicted as shaded regions 
  around the N-body results at $z=3$, $z=1$ and $z=0.5$, respectively. 
  The cosmic-variance error for hexadecapole is not shown here because 
  of the large scatter. Bottom: Fractional differences of the results 
  between N-body simulations and improved PT predictions, 
  $[\xi_{\rm N\mbox{-}body}(s)-\xi_{\rm PT}(s)]/\xi_{\rm PT}(s)$ for 
  different redshifts at $z=0.5$ (open stars), $z=1$ (open squares), 
  $z=2$ (filled triangles), and $z=3$ (crosses).
} 
\label{fig:xi2_red}
\end{figure}

Fig.~\ref{fig:xi2_red} shows the monopole (left), quadrupole 
(middle) and hexadecapole (right) moments of correlation function. 
In each panel, the N-body results are compared with the predictions from 
linear theory (dotted) and the leading-order calculation of improved 
PT (solid) adopting the model (\ref{eq:redshift_model}) 
with linear theory prediction of $\sigma_{\rm v}$. 
Note that the predictions of improved PT are hardly changed 
by including the higher-order corrections and/or using the fitted value 
of $\sigma_{\rm v}$, at least around the baryon acoustic peak, and 
the systematic differences between including and ignoring the 
corrections are well within the error-bars of N-body simulations.

As anticipated from the results in real space, 
the baryon acoustic peaks in the monopole moment tend to be 
smeared as decreasing redshift, but the effect seems little bit 
stronger than those in real space. This is due to the additional 
effect coming from the redshift distortion.  
Although no prominent signal of the BAOs exists in 
the higher-multipole moments,  the same tendencies can be seen in 
the quadrupole and hexadecapole moments. The 
improved PT calculations are broadly consistent with N-body results,  
but small discrepancies manifest around the baryon acoustic peak and trough.   
Lower panels of Fig.~\ref{fig:xi2_red} showing 
the fractional differences imply that these are at most $5\%$ effect 
in amplitude, except for the hexadecapole case with large error-bars of 
simulation. It is interesting to note 
that no noticeable redshift dependence appears in the fractional differences, 
indicating that the discrepancies may be 
attributed to the model of redshift-space distortion. 
Furthermore,  it turns out that these are well 
within the cosmic-variance errors of the ground-based BAO measurement, 
indicated as shaded region. Assuming that the underlying 
density field is well described by a Gaussian random field, 
the cosmic variance for the multipole  
correlation functions $\xi_{\ell}^{(\rm S)}$ can be written as 
(see \cite{Smith:2007gi,Cohn:2005ex,Bernstein:1993nb} 
for cosmic-variance errors in real space)
%%%%%%%%%%%%%%%%%%%%%%%%%%%%%%%%%%%%%%%%%%%%%%%%%%%%%%%%%%%%%%%%%%%%%%
\begin{equation}
\left[\Delta \xi_\ell^{\rm(S)}(s)\right]^2 = \frac{2}{V}\,\int
\frac{dk\,k^2}{2\pi^2}\,\{j_\ell(ks)\,\sigma_{P,\ell}(k)\}^2  
\label{eq:delta_xi}
\end{equation}
%%%%%%%%%%%%%%%%%%%%%%%%%%%%%%%%%%%%%%%%%%%%%%%%%%%%%%%%%%%%%%%%%%%%%%
with $\sigma_{P,\ell}$ given by Eq.~(\ref{eq:sigma_P_ell}). 
Note that the analytic estimation of cosmic-variance errors 
$\Delta \xi_\ell^{\rm(S)}$ shown in Fig.~\ref{fig:xi2_red} 
reproduce the N-body results quite well.

Hence, compared to the power spectrum in redshift space,  
the correlation functions obtained from 
the N-body simulation and analytic calculation can have a better 
agreement. Presumably, this is because the acoustic peak structure 
in the correlation function is mostly attributed 
to the low-k behavior of the BAOs, and 
the power spectrum at low-k modes 
is accurately described by the model (\ref{eq:redshift_model}) 
with the linear theory prediction of $\sigma_{\rm v}$. 
In other words, the baryon acoustic peak would be robust 
against the non-linear effects at high-k modes (see also 
\cite{Matsubara:2008wx, Matsubara:2007wj, Sanchez:2008iw}). This 
implies that even the prediction at the current level 
is sufficient to characterize the acoustic peak 
in the correlation function, and it can be used as an accurate 
theoretical template for future precision BAO measurement.

Note, however, that the measured amplitudes of the two-point 
correlation function are strongly correlated between different 
scales. In practice, not only the diagonal component but also 
the off-diagonal components of the covariance of the correlation 
function must be considered for a reliable estimation of 
cosmological distance, and a more careful study is needed. 

%%%%%%%%%%%%%%%%%%%%%%%%%%%%%%%%%%%%%%%%%%%%%%%%%%%%%%%%%%%
%%%%%%%%%%%%%%%%%%%%%%%%%%%%%%%%%%%%%%%%%%%%%%%%%%%%%%%%%%%

%%%%%%%%%%%%%%%%%%%%%%%%%%%%%%%%%%%%%%%%%%%%%%%%%%%%%%%
%%%%%%%%%%%%%%%%%%%%%%%%%%%%%%%%%%%%%%%%%%%%%%%%%%%%%%%
\section{Discussion and Conclusions}
\label{sec:conclusion}
%%%%%%%%%%%%%%%%%%%%%%%%%%%%%%%%%%%%%%%%%%%%%%%%%%%%%%%
%%%%%%%%%%%%%%%%%%%%%%%%%%%%%%%%%%%%%%%%%%%%%%%%%%%%%%%

In this paper, we have presented the improved PT calculations of 
the matter power spectrum and two-point correlation function in real and 
redshift spaces. Based on the closure approximation of 
the renormalized PT treatment, 
a closed set of the non-perturbative expressions 
for power spectrum and propagator is obtained. 
The resultant expression includes the effect of resummation for 
a class of loop diagrams at infinite order, and thereby 
the convergence of higher-order contributions is expected to be 
improved. Employing the Born approximation, we have analytically 
calculated the non-linear power spectrum, and compared 
the convergence properties of improved PT with those of standard 
PT by explicitly computing the higher-order corrections.

We have also made a detailed comparison between the improved PT result
and N-body simulations. 
With a large boxsize and many realization data of N-body simulations, 
the statistical errors of two-point statistics are greatly reduced 
by the correction of the effect of finite-mode sampling, 
and this enables us to investigate the 
convergence check of numerical and analytic calculations 
at a percent level. Then, specifically focusing on the behaviors of BAOs, 
the power spectrum and two-point correlation functions are calculated in 
both real and redshift spaces. 
In redshift space, the effect of redshift-space distortion which 
changes the clustering pattern of mass distribution should 
be incorporated into the improved PT predictions. In this paper,  
adopting the model proposed by Ref.~\cite{Scoccimarro:2004tg} 
(Eq.~(\ref{eq:redshift_model}) ), we have quantified 
the extent to which the current model description 
faithfully reproduces the N-body results, and clarified the 
key ingredients toward an improved prescription of redshift-space 
distortion.

Our important findings are summarized as follows:

\begin{itemize}
\item The improved PT expansion based on the 
  Born approximation has better convergence properties, 
  in marked contrast with the standard PT expansion. The 
  corrections coming from the mode-coupling term are well-localized 
  positive functions of wavenumbers, and their contributions tend to be 
  shifted to a higher $k$ region as increasing the order of perturbation. 
  Thus, the inclusion of higher-order corrections stably improves the 
  prediction, and the range of agreement with N-body results becomes 
  wider in wavenumber. 

\item In real-space power spectrum, the improved PT prediction 
  including up to the second-order Born correction 
  seems essential for modeling BAO precisely. We estimated the maximum 
  wavenumber $k_{1\%}$, below which the results of both the N-body simulation 
  and improved PT calculation converge well within the $1\%$ accuracy. The 
  resultant value of $k_{1\%}$ can be summarized as Eq.~(\ref{eq:k_criterion}) 
  with the constant value $C=0.7$, which provides a way to 
  estimate $k_{1\%}$ in a cosmology independent manner. 
  On the other hand, if we consider the two-point 
  correlation function in real space, the leading-order calculation 
  turns out to be sufficiently accurate, and no higher-order correction is 
  needed to describe the non-linear evolution of baryon acoustic peak seen 
  in the N-body simulations. 

\item Modeling redshift-space power spectrum with 
  Eq.~(\ref{eq:redshift_model}) gives a broadly consistent result with 
  N-body simulations, if we regard the velocity dispersion $\sigma_{\rm v}$ 
  as a fitting parameter. However, discrepancy between improved 
  PT predictions and N-body results has appeared in the quadrupole power 
  spectrum, and it becomes larger than the statistical errors limited by the 
  cosmic variance of the survey volume $V\sim$ a few $h^{-3}$Gpc$^3$. This 
  is true even in the valid range of improved PT, $k\lesssim k_{1\%}$. On 
  the other hand, while a small descrepancy has been also found in the 
  two-point correlation, it turns out that the discrepancy is well within 
  the cosmic-variance error, and even the leading-order prediction using 
  the linear theory estimate of $\sigma_{\rm v}$ can be used as an accurate 
  theoretical template for future ground-based BAO measurement. 
\end{itemize}

The recently proposed techniques to deal with the non-linear gravitational 
clustering, including the present treatment, have been greatly 
developed, and they would be a promising cosmological tool to precisely model 
the shape and amplitude of the power spectrum and/or the correlation 
functions in an accuracy of sub-percent level. Combining the 
model of redshift-space distortion, we are now able to discuss 
the non-linear clustering in redshift space. Although 
the present paper is especially concerned with the analytical work, 
we note that the non-perturbative formulation with closure approximation 
is suited for forward treatment in time \cite{Hiramatsu:2009ki}, in which 
all orders of Born approximation can be fully incorporated into the 
predictions by numerically solving the evolution equations.  
This approach would be particularly useful to study the non-linear 
matter power spectrum in general cosmological models, including the 
modified theory of gravity \cite{Koyama:2009me}.

Finally,  in practical application to the precision BAO measurements, 
there are several remaining issues to be addressed in the future work. 
The improvement of the model of redshift-space distortion is, 
of course, a very important and urgent task. The effect of galaxy biasing 
is also one of the key ingredients for modeling accurate theoretical 
template, and several attempts to take account of this effect have been 
recently made 
\cite{Heavens:1998es,Matsubara:2008wx,Jeong:2008rj,Smith:2006ne,
  McDonald:2006mx,McDonald:2009dh,Taruya:1999vq}. 
Another interesting direction is to develop a 
fast computation of non-linear power spectrum or correlation function for an 
arbitrary cosmological model. Recently, the statistical sampling method for 
precise power spectrum emulation has been proposed 
\cite{Heitmann:2006hr,Heitmann:2009cu,Habib:2007ca}. 
In this treatment, 
only a limited set of cosmological models can be used to predict power 
spectrum at the required accuracy over the prior parameter ranges. 
The analytic approaches combining this method may provide a fast 
and reliable way to estimate the two-point statistics, and the development
of this method would be valuable.

%%%%%%%%%%%%%%%%%%%%%%%%%%%%%%%%%%%%%%%%%%%%%%%%%%%%
\begin{acknowledgments}
We would like to thank Yasushi Suto and Alan Heavens for 
comments and discussion, Thierry 
Sousbie for teaching us an efficient computational method for two-point 
correlation function. AT is supported by a Grant-in-Aid for Scientific 
Research from the Japan Society for the Promotion of Science (JSPS) 
(No.~21740168). TN and SS acknowledge a support from JSPS fellows. 
This work was supported in part by 
Grant-in-Aid for Scientific Research on Priority Areas No.~467 
``Probing the Dark Energy through an Extremely Wide and Deep Survey with 
Subaru Telescope'', and JSPS Core-to-Core Program ``International 
Research Network for Dark Energy''.
\end{acknowledgments} 
%%%%%%%%%%%%%%%%%%%%%%%%%%%%%%%%%%%%%%%%%%%%%%%%%%%%

\appendix
%%%%%%%%%%%%%%%%%%%%%%%%%%%%%%%%%%%%%%%%%%%%%%%%%%%%%%%
%%%%%%%%%%%%%%%%%%%%%%%%%%%%%%%%%%%%%%%%%%%%%%%%%%%%%%%
\section{Standard Perturbation Theory up to the Two-loop Order}
\label{app:SPT}
%%%%%%%%%%%%%%%%%%%%%%%%%%%%%%%%%%%%%%%%%%%%%%%%%%%%%%%
%%%%%%%%%%%%%%%%%%%%%%%%%%%%%%%%%%%%%%%%%%%%%%%%%%%%%%%

In this Appendix, we briefly summarize the standard PT and derive 
a set of perturbative solutions. Based on these solutions,  
we obtain the analytic expressions for power spectrum up to the 
two-loop order.

As we mentioned in Sec.~\ref{subsec:SPT_vs_RPT}, standard PT is the 
straightforward expansion of the quantity $\Phi_a$, and the perturbative 
solutions are obtained by order-by-order treatment of 
Eq.~(\ref{eq:vec_fluid_eq}). In order to systematically derive the 
solutions, the Einstein-de Sitter (EdS) approximation is often used in the 
literature. In the EdS approximation, the matrix $\Omega_{ab}$ given 
by Eq.~(\ref{eq:matrix_M}) is replaced with the one in the EdS universe, i.e., 
$\Omega_{\rm}(\eta)=1$ and $f=dlnD/dln a= 1$. This means that 
all the non-linear growth factors 
appearing in the higher-order solutions are expressed in 
terms of the linear growth factor $D(t)$. Neglecting the 
contributions from the decaying mode, 
the resultant solution for $\Phi_a$ is then expanded as  
%%%%%%%%%%%%%%%%%%%%%%%%%%%%%%%%%%%%%%%%%%%%%%%%%%%%%%%
\begin{equation}
\Phi_a(\bfk;\eta) = e^{\eta} \Phi_a^{(1)}(\bfk;\eta) 
+ e^{2\eta}  \Phi_a^{(2)}(\bfk;\eta) +  e^{3\eta}  
\Phi_a^{(3)}(\bfk;\eta)+ \cdots, 
\label{eq:expand}
\end{equation}
%%%%%%%%%%%%%%%%%%%%%%%%%%%%%%%%%%%%%%%%%%%%%%%%%%%%%%%
The solution for each order of perturbation is expressed as
%%%%%%%%%%%%%%%%%%%%%%%%%%%%%%%%%%%%%%%%%%%%%%%%%%%%%%%
\begin{equation}
\Phi_a^{(n)}(\bfk)= \int\frac{d^3\bfk_1\cdots d^3\bfk_n}{(2\pi)^{3(n-1)}}\,
\delta_{\rm D}(\bfk-\bfk_1-\cdots-\bfk_n)\,\mathcal{F}_{a}^{(n)}
(\bfk_1,\cdots,\bfk_n)\delta_0(\bfk_1)\cdots\delta_0(\bfk_n), 
\label{eq:PT_sol}
\end{equation}
%%%%%%%%%%%%%%%%%%%%%%%%%%%%%%%%%%%%%%%%%%%%%%%%%%%%%%%
where $\delta_0$ is the initial density field which we assume 
Gaussian statistic. The function $\mathcal{F}_a^{n}$ is 
the symmetrized kernel of the $n$-th order solutions. 
The explicit expressions for the kernel $F_{a}^{(n)}$ 
is obtained from the recursion relation, which can be derived by 
substituting the expansion (\ref{eq:expand}) with (\ref{eq:PT_sol}) 
into Eq.~(\ref{eq:vec_fluid_eq}) (e.g., \cite{Goroff:1986ep,
Bernardeau:2001qr,Crocce:2005xy,Nishimichi:2007xt}):  
%%%%%%%%%%%%%%%%%%%%%%%%%%%%%%%%%%%%%%%%%%%%%%%%%%%%%%%
\begin{eqnarray}
&&F_{a}^{(1)}(\bfk_1)=(1,\,\,1),
\nonumber\\
&&F_{a}^{(n)}(\bfk_1,\cdots,\bfk_n)=
\sigma_{ab}^{(n)}\,\sum_{m=1}^{n-1}\gamma_{bcd}(\bfq_1,\bfq_2)
F_{c}^{(m)}(\bfk_1,\cdots,\bfk_m)F_{d}^{(n-m)}(\bfk_{n-m+1},\cdots,\bfk_n)
\end{eqnarray}
%%%%%%%%%%%%%%%%%%%%%%%%%%%%%%%%%%%%%%%%%%%%%%%%%%%%%%%
with $\bfq_1\equiv\bfk_1+\cdots+\bfk_m$ 
and $\bfq_2\equiv\bfk_{m+1}+\cdots+\bfk_n$. Here, 
the matrix $\sigma_{ab}^{(n)}$ is given by 
%%%%%%%%%%%%%%%%%%%%%%%%%%%%%%%%%%%%%%%%%%%%%%%%%%%%%%%
\begin{equation}
\sigma_{ab}^{(n)}=\frac{1}{(2n+3)(n-1)}
\left(
\begin{array}{cc}
2n+1 & 2 \\
3 & 2n
\end{array}
\right).
\label{eq:recursion}
\end{equation}
%%%%%%%%%%%%%%%%%%%%%%%%%%%%%%%%%%%%%%%%%%%%%%%%%%%%%%%
Note that the kernel $F_a^{(n)}$ given above is not yet symmetric under the 
permutations of arguments, $\bfk_1,\cdots,\bfk_n$, and it should 
be symmetrized: 
%%%%%%%%%%%%%%%%%%%%%%%%%%%%%%%%%%%%%%%%%%%%%%%%%%%%%%%
\begin{equation}
\mathcal{F}_a^{(n)}=\frac{1}{n !}\sum_{\rm permutations} 
F_a^{(n)}(\bfk_1,\cdots,\bfk_n).
\label{eq:symmetrize}
\end{equation}
%%%%%%%%%%%%%%%%%%%%%%%%%%%%%%%%%%%%%%%%%%%%%%%%%%%%%%%

Using the perturbative solutions, the power spectrum 
defined by (\ref{eq:def_Pk}) is expanded as
%%%%%%%%%%%%%%%%%%%%%%%%%%%%%%%%%%%%%%%%%%%%%%%%%%%%%%%
\begin{equation}
P_{ab}(k;\eta) = e^{2\eta}\,\,P_{ab}^{(11)}(k) + e^{4\eta}\,
\Bigl\{\, P_{ab}^{(22)}(k) + P_{ab}^{(13)}(k)\,\Bigr\} + 
e^{6\eta}\,
\Bigl\{\, P_{ab}^{(33)}(k) + P_{ab}^{(24)}(k)+ 
P_{ab}^{(15)}(k)\,\Bigr\} +\cdots.
\end{equation}
%%%%%%%%%%%%%%%%%%%%%%%%%%%%%%%%%%%%%%%%%%%%%%%%%%%%%%%
Here, the quantity $P^{(mn)}$ implies the ensemble average 
obtained from the $m$-th and $n$-th order perturbative solutions. 
In the above expression, the first term at the right-hand side is 
the linear power spectrum, while the second and third terms proportional 
to the growth factors $e^{4\eta}$ and $e^{6\eta}$ are 
respectively the so-called one-loop and two-loop corrections. 
The explicit expressions for these corrections become 
(e.g., \cite{Fry:1993bj,Carlson:2009it})
%%%%%%%%%%%%%%%%%%%%%%%%%%%%%%%%%%%%%%%%%%%%%%%%%%%%%%%
\begin{eqnarray}
P_{ab}^{(11)}(k) &=& u_au_bP_0(k) 
\\
P_{ab}^{(22)}(k) &=& 2\,\,\int \frac{d^3\bfq}{(2\pi)^3}\,
\calF^{(2)}_a(\bfq,\bfk-\bfq)\calF^{(2)}_b(\bfq,\bfk-\bfq)\,
P_0(q)P_0(|\bfk-\bfq|)
\\
P_{ab}^{(13)}(k)&=& 3 \,P_0(k)\,\int \frac{d^3\bfq}{(2\pi)^3}\,
\left\{ \calF^{(3)}_a(\bfk,\bfq,-\bfq) + \calF^{(3)}_b(\bfk,\bfq,-\bfq)\right\}
\,P_0(q)
\\
P_{ab}^{(33)}(k)&=& 9 \,P_0(k)\,\int \frac{d^3\bfp d^3\bfq}{(2\pi)^6}\,
\calF^{(3)}_a(\bfk,\bfp,-\bfp) \calF^{(3)}_b(\bfk,\bfq,-\bfq)\,P_0(p)P_0(q) 
\nonumber\\
&&+ 6 \,\int \frac{d^3\bfp d^3q}{(2\pi)^6}\,
\calF^{(3)}_a(\bfp,\bfq,\bfk-\bfp-\bfq) \calF^{(3)}_b(\bfp,\bfq,\bfk-\bfp-\bfq)
\,P_0(p)\,P_0(q)\,P_0(|\bfk-\bfp-\bfq|)
\\
P_{ab}^{(24)}(k)&=& 12 \,\int \frac{d^3\bfp d^3\bfq}{(2\pi)^6}\,
\Bigl\{ \calF^{(2)}_a(\bfp, \bfk-\bfp) 
\calF^{(4)}_b(\bfp,\bfq,-\bfq,\bfk-\bfp)\,
+ \calF^{(4)}_a(\bfp,\bfq,-\bfq,\bfk-\bfp)\calF^{(2)}_b(\bfp, \bfk-\bfp) 
\Bigr\}\,
\nonumber\\
&&\times \,\,P_0(p)\,P_0(q)\,P_0(|\bfk-\bfp|) 
\\
P_{ab}^{(15)}(k)&=& 15 P_0(k)\,\int \frac{d^3\bfp d^3\bfq}{(2\pi)^6}\,
\Bigl\{ \calF^{(5)}_a(\bfp,\bfq,\bfk,-\bfp,-\bfq)\,
+ \calF^{(5)}_b(\bfp,\bfq,\bfk,-\bfp,-\bfq)\Bigr\}\,
\,\,P_0(p)\,P_0(q),
\end{eqnarray}
%%%%%%%%%%%%%%%%%%%%%%%%%%%%%%%%%%%%%%%%%%%%%%%%%%%%%%%
where $P_0$ is the initial power spectrum of the density field 
$\delta_0$ defined by Eq.~(\ref{eq:def_P_0}), and we set $u_{a}=(1,1)$.

Note that 
the expression for one-loop power spectra can be further reduced to 
the one-dimensional and two-dimensional integral for $P^{(13)}$ and 
$P^{(22)}$, respectively (e.g., \cite{Suto:1990wf,Makino:1991rp,
Scoccimarro:1996se,Nishimichi:2007xt}).  
In the results presented in Sec.~\ref{subsubsec:real_pk}, 
we used the method of Gaussian quadratures for 
numerical integration of one-loop power spectra. On the other hand, 
for the two-loop power spectra, the integration cannot be simplified 
except for the first term in $P^{(33)}$, and we need to directly evaluate 
the six-dimensional integration. We adopted the Monte-Carlo 
integration to the two-loop power spectra. The integration kernels 
for each term are generated numerically using the recursion 
relation (\ref{eq:recursion}) and the condition (\ref{eq:symmetrize}).

%%%%%%%%%%%%%%%%%%%%%%%%%%%%%%%%%%%%%%%%%%%%%%%%%%%%%%%
%%%%%%%%%%%%%%%%%%%%%%%%%%%%%%%%%%%%%%%%%%%%%%%%%%%%%%%
\section{Comparison to other works}
\label{app:comparison}
%%%%%%%%%%%%%%%%%%%%%%%%%%%%%%%%%%%%%%%%%%%%%%%%%%%%%%%
%%%%%%%%%%%%%%%%%%%%%%%%%%%%%%%%%%%%%%%%%%%%%%%%%%%%%%%

%%%%%%%%%%%%%%%%%%%%%%%%%%%%%%%%%%%%%%%%%%%%%%%%%%%%%%%%%%%
\begin{figure}[t]
\begin{center}
\includegraphics[width=2.2cm,angle=-90]{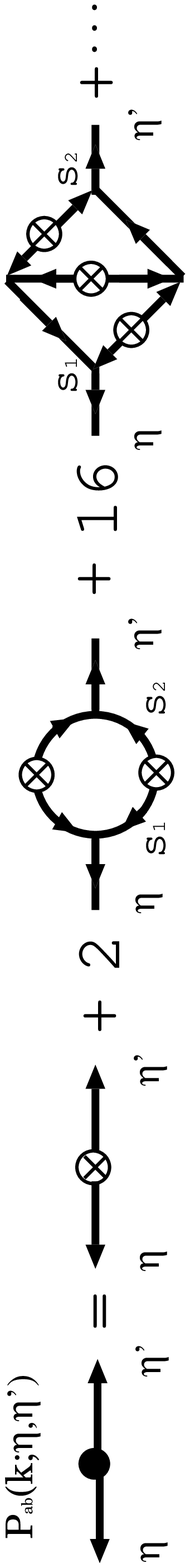}
\end{center}
\caption{Diagrammatic representation for the perturbative treatment of 
  the power spectrum proposed by Crocce \& Scoccimarro (2008), based on 
  the renormalized PT. This is compared with Fig.~\ref{fig:Born}. 
\label{fig:RPT_Born}}
%\end{figure}
%%%%%%%%%%%%%%%%%%%%%%%%%%%%%%%%%%%%%%%%%%%%%%%%%%%%%%%%%%%
%\begin{figure}[h]
\begin{center}
\includegraphics[height=5.8cm,angle=0]{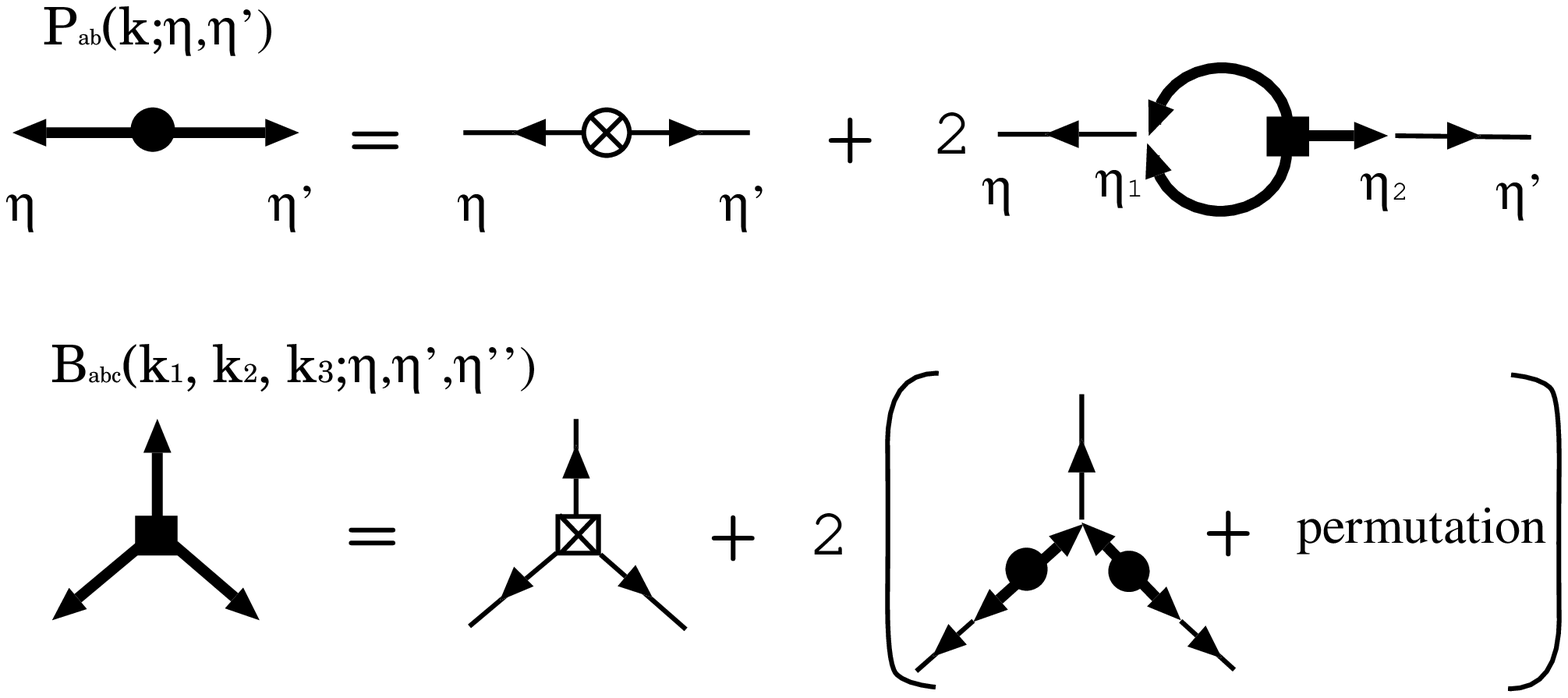}
\end{center}

\vspace*{-0.5cm}

\caption{Diagrammatic representation of the non-perturbative treatment 
  proposed by Pietroni (2008), which can be compared with 
  Fig.~\ref{fig:diagram_CLA}. 
\label{fig:Pietroni}}
\end{figure}
%%%%%%%%%%%%%%%%%%%%%%%%%%%%%%%%%%%%%%%%%%%%%%%%%%%%%%%%%%%

In this Appendix, we collect several recent works that attempt to improve 
the prediction of power spectrum and/or two-point correlation function,  
and discuss their qualitative differences. A 
quantitative aspect of various analytic methods has been recently 
investigated in Ref.~\cite{Carlson:2009it}. 
Here, we specifically comment on the approaches proposed by 
Refs.~\cite{Crocce:2007dt,Valageas:2006bi,Pietroni:2008jx,Matsubara:2007wj}, 
which are very close to our treatment.

\begin{description}
\item[Crocce \& Scoccimarro (2008)]:
First let us mention the work by Ref.~\cite{Crocce:2007dt}.  
Although the treatment presented in the paper are often quoted 
as 'RPT', strictly speaking, this is just the approximate treatment,  
which differs from the renormalized PT \cite{Crocce:2005xy}. 
As we mentioned in Sec.~\ref{subsec:SPT_vs_RPT}, 
renormalized PT is the exact non-perturbative formulation without 
any approximations, and the power spectrum given by 
Eq.~(\ref{eq:RPT_expansion}) is expressed as the infinite series of irreducible 
loop diagrams constructed from the non-linear propagator, full vertex, 
and non-linear power spectrum. To make the analysis tractable, 
they adopted the following approximations:  
(i) the renormalized vertex is well-described by 
 the (linear) vertex function; (ii) the non-linear power spectra 
that enter into the calculation of $P_{ab}^{\rm(MC)}$ are all 
 replaced with the linear-order ones. In our language, this 
 corresponds to the first-order Born approximation.  
 Then, using the approximate solution for propagator in 
 Ref.~\cite{Crocce:2005xz}, they explicitly calculated 
 the power spectrum including the corrections up to the two-loop order. 
 The diagrams that they actually computed are shown in 
 Fig.~\ref{fig:RPT_Born}.

 Compared to our analytical treatment with Born approximation, 
 there are two main differences. One is the higher-order corrections that 
 appear in the diagrams (see Fig.~\ref{fig:Born}). Another important 
 difference is the asymptotic behaviors in the non-linear propagator. 
 At $k\to\infty$, the propagator used in their paper behaves like 
 $G_{ab}\to g_{ab}\exp[-x^2/2]$, which contrasts with 
 $G_{ab}\to g_{ab}\,J_1(2x)/x$ in our closure approximation, where 
 $g_{ab}$ is the linear propagator and $x$ is defined by 
 $x\equiv k\,\sigma_{\rm v}(e^{\eta}-e^{\eta'})$. These distinctive features 
 come from the partial resummation of a different class of higher-order 
 terms when constructing the approximate solution of non-linear propagator 
 (see Ref.~\cite{Crocce:2005xz,Taruya:2007xy} in details). Despite these  
 remarkable differences, it has been shown in Ref.~\cite{Nishimichi:2008ry} 
 that the leading-order calculations neglecting the higher-order terms 
 (two-loop diagram or second-order Born correction) can produce the same 
 results which is indistinguishable from each other. 
 This is true at least on large scales, where the agreement between 
 N-body simulations and improved PT predictions is better than a few percent. 

\item[Pietroni (2008)]: Next consider the method proposed by 
  Ref.~\cite{Pietroni:2008jx}, called time-RG method. 
  This method is based on the 
  moment-based approach, and we first write down the moment equations.  
  In general, this produces an infinite hierarchy of equations, however,  
  Ref.~\cite{Pietroni:2008jx} assumes a vanishing trispectrum in order
  to truncate the hierarchy. As a result, a closed set of equations 
  for power spectrum $P_{ab}$ is obtained, which 
  coulples with the evolution of bispectrum $B_{abc}$ in 
  some non-perturbative ways. Diagrammatic 
  representation of this closed equations is shown in 
  Fig.~\ref{fig:Pietroni},  which can be compared with 
  Fig.~\ref{fig:diagram_CLA} in our treatment. Note that in the subject
  of statistical theory of turbulence, this truncation procedure 
  is referred to as {\it quasi-normal approximation} 
  (e.g.,\cite{Proudman:1954,Tatsumi:1957,Leslie1973}), and it is known 
  to have several drawbacks; positivity of the energy spectrum 
  is not ensured, and it fails to recover the Kolmogolov spectrum 
  in the inertial range of turbulence. 

  Nevertheless, the advantage of this treatment, similar to our closure 
  approximation,  is that the power spectrum can be computed numerically 
  by solving the evolution equations. This forward treatment seems quite 
  efficient to bring out the non-perturbative effects incorporated into 
  the formalism, and it has a wide applicability to include various physical 
  effects. Recently, the formalism has been extended to 
  deal with the effect of massive neutrinos \cite{Lesgourgues:2009am}. 
  
\item[Valageas (2007)]: The method proposed by 
  Ref.~\cite{Valageas:2006bi} 
  is based on the path-integral formalism. Starting from the action for 
  the cosmological fluid equation (\ref{eq:vec_fluid_eq}), which describes 
  the statistical properties of the vector field $\Phi_a$, the large-N 
  expansions as a technique of quantum field theory have been applied to 
  derive the governing equations for power spectrum and propagator. 
  In Ref.~\cite{Valageas:2006bi},  two kinds of expansions have been 
  presented, leading to the two different non-perturbative schemes, i.e., 
  steepest descent method and 2PI effective action method. Although both 
  methods consistently reproduce the standard PT at the one-loop level,  
  the latter includes the non-purturbative contributions  which are not 
  properly taken into account by the former method. Thus, the 2PI 
  effective action method is expected to provide a better result. It is 
  interesting to note that despite the field-theoretical derivation, 
  the resultant governing equations for 
  the 2PI effective action method turn out to be mathematically 
  equivalent to those obtained from the closure approximation 
  \cite{Taruya:2007xy}. 
  Hence, the diagrammatic representation of this formalism is exactly 
  the same as shown in Fig.~\ref{fig:diagram_CLA}.

\item[Matsubara (2008a)]: Finally, we briefly mention the treatment proposed 
by Ref.~\cite{Matsubara:2007wj}. This is the Lagrangian-based approach, and  
we begin by writing down the exact expressions for matter power spectrum
in terms of the displacement vectors. The resultant 
expression is in the exponential form, and the purterbative expansions are 
then applied for the explicit calculation of the ensemble average. 
While a naive expansion of the displacement vectors, together with 
the solutions of Lagrangian PT, merely reproduces the (standard) Eulerian PT 
results, Ref.~\cite{Matsubara:2007wj} has applied a partial expansion,  
and some of the terms have been kept in the exponential form.  
This can be interpreted as the partial resummation of a class of the 
infinite diagrams. 
The resultant expression for power spectrum is quite similar to the 
one-loop result of standard PT, but slightly differs from it 
in the sense that there appears the exponential prefactor. As a 
consequence, the prediction 
reasonably recovers the damping behavior of the BAOs seen in the 
N-body simulations, and it also explains the smearing effect on 
the baryon acoustic peak in the two-point correlation function.

One noticiable point of this method is that 
it is rather straightforward to generalize the calculations in real space 
to those in redshift space, since the displacement vectors in redshift 
space can be simply given by a linear mapping from those in 
real space. Further, 
the computational cost is less expensive compared to the other analytic 
methods. Although the validity range of this method is restricted to a 
narrow range of the low-$k$ modes, it would be very powerful for a fast 
compuation of the two-point correlation function. 

\end{description}

%%%%%%%%%%%%%%%%%%%%%%%%%%%%%%%%%%%%%%%%%%%%%%%%%%%%%%%
%%%%%%%%%%%%%%%%%%%%%%%%%%%%%%%%%%%%%%%%%%%%%%%%%%%%%%%
\section{Convergence of  Different Computational Methods  
  for Two-Point Correlation Function}
\label{app:computing_TPCF}
%%%%%%%%%%%%%%%%%%%%%%%%%%%%%%%%%%%%%%%%%%%%%%%%%%%%%%%
%%%%%%%%%%%%%%%%%%%%%%%%%%%%%%%%%%%%%%%%%%%%%%%%%%%%%%%

%%%%%%%%%%%%%%%%%%%%%%%%%%%%%%%%%%%%%%%%%%%%%%%%%%%%%%%%%%%
\begin{figure}[t]
\begin{center}
\includegraphics[width=7cm,angle=0]{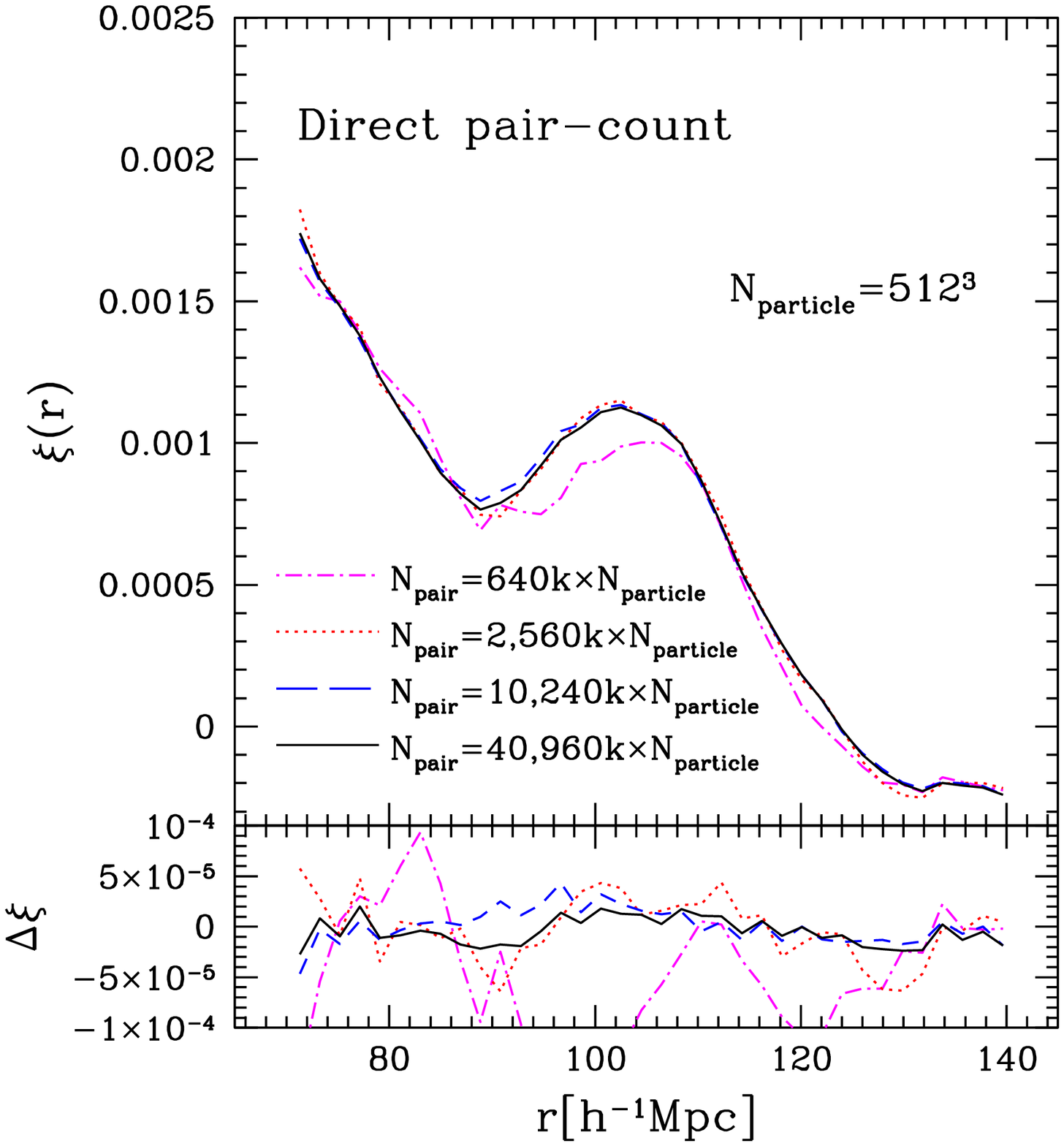}
\includegraphics[width=7cm,angle=0]{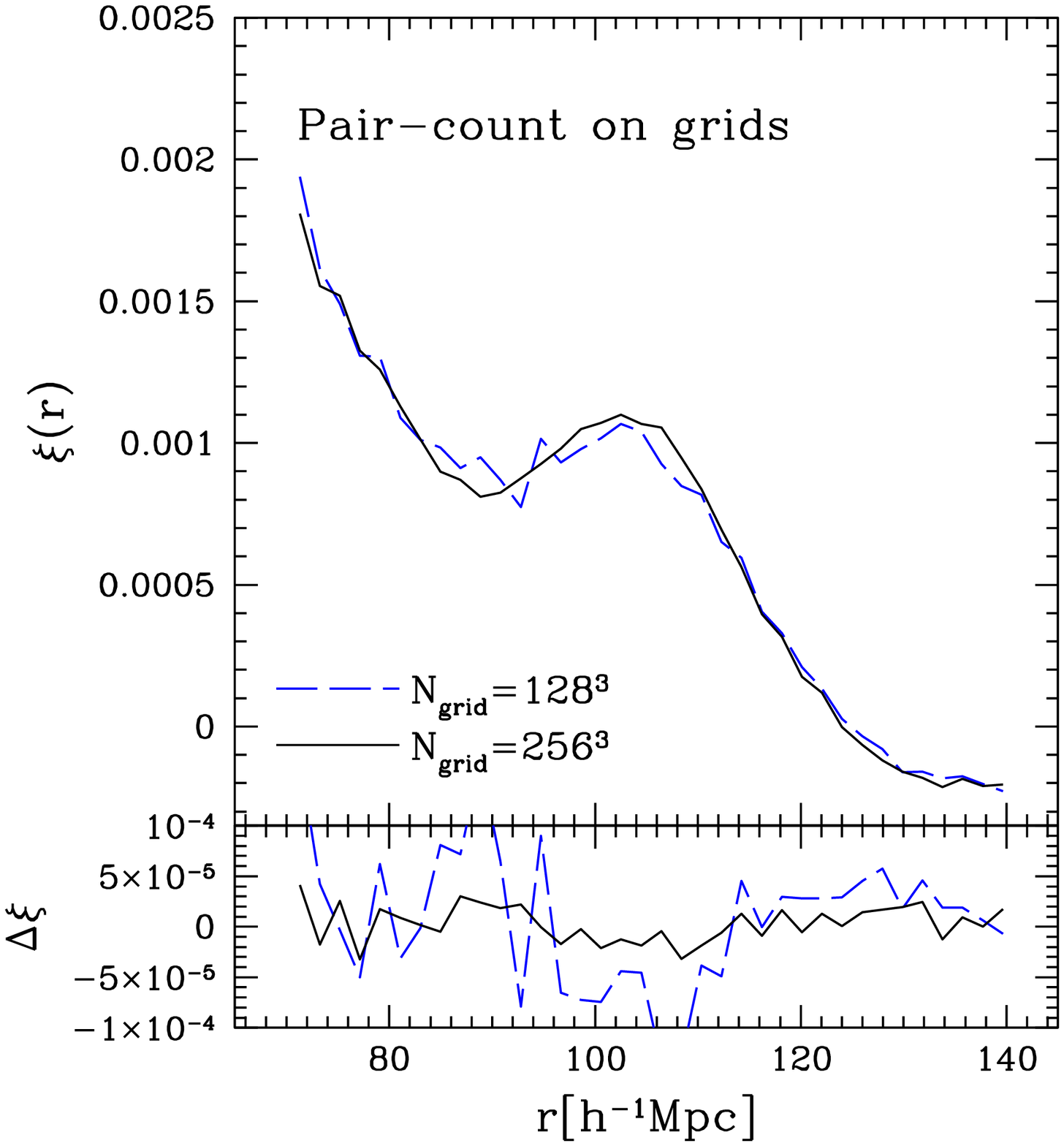}
\includegraphics[width=7cm,angle=0]{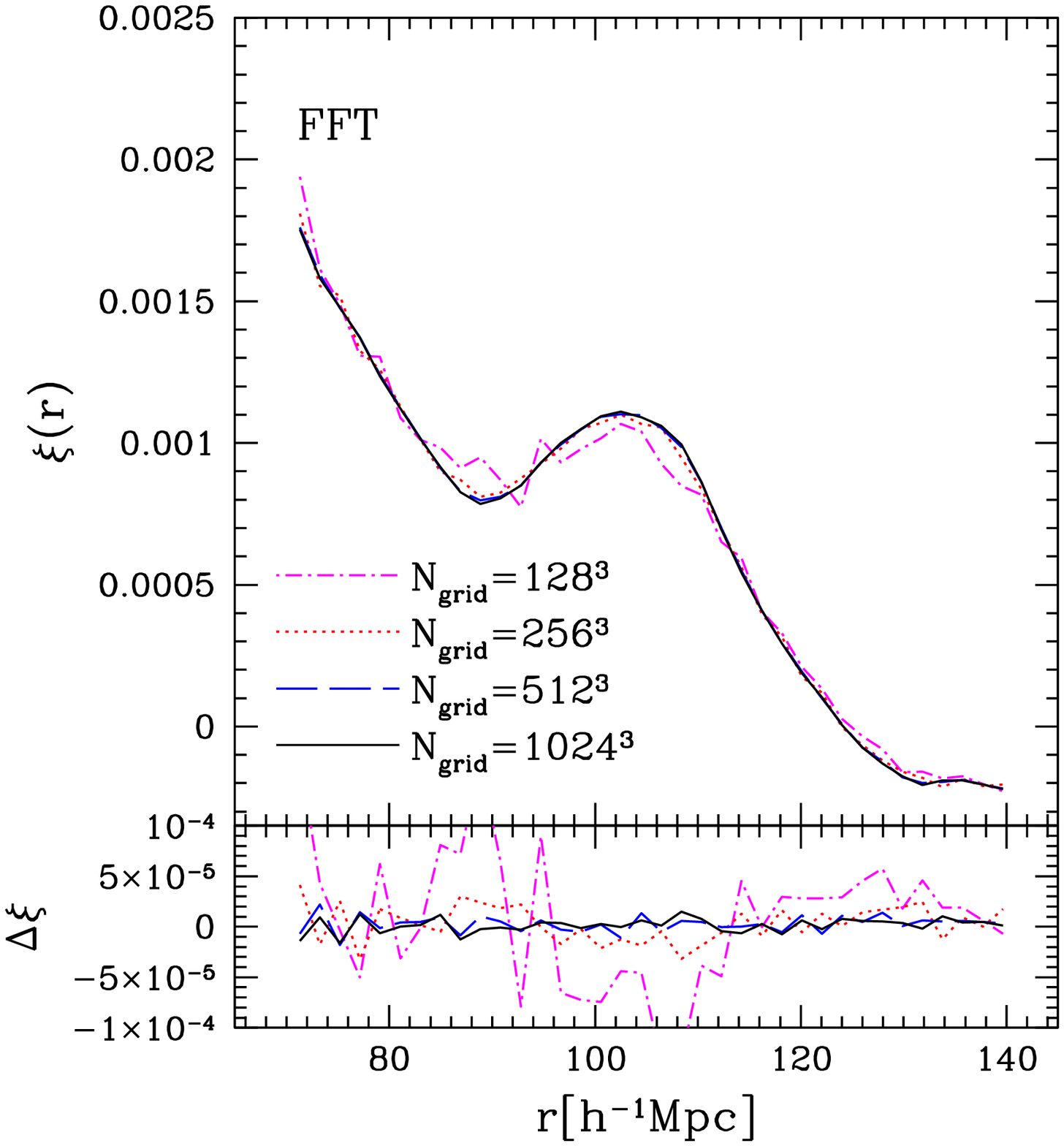}
\includegraphics[width=7cm,angle=0]{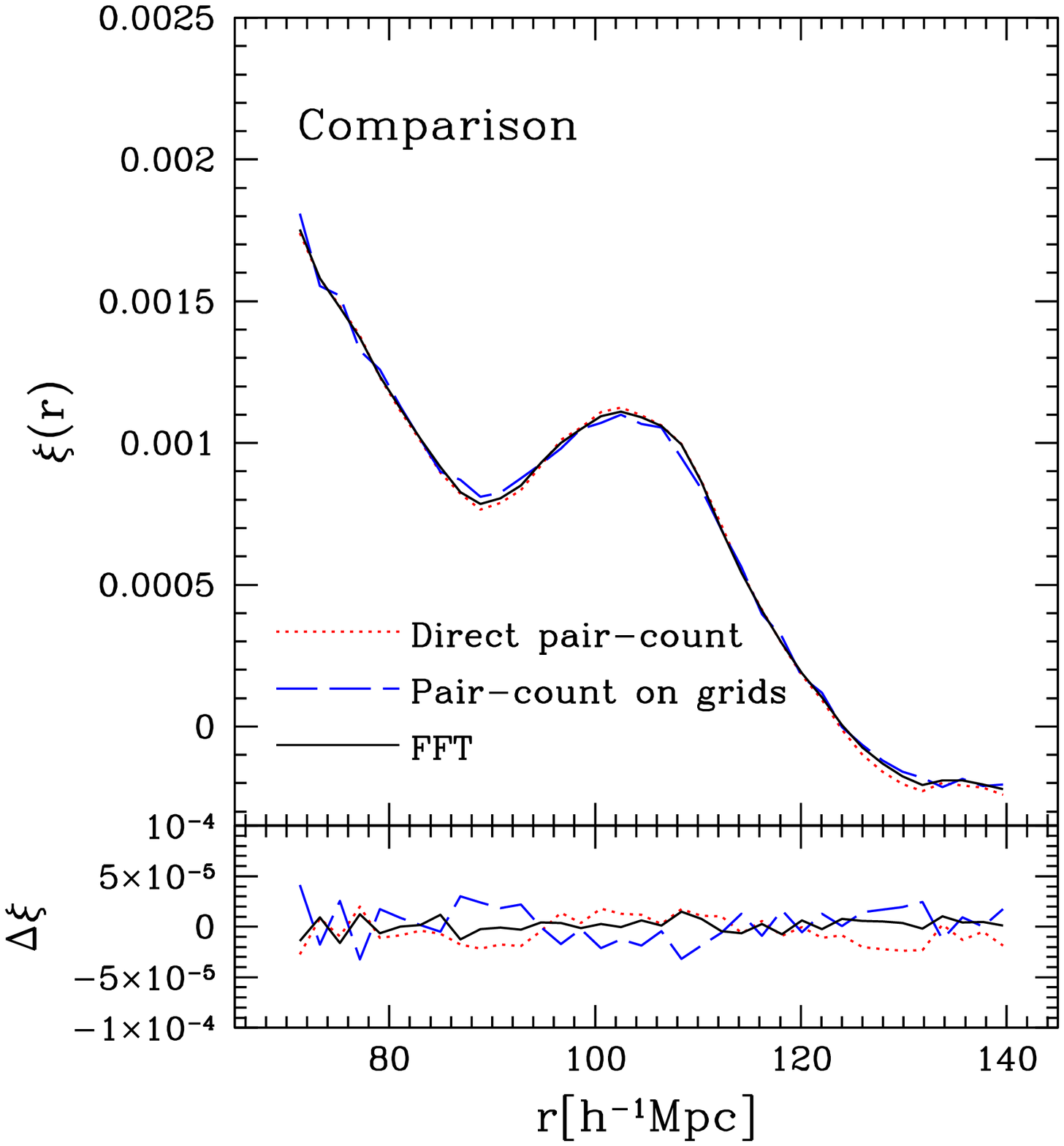}
\end{center}

\vspace*{-0.5cm}

\caption{Comparison between different computational methods for two-point 
  correlation function. 
\label{fig:convergence_xi}}
\end{figure}
%%%%%%%%%%%%%%%%%%%%%%%%%%%%%%%%%%%%%%%%%%%%%%%%%%%%%%%%%%%

In this paper, the grid-based calculation with FFT has been 
used for computing the two-point correlation functions from N-body data. 
Here, we compare it with other computational methods and check 
their convergences.

Fig.~\ref{fig:convergence_xi} shows the two-point correlation functions 
measured at $z=0.5$ from the single realization of \verb|wmap5| simulations. 
Upper-left panel shows the results from the direct pair-counting. For each 
particle, we randomly select pairs, which are accumulated for each bin of 
separations, allowing for oversampling. The estimated values of 
two-point correlation function 
are then plotted for different number of samples $N_{\rm sample}$: 
$N_{\rm sample}=640\mbox{k}$, $2,560\mbox{k}$, 
$10,240\mbox{k}$ and $40,960\mbox{k}$. 
The resultant total number of pairs, $N_{\rm pair}$, indicated in the 
panel is given by $N_{\rm pair}=N_{\rm sample}\times N_{\rm particle}$, 
with $N_{\rm particle}=512^3$ being the total number of particles. 
Note that the actual number of pairs that enters into the plotted range 
is less than $N_{\rm pair}$. On the other hand, upper-right panel shows 
the results from the 
grid-based pair-counting introduced by Barriga \& Gazta\~naga (2002) 
\cite{Barriga:2001wn} (see also Ref.~\cite{Sanchez:2008iw}).  In this 
method, we first construct the density field on a grid of 
$N_{\rm grid}$ cells, and then estimate the correlation function through
the pair count on grids: 
%%%%%%%%%%%%%%%%%%%%%%%%%%%%%%%%%%%%%%%%%%%%%%%%%%%%%%%%%%%
\begin{equation}
\widehat{\xi}(|\bfr_{ij}|)=
\frac{1}{N_{\rm pair}(|\bfr_{ij}|)} \sum_{ij}\delta(\bfr_i)\delta(\bfr_j).
\end{equation}
%%%%%%%%%%%%%%%%%%%%%%%%%%%%%%%%%%%%%%%%%%%%%%%%%%%%%%%%%%%
Compared to the direct pair-counting, 
this method is computationally efficient when we store the list of 
neighbor particles which contribute to a given bin of separation. 
We plot the results adopting the two different number of 
cells, $N_{\rm grid}=128^3$ and $256^3$. In lower-left panel, the grid-based 
calculation with FFT (see Eq.~(\ref{eq:estimator_xi})) is used to 
compute the two-point correlation function, with different numbers of 
cells, $N_{\rm grid}=128^3$, $256^3$, $512^3$ and $1,024^3$. Note that we adopt 
$N_{\rm grid}=1,024^3$ in the analysis presented in Sec.~\ref{sec:PT_vs_N-body}. 
Finally, in lower-right panel, the results for three different methods with the 
largest number of pairs or grids are collected and compared with each other.

To check the convergence, we further evaluate the residuals from the mean 
values, $\Delta\xi\equiv \widehat{\xi}-\overline{\xi}$, and plot the results 
in each panel of Fig.~\ref{fig:convergence_xi}. Here, the mean values 
$\overline{\xi}$ are estimated from the ensemble average over the 
three different results using the largest number of pairs or grids. 
As increasing the numbers $N_{\rm pair}$ or $N_{\rm grid}$, 
the results for three different methods all approach the mean values 
$\overline{\xi}$, and a few percent-level agreement is achieved 
over the range of our interest (except for the vicinity of zero-crossing 
point, $\xi\approx0$). It is interesting to note that residuals 
obtained from the grid-based pair-count and FFT methods almost coincide 
with each other and the differences are hard to distinguish, indicating 
that both methods are equivalent even in the practical situation. 
These experiments suggest that the grid-based calculation with FFT is a 
reliable estimation method comparable to the other methods. It should be 
emphasized that the method using FFT is much more efficient than other 
pair-count methods. For example, using $8$ cores of $3$GHz processors, 
the direct pair-counting with $N_{\rm pair}=10,240\mbox{k}\times N_{\rm particle}$ 
takes about two weeks to get the results 
shown in Fig.~\ref{fig:convergence_xi}. The grid-based pair-counting 
is computationally less expensive than the direct pair-counting, but 
it still needs time-consuming calculations, especially for a 
large number of grids. By contrast, 
the method using FFT only requires few minutes even with 
$N_{\rm grid}=1,024^3$. This can be achieved by a single-node calculation.

%\clearpage
%%%%%%%%%%%%%%%%%%%%%%%%%%%%%%%%%%%%%%%%%%%%%%%%%%%%%%%
% \bibliography{ref}

\begin{thebibliography}{91}
\expandafter\ifx\csname natexlab\endcsname\relax\def\natexlab#1{#1}\fi
\expandafter\ifx\csname bibnamefont\endcsname\relax
  \def\bibnamefont#1{#1}\fi
\expandafter\ifx\csname bibfnamefont\endcsname\relax
  \def\bibfnamefont#1{#1}\fi
\expandafter\ifx\csname citenamefont\endcsname\relax
  \def\citenamefont#1{#1}\fi
\expandafter\ifx\csname url\endcsname\relax
  \def\url#1{\texttt{#1}}\fi
\expandafter\ifx\csname urlprefix\endcsname\relax\def\urlprefix{URL }\fi
\providecommand{\bibinfo}[2]{#2}
\providecommand{\eprint}[2][]{\url{#2}}

\bibitem[{\citenamefont{Spergel et~al.}(2003)}]{Spergel:2003cb}
\bibinfo{author}{\bibfnamefont{D.~N.} \bibnamefont{Spergel}}
  \bibnamefont{et~al.} (\bibinfo{collaboration}{WMAP}),
  \bibinfo{journal}{Astrophys. J. Suppl.} \textbf{\bibinfo{volume}{148}},
  \bibinfo{pages}{175} (\bibinfo{year}{2003}), \eprint{astro-ph/0302209}.

\bibitem[{\citenamefont{Spergel et~al.}(2007)}]{Spergel:2006hy}
\bibinfo{author}{\bibfnamefont{D.~N.} \bibnamefont{Spergel}}
  \bibnamefont{et~al.} (\bibinfo{collaboration}{WMAP}),
  \bibinfo{journal}{Astrophys. J. Suppl.} \textbf{\bibinfo{volume}{170}},
  \bibinfo{pages}{377} (\bibinfo{year}{2007}), \eprint{astro-ph/0603449}.

\bibitem[{\citenamefont{Komatsu et~al.}(2009)}]{Komatsu:2008hk}
\bibinfo{author}{\bibfnamefont{E.}~\bibnamefont{Komatsu}} \bibnamefont{et~al.}
  (\bibinfo{collaboration}{WMAP}), \bibinfo{journal}{Astrophys. J. Suppl.}
  \textbf{\bibinfo{volume}{180}}, \bibinfo{pages}{330} (\bibinfo{year}{2009}),
  \eprint{0803.0547}.

\bibitem[{\citenamefont{Tegmark et~al.}(2004)}]{Tegmark:2003ud}
\bibinfo{author}{\bibfnamefont{M.}~\bibnamefont{Tegmark}} \bibnamefont{et~al.}
  (\bibinfo{collaboration}{SDSS}), \bibinfo{journal}{Phys. Rev.}
  \textbf{\bibinfo{volume}{D69}}, \bibinfo{pages}{103501}
  (\bibinfo{year}{2004}), \eprint{astro-ph/0310723}.

\bibitem[{\citenamefont{Tegmark et~al.}(2006)}]{Tegmark:2006az}
\bibinfo{author}{\bibfnamefont{M.}~\bibnamefont{Tegmark}} \bibnamefont{et~al.}
  (\bibinfo{collaboration}{SDSS}), \bibinfo{journal}{Phys. Rev.}
  \textbf{\bibinfo{volume}{D74}}, \bibinfo{pages}{123507}
  (\bibinfo{year}{2006}), \eprint{astro-ph/0608632}.

\bibitem[{\citenamefont{Perlmutter et~al.}(1999)}]{Perlmutter:1998np}
\bibinfo{author}{\bibfnamefont{S.}~\bibnamefont{Perlmutter}}
  \bibnamefont{et~al.} (\bibinfo{collaboration}{Supernova Cosmology Project}),
  \bibinfo{journal}{Astrophys. J.} \textbf{\bibinfo{volume}{517}},
  \bibinfo{pages}{565} (\bibinfo{year}{1999}), \eprint{astro-ph/9812133}.

\bibitem[{\citenamefont{Riess et~al.}(1998)}]{Riess:1998cb}
\bibinfo{author}{\bibfnamefont{A.~G.} \bibnamefont{Riess}} \bibnamefont{et~al.}
  (\bibinfo{collaboration}{Supernova Search Team}), \bibinfo{journal}{Astron.
  J.} \textbf{\bibinfo{volume}{116}}, \bibinfo{pages}{1009}
  (\bibinfo{year}{1998}), \eprint{astro-ph/9805201}.

\bibitem[{\citenamefont{Nojiri and Odintsov}(2006)}]{Nojiri:2006ri}
\bibinfo{author}{\bibfnamefont{S.}~\bibnamefont{Nojiri}} \bibnamefont{and}
  \bibinfo{author}{\bibfnamefont{S.~D.} \bibnamefont{Odintsov}},
  \bibinfo{journal}{ECONF} \textbf{\bibinfo{volume}{C0602061}},
  \bibinfo{pages}{06} (\bibinfo{year}{2006}), \eprint{hep-th/0601213}.

\bibitem[{\citenamefont{Durrer and
  Maartens}(2008{\natexlab{a}})}]{Durrer:2007re}
\bibinfo{author}{\bibfnamefont{R.}~\bibnamefont{Durrer}} \bibnamefont{and}
  \bibinfo{author}{\bibfnamefont{R.}~\bibnamefont{Maartens}},
  \bibinfo{journal}{Gen. Rel. Grav.} \textbf{\bibinfo{volume}{40}},
  \bibinfo{pages}{301} (\bibinfo{year}{2008}{\natexlab{a}}),
  \eprint{0711.0077}.

\bibitem[{\citenamefont{Durrer and
  Maartens}(2008{\natexlab{b}})}]{Durrer:2008in}
\bibinfo{author}{\bibfnamefont{R.}~\bibnamefont{Durrer}} \bibnamefont{and}
  \bibinfo{author}{\bibfnamefont{R.}~\bibnamefont{Maartens}}
  (\bibinfo{year}{2008}{\natexlab{b}}), \eprint{0811.4132}.

\bibitem[{\citenamefont{Koyama}(2008)}]{Koyama:2007rx}
\bibinfo{author}{\bibfnamefont{K.}~\bibnamefont{Koyama}},
  \bibinfo{journal}{Gen. Rel. Grav.} \textbf{\bibinfo{volume}{40}},
  \bibinfo{pages}{421} (\bibinfo{year}{2008}), \eprint{0706.1557}.

\bibitem[{\citenamefont{Dvali et~al.}(2000)\citenamefont{Dvali, Gabadadze, and
  Porrati}}]{Dvali:2000hr}
\bibinfo{author}{\bibfnamefont{G.~R.} \bibnamefont{Dvali}},
  \bibinfo{author}{\bibfnamefont{G.}~\bibnamefont{Gabadadze}},
  \bibnamefont{and} \bibinfo{author}{\bibfnamefont{M.}~\bibnamefont{Porrati}},
  \bibinfo{journal}{Phys. Lett.} \textbf{\bibinfo{volume}{B485}},
  \bibinfo{pages}{208} (\bibinfo{year}{2000}), \eprint{hep-th/0005016}.

\bibitem[{\citenamefont{Hu and Sawicki}(2007)}]{Hu:2007nk}
\bibinfo{author}{\bibfnamefont{W.}~\bibnamefont{Hu}} \bibnamefont{and}
  \bibinfo{author}{\bibfnamefont{I.}~\bibnamefont{Sawicki}},
  \bibinfo{journal}{Phys. Rev.} \textbf{\bibinfo{volume}{D76}},
  \bibinfo{pages}{064004} (\bibinfo{year}{2007}), \eprint{0705.1158}.

\bibitem[{\citenamefont{Starobinsky}(2007)}]{Starobinsky:2007hu}
\bibinfo{author}{\bibfnamefont{A.~A.} \bibnamefont{Starobinsky}},
  \bibinfo{journal}{JETP Lett.} \textbf{\bibinfo{volume}{86}},
  \bibinfo{pages}{157} (\bibinfo{year}{2007}), \eprint{0706.2041}.

\bibitem[{\citenamefont{Seo and Eisenstein}(2003)}]{Seo:2003pu}
\bibinfo{author}{\bibfnamefont{H.-J.} \bibnamefont{Seo}} \bibnamefont{and}
  \bibinfo{author}{\bibfnamefont{D.~J.} \bibnamefont{Eisenstein}},
  \bibinfo{journal}{Astrophys. J.} \textbf{\bibinfo{volume}{598}},
  \bibinfo{pages}{720} (\bibinfo{year}{2003}), \eprint{astro-ph/0307460}.

\bibitem[{\citenamefont{Blake and Glazebrook}(2003)}]{Blake:2003rh}
\bibinfo{author}{\bibfnamefont{C.}~\bibnamefont{Blake}} \bibnamefont{and}
  \bibinfo{author}{\bibfnamefont{K.}~\bibnamefont{Glazebrook}},
  \bibinfo{journal}{Astrophys. J.} \textbf{\bibinfo{volume}{594}},
  \bibinfo{pages}{665} (\bibinfo{year}{2003}), \eprint{astro-ph/0301632}.

\bibitem[{\citenamefont{Eisenstein et~al.}(2005)}]{Eisenstein:2005su}
\bibinfo{author}{\bibfnamefont{D.~J.} \bibnamefont{Eisenstein}}
  \bibnamefont{et~al.} (\bibinfo{collaboration}{SDSS}),
  \bibinfo{journal}{Astrophys. J.} \textbf{\bibinfo{volume}{633}},
  \bibinfo{pages}{560} (\bibinfo{year}{2005}), \eprint{astro-ph/0501171}.

\bibitem[{\citenamefont{Huetsi}(2006)}]{Huetsi:2005tp}
\bibinfo{author}{\bibfnamefont{G.}~\bibnamefont{Huetsi}},
  \bibinfo{journal}{Astron. Astrophys.} \textbf{\bibinfo{volume}{449}},
  \bibinfo{pages}{891} (\bibinfo{year}{2006}), \eprint{astro-ph/0512201}.

\bibitem[{\citenamefont{Cole et~al.}(2005)}]{Cole:2005sx}
\bibinfo{author}{\bibfnamefont{S.}~\bibnamefont{Cole}} \bibnamefont{et~al.}
  (\bibinfo{collaboration}{The 2dFGRS}), \bibinfo{journal}{Mon. Not. Roy.
  Astron. Soc.} \textbf{\bibinfo{volume}{362}}, \bibinfo{pages}{505}
  (\bibinfo{year}{2005}), \eprint{astro-ph/0501174}.

\bibitem[{\citenamefont{Percival et~al.}(2007{\natexlab{a}})}]{Percival:2006gs}
\bibinfo{author}{\bibfnamefont{W.~J.} \bibnamefont{Percival}}
  \bibnamefont{et~al.}, \bibinfo{journal}{Astrophys. J.}
  \textbf{\bibinfo{volume}{657}}, \bibinfo{pages}{51}
  (\bibinfo{year}{2007}{\natexlab{a}}), \eprint{astro-ph/0608635}.

\bibitem[{\citenamefont{Percival et~al.}(2007{\natexlab{b}})}]{Percival:2007yw}
\bibinfo{author}{\bibfnamefont{W.~J.} \bibnamefont{Percival}}
  \bibnamefont{et~al.}, \bibinfo{journal}{Mon. Not. Roy. Astron. Soc.}
  \textbf{\bibinfo{volume}{381}}, \bibinfo{pages}{1053}
  (\bibinfo{year}{2007}{\natexlab{b}}), \eprint{0705.3323}.

\bibitem[{\citenamefont{Hu and Sugiyama}(1996)}]{Hu:1995en}
\bibinfo{author}{\bibfnamefont{W.}~\bibnamefont{Hu}} \bibnamefont{and}
  \bibinfo{author}{\bibfnamefont{N.}~\bibnamefont{Sugiyama}},
  \bibinfo{journal}{Astrophys. J.} \textbf{\bibinfo{volume}{471}},
  \bibinfo{pages}{542} (\bibinfo{year}{1996}), \eprint{astro-ph/9510117}.

\bibitem[{\citenamefont{Eisenstein and Hu}(1998)}]{Eisenstein:1997ik}
\bibinfo{author}{\bibfnamefont{D.~J.} \bibnamefont{Eisenstein}}
  \bibnamefont{and} \bibinfo{author}{\bibfnamefont{W.}~\bibnamefont{Hu}},
  \bibinfo{journal}{Astrophys. J.} \textbf{\bibinfo{volume}{496}},
  \bibinfo{pages}{605} (\bibinfo{year}{1998}), \eprint{astro-ph/9709112}.

\bibitem[{\citenamefont{Meiksin et~al.}(1998)\citenamefont{Meiksin, White, and
  Peacock}}]{Meiksin:1998ra}
\bibinfo{author}{\bibfnamefont{A.}~\bibnamefont{Meiksin}},
  \bibinfo{author}{\bibfnamefont{M.~J.} \bibnamefont{White}}, \bibnamefont{and}
  \bibinfo{author}{\bibfnamefont{J.~A.} \bibnamefont{Peacock}}
  (\bibinfo{year}{1998}), \eprint{astro-ph/9812214}.

\bibitem[{\citenamefont{Seo and Eisenstein}(2005)}]{Seo:2005ys}
\bibinfo{author}{\bibfnamefont{H.-J.} \bibnamefont{Seo}} \bibnamefont{and}
  \bibinfo{author}{\bibfnamefont{D.~J.} \bibnamefont{Eisenstein}},
  \bibinfo{journal}{Astrophys. J.} \textbf{\bibinfo{volume}{633}},
  \bibinfo{pages}{575} (\bibinfo{year}{2005}), \eprint{astro-ph/0507338}.

\bibitem[{\citenamefont{Albrecht et~al.}(2006)}]{Albrecht:2006um}
\bibinfo{author}{\bibfnamefont{A.~J.} \bibnamefont{Albrecht}}
  \bibnamefont{et~al.} (\bibinfo{year}{2006}), \eprint{astro-ph/0609591}.

\bibitem[{\citenamefont{Peacock et~al.}(2006)}]{Peacock:2006kj}
\bibinfo{author}{\bibfnamefont{J.~A.} \bibnamefont{Peacock}}
  \bibnamefont{et~al.} (\bibinfo{year}{2006}), \eprint{astro-ph/0610906}.

\bibitem[{\citenamefont{Bassett et~al.}(2005)\citenamefont{Bassett, Nichol, and
  Eisenstein}}]{Bassett:2005kn}
\bibinfo{author}{\bibfnamefont{B.~A.} \bibnamefont{Bassett}},
  \bibinfo{author}{\bibfnamefont{R.~C.} \bibnamefont{Nichol}},
  \bibnamefont{and} \bibinfo{author}{\bibfnamefont{D.~J.}
  \bibnamefont{Eisenstein}} (\bibinfo{collaboration}{WFMOS})
  (\bibinfo{year}{2005}), \eprint{astro-ph/0510272}.

\bibitem[{\citenamefont{Schlegel et~al.}(2009)\citenamefont{Schlegel, White,
  and Eisenstein}}]{Schlegel:2009hj}
\bibinfo{author}{\bibfnamefont{D.}~\bibnamefont{Schlegel}},
  \bibinfo{author}{\bibfnamefont{M.}~\bibnamefont{White}}, \bibnamefont{and}
  \bibinfo{author}{\bibfnamefont{D.}~\bibnamefont{Eisenstein}}
  (\bibinfo{collaboration}{with input from the SDSS-III})
  (\bibinfo{year}{2009}), \eprint{0902.4680}.

\bibitem[{\citenamefont{Hill et~al.}(2008)}]{Hill:2008mv}
\bibinfo{author}{\bibfnamefont{G.~J.} \bibnamefont{Hill}} \bibnamefont{et~al.}
  (\bibinfo{year}{2008}), \eprint{0806.0183}.

\bibitem[{\citenamefont{Wang et~al.}(2008)}]{Wang:2008hgb}
\bibinfo{author}{\bibfnamefont{X.}~\bibnamefont{Wang}} \bibnamefont{et~al.}
  (\bibinfo{year}{2008}), \eprint{0809.3002}.

\bibitem[{\citenamefont{Crocce and
  Scoccimarro}(2006{\natexlab{a}})}]{Crocce:2005xy}
\bibinfo{author}{\bibfnamefont{M.}~\bibnamefont{Crocce}} \bibnamefont{and}
  \bibinfo{author}{\bibfnamefont{R.}~\bibnamefont{Scoccimarro}},
  \bibinfo{journal}{Phys. Rev.} \textbf{\bibinfo{volume}{D73}},
  \bibinfo{pages}{063519} (\bibinfo{year}{2006}{\natexlab{a}}),
  \eprint{astro-ph/0509418}.

\bibitem[{\citenamefont{Crocce and
  Scoccimarro}(2006{\natexlab{b}})}]{Crocce:2005xz}
\bibinfo{author}{\bibfnamefont{M.}~\bibnamefont{Crocce}} \bibnamefont{and}
  \bibinfo{author}{\bibfnamefont{R.}~\bibnamefont{Scoccimarro}},
  \bibinfo{journal}{Phys. Rev.} \textbf{\bibinfo{volume}{D73}},
  \bibinfo{pages}{063520} (\bibinfo{year}{2006}{\natexlab{b}}),
  \eprint{astro-ph/0509419}.

\bibitem[{\citenamefont{Crocce and Scoccimarro}(2008)}]{Crocce:2007dt}
\bibinfo{author}{\bibfnamefont{M.}~\bibnamefont{Crocce}} \bibnamefont{and}
  \bibinfo{author}{\bibfnamefont{R.}~\bibnamefont{Scoccimarro}},
  \bibinfo{journal}{Phys. Rev.} \textbf{\bibinfo{volume}{D77}},
  \bibinfo{pages}{023533} (\bibinfo{year}{2008}), \eprint{0704.2783}.

\bibitem[{\citenamefont{Matsubara}(2008{\natexlab{a}})}]{Matsubara:2007wj}
\bibinfo{author}{\bibfnamefont{T.}~\bibnamefont{Matsubara}},
  \bibinfo{journal}{Phys. Rev.} \textbf{\bibinfo{volume}{D77}},
  \bibinfo{pages}{063530} (\bibinfo{year}{2008}{\natexlab{a}}),
  \eprint{0711.2521}.

\bibitem[{\citenamefont{Matsubara}(2008{\natexlab{b}})}]{Matsubara:2008wx}
\bibinfo{author}{\bibfnamefont{T.}~\bibnamefont{Matsubara}},
  \bibinfo{journal}{Phys. Rev.} \textbf{\bibinfo{volume}{D78}},
  \bibinfo{pages}{083519} (\bibinfo{year}{2008}{\natexlab{b}}),
  \eprint{0807.1733}.

\bibitem[{\citenamefont{McDonald}(2007)}]{McDonald:2006hf}
\bibinfo{author}{\bibfnamefont{P.}~\bibnamefont{McDonald}},
  \bibinfo{journal}{Phys. Rev.} \textbf{\bibinfo{volume}{D75}},
  \bibinfo{pages}{043514} (\bibinfo{year}{2007}), \eprint{astro-ph/0606028}.

\bibitem[{\citenamefont{Izumi and Soda}(2007)}]{Izumi:2007su}
\bibinfo{author}{\bibfnamefont{K.}~\bibnamefont{Izumi}} \bibnamefont{and}
  \bibinfo{author}{\bibfnamefont{J.}~\bibnamefont{Soda}},
  \bibinfo{journal}{Phys. Rev.} \textbf{\bibinfo{volume}{D76}},
  \bibinfo{pages}{083517} (\bibinfo{year}{2007}), \eprint{0706.1604}.

\bibitem[{\citenamefont{Pietroni}(2008)}]{Pietroni:2008jx}
\bibinfo{author}{\bibfnamefont{M.}~\bibnamefont{Pietroni}},
  \bibinfo{journal}{JCAP} \textbf{\bibinfo{volume}{0810}}, \bibinfo{pages}{036}
  (\bibinfo{year}{2008}), \eprint{0806.0971}.

\bibitem[{\citenamefont{Matarrese and Pietroni}(2007)}]{Matarrese:2007wc}
\bibinfo{author}{\bibfnamefont{S.}~\bibnamefont{Matarrese}} \bibnamefont{and}
  \bibinfo{author}{\bibfnamefont{M.}~\bibnamefont{Pietroni}},
  \bibinfo{journal}{JCAP} \textbf{\bibinfo{volume}{0706}}, \bibinfo{pages}{026}
  (\bibinfo{year}{2007}), \eprint{astro-ph/0703563}.

\bibitem[{\citenamefont{Valageas}(2004)}]{Valageas:2003gm}
\bibinfo{author}{\bibfnamefont{P.}~\bibnamefont{Valageas}},
  \bibinfo{journal}{Astron. Astrophys.} \textbf{\bibinfo{volume}{421}},
  \bibinfo{pages}{23} (\bibinfo{year}{2004}), \eprint{astro-ph/0307008}.

\bibitem[{\citenamefont{Valageas}(2007)}]{Valageas:2006bi}
\bibinfo{author}{\bibfnamefont{P.}~\bibnamefont{Valageas}},
  \bibinfo{journal}{Astron. Astrophys.} \textbf{\bibinfo{volume}{465}},
  \bibinfo{pages}{725} (\bibinfo{year}{2007}), \eprint{astro-ph/0611849}.

\bibitem[{\citenamefont{Taruya and Hiramatsu}(2008)}]{Taruya:2007xy}
\bibinfo{author}{\bibfnamefont{A.}~\bibnamefont{Taruya}} \bibnamefont{and}
  \bibinfo{author}{\bibfnamefont{T.}~\bibnamefont{Hiramatsu}},
  \bibinfo{journal}{Astrophys.J.} \textbf{\bibinfo{volume}{674}},
  \bibinfo{pages}{617} (\bibinfo{year}{2008}), \eprint{0708.1367}.

\bibitem[{\citenamefont{Leslie}(Clarendon Press, Oxford, 1973)}]{Leslie1973}
\bibinfo{author}{\bibfnamefont{D.}~\bibnamefont{Leslie}},
  \bibinfo{journal}{{\it Developments in the Theory of Turbulence}}
  (\bibinfo{year}{Clarendon Press, Oxford, 1973}).

\bibitem[{\citenamefont{Nishimichi et~al.}(2009)}]{Nishimichi:2008ry}
\bibinfo{author}{\bibfnamefont{T.}~\bibnamefont{Nishimichi}}
  \bibnamefont{et~al.}, \bibinfo{journal}{Publ. Astron. Soc. Jap.}
  \textbf{\bibinfo{volume}{61}}, \bibinfo{pages}{321} (\bibinfo{year}{2009}),
  \eprint{0810.0813}.

\bibitem[{\citenamefont{Carlson et~al.}(2009)\citenamefont{Carlson, White, and
  Padmanabhan}}]{Carlson:2009it}
\bibinfo{author}{\bibfnamefont{J.}~\bibnamefont{Carlson}},
  \bibinfo{author}{\bibfnamefont{M.}~\bibnamefont{White}}, \bibnamefont{and}
  \bibinfo{author}{\bibfnamefont{N.}~\bibnamefont{Padmanabhan}}
  (\bibinfo{year}{2009}), \eprint{0905.0479}.

\bibitem[{\citenamefont{Bernardeau et~al.}(2002)\citenamefont{Bernardeau,
  Colombi, Gaztanaga, and Scoccimarro}}]{Bernardeau:2001qr}
\bibinfo{author}{\bibfnamefont{F.}~\bibnamefont{Bernardeau}},
  \bibinfo{author}{\bibfnamefont{S.}~\bibnamefont{Colombi}},
  \bibinfo{author}{\bibfnamefont{E.}~\bibnamefont{Gaztanaga}},
  \bibnamefont{and}
  \bibinfo{author}{\bibfnamefont{R.}~\bibnamefont{Scoccimarro}},
  \bibinfo{journal}{Phys. Rept.} \textbf{\bibinfo{volume}{367}},
  \bibinfo{pages}{1} (\bibinfo{year}{2002}), \eprint{astro-ph/0112551}.

\bibitem[{\citenamefont{Hiramatsu and Taruya}(2009)}]{Hiramatsu:2009ki}
\bibinfo{author}{\bibfnamefont{T.}~\bibnamefont{Hiramatsu}} \bibnamefont{and}
  \bibinfo{author}{\bibfnamefont{A.}~\bibnamefont{Taruya}}
  (\bibinfo{year}{2009}), \eprint{0902.3772}.

\bibitem[{\citenamefont{Koyama et~al.}(2009)\citenamefont{Koyama, Taruya, and
  Hiramatsu}}]{Koyama:2009me}
\bibinfo{author}{\bibfnamefont{K.}~\bibnamefont{Koyama}},
  \bibinfo{author}{\bibfnamefont{A.}~\bibnamefont{Taruya}}, \bibnamefont{and}
  \bibinfo{author}{\bibfnamefont{T.}~\bibnamefont{Hiramatsu}}
  (\bibinfo{year}{2009}), \eprint{0902.0618}.

\bibitem[{\citenamefont{Smith et~al.}(2003)}]{Smith:2002dz}
\bibinfo{author}{\bibfnamefont{R.~E.} \bibnamefont{Smith}} \bibnamefont{et~al.}
  (\bibinfo{collaboration}{The Virgo Consortium}), \bibinfo{journal}{Mon. Not.
  Roy. Astron. Soc.} \textbf{\bibinfo{volume}{341}}, \bibinfo{pages}{1311}
  (\bibinfo{year}{2003}), \eprint{astro-ph/0207664}.

\bibitem[{\citenamefont{Hahn}(2005)}]{Hahn:2004fe}
\bibinfo{author}{\bibfnamefont{T.}~\bibnamefont{Hahn}},
  \bibinfo{journal}{Comput. Phys. Commun.} \textbf{\bibinfo{volume}{168}},
  \bibinfo{pages}{78} (\bibinfo{year}{2005}), \eprint{hep-ph/0404043}.

\bibitem[{\citenamefont{Springel}(2005)}]{Springel:2005mi}
\bibinfo{author}{\bibfnamefont{V.}~\bibnamefont{Springel}},
  \bibinfo{journal}{Mon. Not. Roy. Astron. Soc.}
  \textbf{\bibinfo{volume}{364}}, \bibinfo{pages}{1105} (\bibinfo{year}{2005}),
  \eprint{astro-ph/0505010}.

\bibitem[{\citenamefont{Crocce et~al.}(2006)\citenamefont{Crocce, Pueblas, and
  Scoccimarro}}]{Crocce:2006ve}
\bibinfo{author}{\bibfnamefont{M.}~\bibnamefont{Crocce}},
  \bibinfo{author}{\bibfnamefont{S.}~\bibnamefont{Pueblas}}, \bibnamefont{and}
  \bibinfo{author}{\bibfnamefont{R.}~\bibnamefont{Scoccimarro}},
  \bibinfo{journal}{Mon. Not. Roy. Astron. Soc.}
  \textbf{\bibinfo{volume}{373}}, \bibinfo{pages}{369} (\bibinfo{year}{2006}),
  \eprint{astro-ph/0606505}.

\bibitem[{\citenamefont{Lewis et~al.}(2000)\citenamefont{Lewis, Challinor, and
  Lasenby}}]{Lewis:1999bs}
\bibinfo{author}{\bibfnamefont{A.}~\bibnamefont{Lewis}},
  \bibinfo{author}{\bibfnamefont{A.}~\bibnamefont{Challinor}},
  \bibnamefont{and} \bibinfo{author}{\bibfnamefont{A.}~\bibnamefont{Lasenby}},
  \bibinfo{journal}{Astrophys. J.} \textbf{\bibinfo{volume}{538}},
  \bibinfo{pages}{473} (\bibinfo{year}{2000}), \eprint{astro-ph/9911177}.

\bibitem[{\citenamefont{Takahashi et~al.}(2008)}]{Takahashi:2008wn}
\bibinfo{author}{\bibfnamefont{R.}~\bibnamefont{Takahashi}}
  \bibnamefont{et~al.}, \bibinfo{journal}{Mon. Not. Roy. Astron. Soc.}
  \textbf{\bibinfo{volume}{389}}, \bibinfo{pages}{1675} (\bibinfo{year}{2008}),
  \eprint{0802.1808}.

\bibitem[{\citenamefont{Eisenstein et~al.}(2007)\citenamefont{Eisenstein, Seo,
  and White}}]{Eisenstein:2006nj}
\bibinfo{author}{\bibfnamefont{D.~J.} \bibnamefont{Eisenstein}},
  \bibinfo{author}{\bibfnamefont{H.-j.} \bibnamefont{Seo}}, \bibnamefont{and}
  \bibinfo{author}{\bibfnamefont{.}~\bibnamefont{White},
  \bibfnamefont{Martin~J.}}, \bibinfo{journal}{Astrophys. J.}
  \textbf{\bibinfo{volume}{664}}, \bibinfo{pages}{660} (\bibinfo{year}{2007}),
  \eprint{astro-ph/0604361}.

\bibitem[{\citenamefont{Smith et~al.}(2008)\citenamefont{Smith, Scoccimarro,
  and Sheth}}]{Smith:2007gi}
\bibinfo{author}{\bibfnamefont{R.~E.} \bibnamefont{Smith}},
  \bibinfo{author}{\bibfnamefont{R.}~\bibnamefont{Scoccimarro}},
  \bibnamefont{and} \bibinfo{author}{\bibfnamefont{R.~K.} \bibnamefont{Sheth}},
  \bibinfo{journal}{Phys. Rev.} \textbf{\bibinfo{volume}{D77}},
  \bibinfo{pages}{043525} (\bibinfo{year}{2008}), \eprint{astro-ph/0703620}.

\bibitem[{\citenamefont{Hamilton}(1997)}]{Hamilton:1997zq}
\bibinfo{author}{\bibfnamefont{A.~J.~S.} \bibnamefont{Hamilton}}
  (\bibinfo{year}{1997}), \eprint{astro-ph/9708102}.

\bibitem[{\citenamefont{Kaiser}(1987)}]{Kaiser:1987qv}
\bibinfo{author}{\bibfnamefont{N.}~\bibnamefont{Kaiser}},
  \bibinfo{journal}{Mon. Not. Roy. Astron. Soc.}
  \textbf{\bibinfo{volume}{227}}, \bibinfo{pages}{1} (\bibinfo{year}{1987}).

\bibitem[{\citenamefont{Peacock and Dodds}(1994)}]{Peacock:1993xg}
\bibinfo{author}{\bibfnamefont{J.~A.} \bibnamefont{Peacock}} \bibnamefont{and}
  \bibinfo{author}{\bibfnamefont{S.~J.} \bibnamefont{Dodds}},
  \bibinfo{journal}{Mon. Not. Roy. Astron. Soc.}
  \textbf{\bibinfo{volume}{267}}, \bibinfo{pages}{1020} (\bibinfo{year}{1994}),
  \eprint{astro-ph/9311057}.

\bibitem[{\citenamefont{Cole et~al.}(1994)\citenamefont{Cole, Fisher, and
  Weinberg}}]{Cole:1993kh}
\bibinfo{author}{\bibfnamefont{S.}~\bibnamefont{Cole}},
  \bibinfo{author}{\bibfnamefont{K.~B.} \bibnamefont{Fisher}},
  \bibnamefont{and} \bibinfo{author}{\bibfnamefont{D.~H.}
  \bibnamefont{Weinberg}}, \bibinfo{journal}{Mon. Not. Roy. Astron. Soc.}
  \textbf{\bibinfo{volume}{267}}, \bibinfo{pages}{785} (\bibinfo{year}{1994}),
  \eprint{astro-ph/9308003}.

\bibitem[{\citenamefont{Park et~al.}(1994)\citenamefont{Park, Vogeley, Geller,
  and Huchra}}]{Park:1994fa}
\bibinfo{author}{\bibfnamefont{C.}~\bibnamefont{Park}},
  \bibinfo{author}{\bibfnamefont{M.~S.} \bibnamefont{Vogeley}},
  \bibinfo{author}{\bibfnamefont{M.~J.} \bibnamefont{Geller}},
  \bibnamefont{and} \bibinfo{author}{\bibfnamefont{J.~P.}
  \bibnamefont{Huchra}}, \bibinfo{journal}{Astrophys. J.}
  \textbf{\bibinfo{volume}{431}}, \bibinfo{pages}{569} (\bibinfo{year}{1994}).

\bibitem[{\citenamefont{Ballinger et~al.}(1996)\citenamefont{Ballinger,
  Peacock, and Heavens}}]{Ballinger:1996cd}
\bibinfo{author}{\bibfnamefont{W.~E.} \bibnamefont{Ballinger}},
  \bibinfo{author}{\bibfnamefont{J.~A.} \bibnamefont{Peacock}},
  \bibnamefont{and} \bibinfo{author}{\bibfnamefont{A.~F.}
  \bibnamefont{Heavens}}, \bibinfo{journal}{Mon. Not. Roy. Astron. Soc.}
  \textbf{\bibinfo{volume}{282}}, \bibinfo{pages}{877} (\bibinfo{year}{1996}),
  \eprint{astro-ph/9605017}.

\bibitem[{\citenamefont{Scoccimarro}(2004)}]{Scoccimarro:2004tg}
\bibinfo{author}{\bibfnamefont{R.}~\bibnamefont{Scoccimarro}},
  \bibinfo{journal}{Phys. Rev.} \textbf{\bibinfo{volume}{D70}},
  \bibinfo{pages}{083007} (\bibinfo{year}{2004}), \eprint{astro-ph/0407214}.

\bibitem[{\citenamefont{Percival and White}(2009)}]{Percival:2008sh}
\bibinfo{author}{\bibfnamefont{W.~J.} \bibnamefont{Percival}} \bibnamefont{and}
  \bibinfo{author}{\bibfnamefont{M.}~\bibnamefont{White}},
  \bibinfo{journal}{Mon. Not. Roy. Astron. Soc.}
  \textbf{\bibinfo{volume}{393}}, \bibinfo{pages}{297} (\bibinfo{year}{2009}),
  \eprint{0808.0003}.

\bibitem[{\citenamefont{Shaw and Lewis}(2008)}]{:2008yg}
\bibinfo{author}{\bibfnamefont{J.~R.} \bibnamefont{Shaw}} \bibnamefont{and}
  \bibinfo{author}{\bibfnamefont{A.}~\bibnamefont{Lewis}},
  \bibinfo{journal}{Phys. Rev.} \textbf{\bibinfo{volume}{D78}},
  \bibinfo{pages}{103512} (\bibinfo{year}{2008}), \eprint{0808.1724}.

\bibitem[{\citenamefont{Meiksin and White}(1999)}]{Meiksin:1998mu}
\bibinfo{author}{\bibfnamefont{A.}~\bibnamefont{Meiksin}} \bibnamefont{and}
  \bibinfo{author}{\bibfnamefont{M.~J.} \bibnamefont{White}},
  \bibinfo{journal}{Mon. Not. Roy. Astron. Soc.}
  \textbf{\bibinfo{volume}{308}}, \bibinfo{pages}{1179} (\bibinfo{year}{1999}),
  \eprint{astro-ph/9812129}.

\bibitem[{\citenamefont{Scoccimarro et~al.}(1999)\citenamefont{Scoccimarro,
  Zaldarriaga, and Hui}}]{Scoccimarro:1999kp}
\bibinfo{author}{\bibfnamefont{R.}~\bibnamefont{Scoccimarro}},
  \bibinfo{author}{\bibfnamefont{M.}~\bibnamefont{Zaldarriaga}},
  \bibnamefont{and} \bibinfo{author}{\bibfnamefont{L.}~\bibnamefont{Hui}},
  \bibinfo{journal}{Astrophys. J.} \textbf{\bibinfo{volume}{527}},
  \bibinfo{pages}{1} (\bibinfo{year}{1999}), \eprint{astro-ph/9901099}.

\bibitem[{\citenamefont{Takahashi et~al.}(2009)}]{Takahashi:2009bq}
\bibinfo{author}{\bibfnamefont{R.}~\bibnamefont{Takahashi}}
  \bibnamefont{et~al.}, \bibinfo{journal}{Astrophys. J.} p. \bibinfo{pages}{in
  press} (\bibinfo{year}{2009}), \eprint{0902.0371}.

\bibitem[{\citenamefont{Cohn}(2006)}]{Cohn:2005ex}
\bibinfo{author}{\bibfnamefont{J.~D.} \bibnamefont{Cohn}},
  \bibinfo{journal}{New Astron.} \textbf{\bibinfo{volume}{11}},
  \bibinfo{pages}{226} (\bibinfo{year}{2006}), \eprint{astro-ph/0503285}.

\bibitem[{\citenamefont{Bernstein}(1994)}]{Bernstein:1993nb}
\bibinfo{author}{\bibfnamefont{G.~M.} \bibnamefont{Bernstein}},
  \bibinfo{journal}{Astrophys. J.} \textbf{\bibinfo{volume}{424}},
  \bibinfo{pages}{569} (\bibinfo{year}{1994}).

\bibitem[{\citenamefont{Sanchez et~al.}(2008)\citenamefont{Sanchez, Baugh, and
  Angulo}}]{Sanchez:2008iw}
\bibinfo{author}{\bibfnamefont{A.~G.} \bibnamefont{Sanchez}},
  \bibinfo{author}{\bibfnamefont{C.~M.} \bibnamefont{Baugh}}, \bibnamefont{and}
  \bibinfo{author}{\bibfnamefont{R.}~\bibnamefont{Angulo}},
  \bibinfo{journal}{Mon. Not. Roy. Astron. Soc.}
  \textbf{\bibinfo{volume}{390}}, \bibinfo{pages}{1470} (\bibinfo{year}{2008}),
  \eprint{0804.0233}.

\bibitem[{\citenamefont{Heavens et~al.}(1998)\citenamefont{Heavens, Matarrese,
  and Verde}}]{Heavens:1998es}
\bibinfo{author}{\bibfnamefont{A.~F.} \bibnamefont{Heavens}},
  \bibinfo{author}{\bibfnamefont{S.}~\bibnamefont{Matarrese}},
  \bibnamefont{and} \bibinfo{author}{\bibfnamefont{L.}~\bibnamefont{Verde}},
  \bibinfo{journal}{Mon. Not. Roy. Astron. Soc.}
  \textbf{\bibinfo{volume}{301}}, \bibinfo{pages}{797} (\bibinfo{year}{1998}),
  \eprint{astro-ph/9808016}.

\bibitem[{\citenamefont{Jeong and Komatsu}(2009)}]{Jeong:2008rj}
\bibinfo{author}{\bibfnamefont{D.}~\bibnamefont{Jeong}} \bibnamefont{and}
  \bibinfo{author}{\bibfnamefont{E.}~\bibnamefont{Komatsu}},
  \bibinfo{journal}{Astrophys. J.} \textbf{\bibinfo{volume}{691}},
  \bibinfo{pages}{569} (\bibinfo{year}{2009}), \eprint{0805.2632}.

\bibitem[{\citenamefont{Smith et~al.}(2007)\citenamefont{Smith, Scoccimarro,
  and Sheth}}]{Smith:2006ne}
\bibinfo{author}{\bibfnamefont{R.~E.} \bibnamefont{Smith}},
  \bibinfo{author}{\bibfnamefont{R.}~\bibnamefont{Scoccimarro}},
  \bibnamefont{and} \bibinfo{author}{\bibfnamefont{R.~K.} \bibnamefont{Sheth}},
  \bibinfo{journal}{Phys. Rev.} \textbf{\bibinfo{volume}{D75}},
  \bibinfo{pages}{063512} (\bibinfo{year}{2007}), \eprint{astro-ph/0609547}.

\bibitem[{\citenamefont{McDonald}(2006)}]{McDonald:2006mx}
\bibinfo{author}{\bibfnamefont{P.}~\bibnamefont{McDonald}},
  \bibinfo{journal}{Phys. Rev.} \textbf{\bibinfo{volume}{D74}},
  \bibinfo{pages}{103512} (\bibinfo{year}{2006}), \eprint{astro-ph/0609413}.

\bibitem[{\citenamefont{McDonald and Roy}(2009)}]{McDonald:2009dh}
\bibinfo{author}{\bibfnamefont{P.}~\bibnamefont{McDonald}} \bibnamefont{and}
  \bibinfo{author}{\bibfnamefont{A.}~\bibnamefont{Roy}} (\bibinfo{year}{2009}),
  \eprint{0902.0991}.

\bibitem[{\citenamefont{Taruya}(1999)}]{Taruya:1999vq}
\bibinfo{author}{\bibfnamefont{A.}~\bibnamefont{Taruya}},
  \bibinfo{journal}{Astrophys. J.} \textbf{\bibinfo{volume}{537}},
  \bibinfo{pages}{37} (\bibinfo{year}{1999}), \eprint{astro-ph/9909124}.

\bibitem[{\citenamefont{Heitmann et~al.}(2006)\citenamefont{Heitmann, Higdon,
  Nakhleh, and Habib}}]{Heitmann:2006hr}
\bibinfo{author}{\bibfnamefont{K.}~\bibnamefont{Heitmann}},
  \bibinfo{author}{\bibfnamefont{D.}~\bibnamefont{Higdon}},
  \bibinfo{author}{\bibfnamefont{C.}~\bibnamefont{Nakhleh}}, \bibnamefont{and}
  \bibinfo{author}{\bibfnamefont{S.}~\bibnamefont{Habib}},
  \bibinfo{journal}{Astrophys. J.} \textbf{\bibinfo{volume}{646}},
  \bibinfo{pages}{L1} (\bibinfo{year}{2006}), \eprint{astro-ph/0606154}.

\bibitem[{\citenamefont{Heitmann et~al.}(2009)}]{Heitmann:2009cu}
\bibinfo{author}{\bibfnamefont{K.}~\bibnamefont{Heitmann}} \bibnamefont{et~al.}
  (\bibinfo{year}{2009}), \eprint{0902.0429}.

\bibitem[{\citenamefont{Habib et~al.}(2007)\citenamefont{Habib, Heitmann,
  Higdon, Nakhleh, and Williams}}]{Habib:2007ca}
\bibinfo{author}{\bibfnamefont{S.}~\bibnamefont{Habib}},
  \bibinfo{author}{\bibfnamefont{K.}~\bibnamefont{Heitmann}},
  \bibinfo{author}{\bibfnamefont{D.}~\bibnamefont{Higdon}},
  \bibinfo{author}{\bibfnamefont{C.}~\bibnamefont{Nakhleh}}, \bibnamefont{and}
  \bibinfo{author}{\bibfnamefont{B.}~\bibnamefont{Williams}},
  \bibinfo{journal}{Phys. Rev.} \textbf{\bibinfo{volume}{D76}},
  \bibinfo{pages}{083503} (\bibinfo{year}{2007}), \eprint{astro-ph/0702348}.

\bibitem[{\citenamefont{Goroff et~al.}(1986)\citenamefont{Goroff, Grinstein,
  Rey, and Wise}}]{Goroff:1986ep}
\bibinfo{author}{\bibfnamefont{M.~H.} \bibnamefont{Goroff}},
  \bibinfo{author}{\bibfnamefont{B.}~\bibnamefont{Grinstein}},
  \bibinfo{author}{\bibfnamefont{S.~J.} \bibnamefont{Rey}}, \bibnamefont{and}
  \bibinfo{author}{\bibfnamefont{M.~B.} \bibnamefont{Wise}},
  \bibinfo{journal}{Astrophys. J.} \textbf{\bibinfo{volume}{311}},
  \bibinfo{pages}{6} (\bibinfo{year}{1986}).

\bibitem[{\citenamefont{Nishimichi et~al.}(2007)}]{Nishimichi:2007xt}
\bibinfo{author}{\bibfnamefont{T.}~\bibnamefont{Nishimichi}}
  \bibnamefont{et~al.}, \bibinfo{journal}{Publ. Astron. Soc. Jap.}
  \textbf{\bibinfo{volume}{59}}, \bibinfo{pages}{1049} (\bibinfo{year}{2007}),
  \eprint{0705.1589}.

\bibitem[{\citenamefont{Fry}(1994)}]{Fry:1993bj}
\bibinfo{author}{\bibfnamefont{J.~N.} \bibnamefont{Fry}},
  \bibinfo{journal}{Astrophys. J.} \textbf{\bibinfo{volume}{421}},
  \bibinfo{pages}{21} (\bibinfo{year}{1994}).

\bibitem[{\citenamefont{Suto and Sasaki}(1991)}]{Suto:1990wf}
\bibinfo{author}{\bibfnamefont{Y.}~\bibnamefont{Suto}} \bibnamefont{and}
  \bibinfo{author}{\bibfnamefont{M.}~\bibnamefont{Sasaki}},
  \bibinfo{journal}{Phys. Rev. Lett.} \textbf{\bibinfo{volume}{66}},
  \bibinfo{pages}{264} (\bibinfo{year}{1991}).

\bibitem[{\citenamefont{Makino et~al.}(1992)\citenamefont{Makino, Sasaki, and
  Suto}}]{Makino:1991rp}
\bibinfo{author}{\bibfnamefont{N.}~\bibnamefont{Makino}},
  \bibinfo{author}{\bibfnamefont{M.}~\bibnamefont{Sasaki}}, \bibnamefont{and}
  \bibinfo{author}{\bibfnamefont{Y.}~\bibnamefont{Suto}},
  \bibinfo{journal}{Phys. Rev.} \textbf{\bibinfo{volume}{D46}},
  \bibinfo{pages}{585} (\bibinfo{year}{1992}).

\bibitem[{\citenamefont{Scoccimarro and Frieman}(1996)}]{Scoccimarro:1996se}
\bibinfo{author}{\bibfnamefont{R.}~\bibnamefont{Scoccimarro}} \bibnamefont{and}
  \bibinfo{author}{\bibfnamefont{J.}~\bibnamefont{Frieman}},
  \bibinfo{journal}{Astrophys. J.} \textbf{\bibinfo{volume}{473}},
  \bibinfo{pages}{620} (\bibinfo{year}{1996}), \eprint{astro-ph/9602070}.

\bibitem[{\citenamefont{Proudman and Reid}(1954)}]{Proudman:1954}
\bibinfo{author}{\bibfnamefont{I.}~\bibnamefont{Proudman}} \bibnamefont{and}
  \bibinfo{author}{\bibfnamefont{W.~H.} \bibnamefont{Reid}},
  \bibinfo{journal}{Phil. Trans. Roy. Soc. London. Ser.}
  \textbf{\bibinfo{volume}{A247}}, \bibinfo{pages}{163–189}
  (\bibinfo{year}{1954}).

\bibitem[{\citenamefont{Tatsumi}(1957)}]{Tatsumi:1957}
\bibinfo{author}{\bibfnamefont{T.}~\bibnamefont{Tatsumi}},
  \bibinfo{journal}{Proc. Roy. Soc. London. Ser.}
  \textbf{\bibinfo{volume}{A239}}, \bibinfo{pages}{16} (\bibinfo{year}{1957}).

\bibitem[{\citenamefont{Lesgourgues et~al.}(2009)\citenamefont{Lesgourgues,
  Matarrese, Pietroni, and Riotto}}]{Lesgourgues:2009am}
\bibinfo{author}{\bibfnamefont{J.}~\bibnamefont{Lesgourgues}},
  \bibinfo{author}{\bibfnamefont{S.}~\bibnamefont{Matarrese}},
  \bibinfo{author}{\bibfnamefont{M.}~\bibnamefont{Pietroni}}, \bibnamefont{and}
  \bibinfo{author}{\bibfnamefont{A.}~\bibnamefont{Riotto}}
  (\bibinfo{year}{2009}), \eprint{0901.4550}.

\bibitem[{\citenamefont{Barriga and Gaztanaga}(2002)}]{Barriga:2001wn}
\bibinfo{author}{\bibfnamefont{J.}~\bibnamefont{Barriga}} \bibnamefont{and}
  \bibinfo{author}{\bibfnamefont{E.}~\bibnamefont{Gaztanaga}},
  \bibinfo{journal}{Mon. Not. Roy. Astron. Soc.}
  \textbf{\bibinfo{volume}{333}}, \bibinfo{pages}{443} (\bibinfo{year}{2002}),
  \eprint{astro-ph/0112278}.

\end{thebibliography}
% \bibliographystyle{apsrev}

%%%%%%%%%%%%%%%%%%%%%%%%%%%%%%%%%%%%%%%%%%%%%%%%%%%%%%%
\end{document}